\documentclass[12pt, preprint]{aastex}

\usepackage{graphicx}
\usepackage{amssymb}
\usepackage{amsmath}
\usepackage{natbib}
\usepackage{mathrsfs}
\usepackage{rotating}
\usepackage{color}

\shorttitle{HCO$^+$ Detection of Dust-Depleted Gas in the Cavity of the LkCa~15 Pre-Transitional Disk}
\shortauthors{Drabek-Maunder et al.}

\begin{document}

\def\dgr{$d$:$g$}
\def\msun{$M_{\odot}$}

\title{HCO$^+$ Detection of Dust-Depleted Gas in the Inner Hole \\
    of the LkCa~15 Pre-Transitional Disk}

    \author{E. Drabek-Maunder\altaffilmark{1}, S. Mohanty\altaffilmark{1}, J. Greaves\altaffilmark{2}, I. Kamp\altaffilmark{3}, R. Meijerink\altaffilmark{4}, M. Spaans\altaffilmark{3}, W.-F. Thi\altaffilmark{5}, P. Woitke\altaffilmark{6}}
\altaffiltext{1}{Imperial College London, Blackett Lab., Prince Consort Rd, London SW7 2AZ, UK. e.drabek-maunder@imperial.ac.uk}
\altaffiltext{2}{School of Physics and Astronomy, Cardiff University, Cardiff CF24 3AA}
\altaffiltext{3}{Kapteyn Institute, PO Box 800, 9700 AV Groningen, The Netherlands}
\altaffiltext{4}{Leiden Observatory, Leiden University, PO Box, 2300 RA Leiden, The Netherlands}
\altaffiltext{5}{Max-Planck-Institut fur extraterrestrische Physisk, Giessenbachstrasse 1, 85748 Garching, Germany}
\altaffiltext{6}{St. Andrews University, School of Physics and Astronomy, St. Andrews KY16 9SS, UK}

\begin{abstract}

LkCa~15 is an extensively studied star in the Taurus region known for its pre-transitional disk with a large inner cavity in dust continuum and normal gas accretion rate.  The most popular hypothesis to explain the LkCa~15 data invokes one or more planets to carve out the inner cavity, while gas continues to flow across the gap from the outer disk onto the central star.  %Therefore, the absence of dust emission in the cavity can occur from dust filtration at the gap outer boundary or from accreting material funnelled into `streamers' with small areal coverage.  
We present spatially unresolved HCO$^+$~$J = 4\rightarrow3$ observations of the LkCa\,15 disk from the JCMT and model the data with the \textsc{ProDiMo} code.  We find that: {\it (1)} HCO$^+$ line-wings are clearly detected, certifying the presence of gas in the cavity within $\lesssim$\,50\,AU of the star.  {\it (2)} Reproducing the observed line-wing flux requires both a significant suppression of cavity dust (by a factor $\gtrsim10^4$ compared to the ISM) and a substantial increase in the gas scale-height within the cavity ($H_0/R_0 \sim 0.6$). An ISM dust-to-gas ratio (d:g$=$10$^{-2}$) yields too little line-wing flux regardless of the scale-height or cavity gas geometry, while a smaller scale-height also under-predicts the flux even with a reduced d:g.  {\it (3)} The cavity gas mass is consistent with the surface density profile of the outer disk extended inwards to the sublimation radius (corresponding to mass $M_d \sim 0.03$\,\msun), and masses lower by a factor $\gtrsim$10 appear to be ruled out.

%These results suggest the gas should be hotter (and more puffed-up) due to reduced cooling by dust. {\it (3)} The cavity gas mass is consistent with a standard disk surface density (with corresponding mass $M_d \sim 0.03$\,\msun), and masses lower by a factor $\gtrsim$10 appear to be ruled out.  %If planets sculpt the dust cavity, then they do not diminish the LkCa~15 cavity mass.  

\end{abstract}




\section{Introduction}
\label{intro}

The early stages of planet formation are thought to begin in the disks rotating around T Tauri and Herbig stars.  However, the processes dust and gas undergo to form planetary systems are not well understood.  The most natural candidates for sites of planetary formation are transitional disks.  A transitional disk is defined as a primordial or `protoplanetary' disk with little to no near-infrared (near-IR) and mid-infrared (mid-IR) emission in the disk SED and strong dust continuum emission at wavelengths $\geq 10$~$\micron$, where the disk has an inner hole of (presumed) dust depletion.  Similarly, `pre-transitional' disks have an optically thick inner disk that is separated from the optically thick outer disk by an optically thin gap or cavity in dust continuum emission.  Pre-transitional disk SEDs have been fit with evidence to support near-IR dust emission from the thick inner disk in combination with a reduction in mid-IR \citep{2005ApJ...630L.185C, 2010ApJ...717..441E}.

While observational techniques (e.g. interferometry) can be used to resolve transitional disks, molecular line spectroscopy can investigate other disk properties.  Line profiles of Keplerian disks have double-peaked emission due to rotation and higher velocity line-wings that trace the inner disk radii.  Fitting disk models to molecular line profiles can place constraints on the disk mass, extent (radii), inclination and other properties of the disk (e.g. \citealt{2010A&A...518L.124M, 2010A&A...510A..18K, 2009A&A...501L...5W, 2010A&A...518L.125T, 2010A&A...518L.127M, 2011A&A...530L...2T, 2011A&A...534A..44W, 2012A&A...538A..20T}).  \citet{2004MNRAS.351L..99G} use this technique with HCO$^+$~$J=4\rightarrow3$ on six T Tauri stars with circumstellar disks, including LkCa~15.  HCO$^+$ is a useful tracer of the dense gas in disks with a high critical density $n_{crit}\sim2\times10^{6}$~cm$^{-3}$ \citep{2011piim.book.....D} and varies gradually with disk radius \citep{2002A&A...386..622A}.  Results from \citet{2004MNRAS.351L..99G} indicate a lack of HCO$^+$ line-wing emission in the LkCa~15 disk.  \citet{2004MNRAS.351L..99G} placed the outer radius of the disk gap at $\sim200$~AU, which was consistent with marginally resolved HCO$^+$~$J=1\rightarrow0$ interferometric images of LkCa 15 from \citet{2003ApJ...597..986Q}.  The dust-hole size is estimated as 58~AU from IR emission \citep{2010ApJ...717..441E}  and 50~AU from millimetre emission (\citealt{2011ApJ...732...42A}, hereafter A11).  However, there is evidence that gas is present in the dust continuum cavity from observations of $^{12}$CO and $^{13}$CO (e.g. \citealt{2007A&A...467..163P, 2015A&A...579A.106V}), where gas is found at a radii 13$\pm5$~AU and 23$\pm$8~AU for $^{12}$CO and $^{13}$CO respectively.  To better understand the mechanism behind accretion onto LkCa~15, more rigid constraints need to be placed on gas mass in the inner disk cavity using high density tracers like HCO$^+$.  Even though CO is an abundant molecule that can trace low-mass material in a disk, it often becomes optically thick at low density (i.e. $\sim1$~M$_\mathrm{Jupiter}$ of gas; \citealt{2001ApJ...561.1074T}) and is not necessarily suited to tracing the region of the disk forming Jupiter-mass exoplanets.

In this paper, HCO$^+$~$J=4\rightarrow3$ line observations are used to trace the dense gas in the LkCa~15 disk.  We use this spatially unresolved spectrum and a chemical disk model to study the properties (mass, dust-to-gas ratio or `d:g' and scale height) of the central cavity and outer disk.  This work improves on the past study from \citet{2004MNRAS.351L..99G} by observing HCO$^+$ in the LkCa~15 disk at a factor $\sim10$ deeper.  \S\ref{observations} details the observations HCO$^+$~$J=4\rightarrow3$ emission of the LkCa~15 disk.  \S\ref{modelling} describes the modelling parameters we use to fit the data, including the disk surface density, scale height and grain settling.  We present the results from the model fits in \S\ref{results}, detailing how the models are developed and improved to fit the HCO$^+$ line.  Lastly, \S\ref{discussion} discusses the results and the implications for accretion in LkCa~15.  

\section{LkCa~15 Observations}
\label{observations}

HCO$^+$~$J=4\rightarrow3$ observations (356.7343~GHz) were obtained with the Heterodyne Array Receiver Programme (HARP; \citealt{2009MNRAS.399.1026B}) at the James Clerk Maxwell Telescope (JCMT).  Observations were carried out over 8 nights in September 2011 to January 2012, totalling 10 hours in `stare' mode.  A spectrum was reduced from receptor H05.  Pointing calibrations were typically acquired after every four frames.  Frames with offsets $>6''$ between calibrations were examined to ensure the receptor was centred on the source.  The last frame before a poor pointing calibration was discarded, as are all frames preceding a poor calibration that were either excessively noisy or show no significant detection.  Lastly, both the baseline and continuum were subtracted from the final reduced spectrum.

The HCO$^+$ spectrum was initially reduced in units of antenna temperature $T_A^\ast$ versus velocity.  The spectrum was rebinned to 0.3~km~s$^{-1}$ channels with a root mean square (RMS) noise of 0.005\,K.  To compare the data to axisymmetric models, the spectrum was folded along the line centre (defined as the midpoint between the double-peaked Keplerian disk profile, $\mathrm{v_\mathrm{lc}}=6.4 \ \mathrm{km s^{-1}}$), decreasing the noise by $\sqrt{2}$ to 0.003\,K.  This is analogous to the procedure used by \citet{2004MNRAS.351L..99G}, except their data were limited to rms values $\sim10\times$ higher than ours.  The folded and unfolded spectra are compared in Fig.\,\ref{fig:folded_unfolded}.

The spectrum was then converted from temperature $T_A^\ast$ to flux density $S_{\nu}$ units using\footnote{\url{http://docs.jach.hawaii.edu/JCMT/HET/GUIDE/het\_guide.pdf}} 
\begin{equation} 
S_\nu = \frac{2 k T_A^\ast}{\eta_a a_p} = 20.4 \left(\frac{T_A^\ast}{\eta_a}\right) \,\mathrm{[Jy]}
\end{equation} 
where $k$ is the Boltzmann constant, $a_p$ the physical area of the telescope aperture and $\eta_a$ the aperture efficiency (so that $\eta_a a_p$ is the effective area).  Flux ($S_{\nu}$) is in Jansky for the JCMT beam (FWHM$\sim16''$) at 356\,GHz and $\eta_A = 0.56$ (at 345\,GHz).

\section{Modeling Technique}
\label{modelling}
\subsection{\textsc{ProDiMo} Parameters}
\label{prodimo_parameters}

HCO$^+$ is formed by ion-molecule reactions, which are directly influenced by stellar X-ray and UV luminosities, cosmic ray (cr) ionisation rates and PAH abundance.  HCO$^+$ production typically follows:  \begin{equation} \mathrm{ H_2 + cr \rightarrow H_2^+ + e^- \tag{a} }\end{equation} \begin{equation}\mathrm{H_2^+ + H_2 \rightarrow H_3^+ + H \tag{b} }\end{equation} \begin{equation}\mathrm{ H_3^+ + CO \rightarrow HCO^+ + H_2 \tag{c}}, \end{equation} 
where HCO$^+$ and CO abundances relative to H$_2$ are $X(\mathrm{HCO^+})=5\times10^{-9}$ and $X(\mathrm{CO})=10^{-4}$ respectively \citep{2001A&A...377..566V}.  HCO$^+$ recombination is primarily triggered by an increase of electrons from UV emission (i.e. photoelectric effect) and an increase in the abundance of metals (e.g. Na, Mg) which act as electron donors: \begin{equation} \mathrm{ HCO^+ + e^- \rightarrow H + CO} \tag{d}  \end{equation} \begin{equation} \mathrm{ HCO^+ + Na, Mg \rightarrow HCO + Na^+, Mg^+.} \tag{e} \end{equation}  PAHs can destroy HCO$^+$ via the reactions: \begin{equation} \mathrm{PAH^-} + \mathrm{HCO^+} \rightarrow \mathrm{PAH} + \mathrm{CO} + \mathrm{H} \tag{f} \end{equation} \begin{equation} \mathrm{PAH + HCO^+} \rightarrow \mathrm{PAH^+ + CO + H}.\tag{g}\end{equation}  Additionally in warmer temperatures, H$_2$O can be important in both the formation and destruction of HCO$^+$:   \begin{equation} \mathrm{ C^+ + H_2O \rightarrow HCO^+ + H \tag{h} }\end{equation} \begin{equation}\mathrm{H_2O + HCO^+ \rightarrow CO + H_3O^+ \tag{i} }\end{equation}

We model the LkCa\,15 HCO$^+$ data using the disk thermochemical code \textsc{ProDiMo} \citep{2009A&A...501..383W, 2010A&A...510A..18K, 2011A&A...534A..44W}, which solves for 2D dust continuum radiative transfer, gas phase- and photo-chemistry, thermal balance and hydrostatic disk structure assuming axisymmetry. We use the most extensive chemical network in \textsc{ProDiMo}, involving 13 elements, 237 species (atoms, molecules and PAHs) and over 1500 reactions.  Collisions with electrons liberated from PAHs by UV photons are one of the main sources of heating for the disk gas.  We set our model PAH abundance at 10$^{-2}$ w.r.t ISM, which is the standard assumption for T Tauri disks \citep{2006A&A...459..545G}.  The UV opacities for the disk incorporate dust, PAHs and gas, as described in Appendix \ref{disk_opacity}.  Stellar X-rays, included in the photochemistry, are assigned a luminosity of $L_X = 3\times10^{30}$\,erg\,s$^{-1}$ \citep{2013ApJ...765....3S}, with photon energies spanning 0.1--70\,keV and an emission temperature of 10$^7$\,K.  Cosmic ray ionisation rates are set to the standard value of $\zeta_{CR}=10^{-17}$~s$^{-1}$.  Finally, we set a UV excess $L_{UV}/L_\ast = 10^{-2}$ (Henning et al. 2010), where $L_{UV}$\footnote{We note the value for $L_{UV}$ from Henning et al. (2010) has been calculated using observations from 110--207~nm, which is adequate for an order of magnitude estimate for the \textsc{ProDiMo} $L_{UV}$ parameter set between 90--250~nm. } is the UV luminosity between 90-250~nm.  The stellar parameters adopted for LkCa~15 are listed in Table~\ref{table1}.  We note that all models have had the continuum subtracted from the final molecular line spectrum.

\subsection{Disk Structure}
\label{disk_structure}

The LkCa~15 disk SED has been fit in IR wavelengths \citep{2007ApJ...670L.135E,2008ApJ...682L.125E, 2010ApJ...717..441E} and a hole has been revealed in millimetre dust continuum \citep{2006A&A...460L..43P,2011ApJ...732...42A} and IR scattered light ($H$- and $K_s $-band imaging data; \citealt{2010ApJ...718L..87T}).  The disk structure is divided into three to four regions.  The innermost part comprises a small optically thick disk at radii $\sim$0.1--0.2~AU; this is surrounded by an optically thin region extending from $\sim$4~AU \citep{2008ApJ...682L.125E, 2010ApJ...717..441E} to 10~AU (A11).  From $\sim$4 (or 10) to 50~AU, there is little dust emission, indicating a gap or cavity in the disk.  The cavity is encircled by an outer optically thick disk at $\geq50$~AU (A11).  

\begin{figure}
\centering
\includegraphics[width=3in]{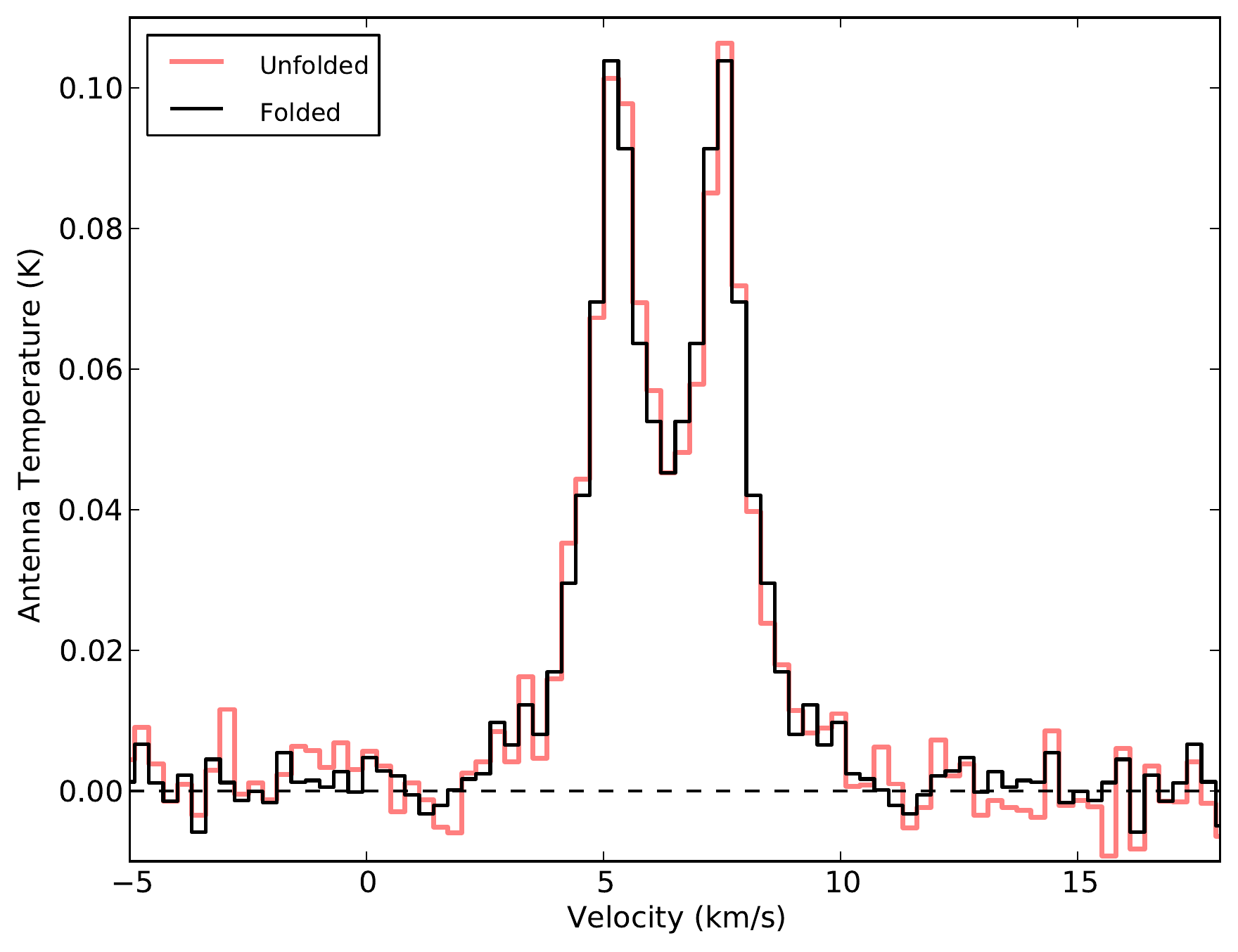}
\caption{The HCO$^+$~$J=4\rightarrow3$ spectrum initial reduction binned to 0.3~km~s$^{-1}$ velocity channels (red) and the spectrum folded symmetrically about the line centre at 6.4~km~s$^{-1}$ (black).  The unfolded spectrum has a 1$\sigma$ RMS of 0.005~K and the folded spectrum has a 1$\sigma$ RMS of 0.003~K. }
\label{fig:folded_unfolded}
\end{figure}

We initially use a benchmark disk with three radial zones to test the consistency of \textsc{ProDiMo} with A11 (\S\ref{andrews_sed}).  However, any gas within the $\sim$0.1--10~AU dusty annuli does not significantly contribute to the HCO$^+$ emission, indicated by the negligible flux at velocities $ \mathrm{v_\mathrm{lc}}\pm \gtrsim10$~km~s$^{-1}$ (Figure~\ref{fig:folded_unfolded}).  Not only will there be relatively less HCO$^+$ in this smaller region of the disk, but higher temperatures at smaller radii can cause higher H$_2$O densities that can dissociate the HCO$^+$ emission (as in \S\ref{prodimo_parameters}).  Therefore from Section~\ref{outer_only_section} on, we adopt a simplified two-component model for the LkCa~15 disk:  an inner region extending from the dust sublimation radius $R_\mathrm{sub}$ to the outer edge of the cavity ($\sim$0.1--50AU) and an outer disk (radii $>50$~AU; see \S\ref{dust_gas_results_same}). %A more complex geometry is explored in Section~\ref{final_model} to compare our results to those from dust continuum data (CHANGE THIS?)

We use the surface density profile described from a power-law with exponential tapering (see A11).  The mass in the disk component is defined as the radial integral of the surface density, \begin{equation} M_{disk} = \int_{R_{in}} ^{R_{out}} 2 \pi \ r  \ \Sigma(r) \  \mathrm{d}r = 2\pi R^2_c \Sigma_0 \frac{1}{2-\lambda} \left[ \exp\left( \frac{-R_{in}}{R_c}\right)^{2-\lambda} - \exp\left( \frac{-R_{out}}{R_c}\right)^{2-\lambda} \right],  \label{eq:surface_density} \end{equation}  where $R_{in}$ and $R_{out}$ are the inner and outer radii, $R_c$ is the characteristic scaling and tapering radius where the surface density decreases at radii $r>>R_c$, $\lambda$ is the power-law exponent defined as $\lambda=1$ \citep{2009ApJ...700.1502A, 2010ApJ...723.1241A} and $\Sigma_0$ is the surface density normalisation.  

A11 analysed the SED and 880~$\micron$ visibility profile of the LkCa~15 disk to constrain the outer disk properties and found $R_{in}=50$~AU and $R_c=85$~AU.  The models were insensitive to the outer radius $R_{out}$ since the dust became optically thin at $R_{out}>>R_c$.  Assuming an ISM d:g ($10^{-2}$), the normalised disk surface density at $R_c$ was $\Sigma_{0,\mathrm{85 \ AU}}=$ 10.8~g~cm$^{-2}$ for both the dust and gas, corresponding to disk mass $M_{disk}=0.055$~M$_\odot$. 

We adopt the A11 values $R_{in}=50$~AU for the outer disk, but our fits to the HCO$^+$ data indicates that $R_c$ for the gas is substantially greater than 85~AU (see \S\ref{dust_gas_results_same}).  A larger radial extent of gas relative to the dust has been noted in other transitional and pre-transitional disks (e.g. \citealt{2009A&A...501..269P, 2012ApJ...744..162A, 2013ApJ...775..136R}), and is potentially due to the effects of radial drift and viscous gas drag on grains \citep{2014ApJ...780..153B}.  As such, we keep the outer disk $R_c$ (and $R_{out}$) as a free parameter determined from fitting the HCO$^+$ line (see \S\ref{varying_rc}).  We set the surface density normalisation for any characteristic radius from Equation~\ref{eq:surface_density} to the value in A11 $\Sigma_0=\Sigma_{0,85 \ \mathrm{AU}}=10.8$~g~cm$^{-2}$.  Therefore, our surface density profile matches up smoothly with the radii from A11 (50~AU $\leq r \leq 85$~AU).  

Additionally, the vertical distribution of the gas and dust in the models is designed to match A11.  The gas scale height $H_g$ at disk radius $r$ follows the relation $H_g=H_0 \left( r/R_0\right)^{\beta},$ where $H_0$ is the reference scale height at radius $R_0$ and $\beta$ is the flaring index ($\beta=1.2$).  The dust grain distribution can be categorised into small and large grain sizes.  For grains below the minimum grain size $a_s$ ($a<a_s$), the dust is well-mixed and the scale height is equivalent to the gas ($H_d = H_g$).  Conversely for larger grains ($a>a_s$), the scale height is decreased according to the  relation $H_d^2 = H_g^2 \left( a/a_s\right)^{-\delta}$, where $\delta$ is the dust settling exponent.  A more simplified vertical dust distribution has been implemented in A11 using a reference scale height $H_0=2.9$~AU at reference radius $R_0=100$~AU for gas and small dust grains ($0.005\micron \leq a \leq 1\micron$).  Similarly, the population of large dust grains ($1\micron < a \leq1$~mm) in A11 has a scale height of 0.6~AU at the same reference radius.  

For consistency with A11, we adopt $a_s=0.1\micron$, $\delta=1.0$, a reference scale height $H_0 =10$~AU at radius $R_0=100$~AU, grain size distribution $n(a) \propto a^{-p}$ with power-law index $p=3.5$ and grain size range from 0.005$\micron$ to 1~mm.  At $1\micron$ we find the dust scale height is $H_{d,1\micron}=3.2$~AU which is comparable to the $1\micron$ boundary between small and large grain sizes from A11.  Similarly in larger grains, we find $H_{d,10\micron}=1.0$~AU, $H_{d,100\micron}=0.32$~AU and $H_{d,1\mathrm{mm}}=0.1$~AU which is comparable to the small grain scale height found in A11. 

%{\color{red}{We note A11 used an inner disk set from 0.1 to 10~AU with with $H_0=30.5$~AU at a reference radius $R_0=100$~AU, maximum grain size $a_{\mathrm{max}}=0.25\micron$ and mass depleted by 10$^{-6}$.  As explained above, we adopt a two-component model for the LkCa~15 disk and do not set an independent inner disk.}}

\subsection{Model Fitting}
\label{fitting}
Our focus in this paper is on modelling the properties of the inner disk cavity (i.e. $r<50$~AU) by fitting the HCO$^+$ high velocity line-wing emission.  Peak emission at low velocities in the HCO$^+$ profile correspond to the outer portion of the disk at radii $r>50$~AU.  Certain features of the outer disk can be difficult to model, including the radial separation between the dust and gas (see \S\ref{dust_gas_results_same} and \ref{varying_rc}) which indicates the gas extends to larger radii than the dust grains in the disk.  Future work will focus on fitting the outer disk properties to address the line flanks and dip in flux at the centre of the double-peaked Keplerian profile (\textsc{DIANA} project\footnote{\url{http://www.diana-project.com}}).

The models are compared to the folded observed profile (from \S\ref{observations}), and a reduced chi-squared ($\chi_{red}^2$) criterion is used for a comparative estimate in assessing how well the models fit.  Since the $\chi _{red}^2$ values for the full spectral profile can be biased by the line peak (which is controlled by the outer disk parameters, mainly the disk outer radius) and our primary goal is to fit the line-wings of the data corresponding to the disk cavity, we calculate $\chi_{red} ^2$ for the line-wings alone at a velocity range $\pm 2.4$--4.6~km~s$^{-1}$ relative to the line centre.  The $\chi_{red}^2$ values are calculated for 4 degrees of freedom, defined by the number of spectral channels in the line-wing of the folded spectrum (7) minus the number of free parameters (3).  Nominally, there are a total of 4 parameters for fitting the inner disk: the minimum grain size for settling $a_s$, the dust settling exponents $\delta_s$, the scale height of the cavity $H_0$ and the cavity dust-to-gas ratio (d:g).  As we will see in Section~\ref{settling_effects}, plausible variations in $a_s$ and $\delta_s$ have hardly any effect on the line-wings.  Formally, if the parameters had no effect on the line-wings, we would not consider them free parameters for the fits.  Under the circumstances, we conservatively adopt their combined effect as a single free parameter.  The probability density function (PDF) for the $\chi^2$ distribution is shown in Figure~\ref{fig:dof}.  The $\chi^2$ mean and limits equivalent to 1$\sigma$ are also shown (i.e. the probability that $\chi^2$ will surpass this limit should not be greater than $\sim 34 \%$).  Best-fit models are chosen by minimising the $\chi^2$ and $\chi_{red} ^2$ values for individual parameters.  We indicate the velocity channels corresponding to the HCO$^+$ line-wings in the figures below using grey boxes and the HCO$^+$ folded spectrum (\S\ref{observations}) is shown with $1\sigma$~r.m.s. error bars.  Model results are summarised in Table~\ref{table_param}.  

Once a final disk model is found, we also make a comparison between the final model and the line-wings in the unfolded observed HCO$^+$ profile to demonstrate the folding process has not biased the fits.  The $\chi_{red}^2$ values are calculated for 11 degrees of freedom (14 spectral channels and 3 free parameters), where the PDF for the $\chi^2$ distribution is shown in Figure~\ref{fig:dof}.

%{\color{red}{The $\chi_{red}^2$ values are calculated for 4 degrees of freedom, defined by the number of spectral channels included per half-line minus three free parameters:  the minimum grain size for settling a$_\mathrm{settle}$ and the dust settling exponents $\delta_s$ are considered one parameter along with the scale height of the cavity $H_0$ and the cavity dust-to-gas ratio.  We indicate the velocity channels corresponding to the HCO$^+$ line-wings in the figures below using grey boxes and the HCO$^+$ folded spectrum (\S\ref{observations}) is shown with $1\sigma$~RMS error bars.  Model results are summarised in Table~\ref{table_param}.}}  %The degrees of freedom is defined by the number of spectral channels minus the six free parameters:  $R_{in}$, $R_{out}$, $R_c$, $a_s$, $\delta_s$ and the dust-to-gas ratio of the inner disc.  

\section{Model Results}
\label{results}

\subsection{Disk with {\it {\bf Empty}} Inner Cavity}
\label{with_innercav}

We first investigate if the HCO$^+$ data and line-wings are consistent with a gas distribution similar to the dust distribution:  an optically thick outer disk surrounding a large, empty inner cavity.

\subsubsection{Comparison to the Dust Results of A11}
\label{dust_gas_results_same}

\subsubsubsection{SED Fitting}
\label{andrews_sed}

To test the consistency of \textsc{ProDiMo} with past dust continuum models of LkCa~15, we first use a benchmark disk model as described in A11.  In Section~\ref{disk_structure}, the A11 model structure includes three radial zones: a dust-depleted inner disk (from the sublimation radius $R_\mathrm{sub}$ to 10~AU), a cavity that is void of material from 10~AU to 50~AU and an outer disk ($r>50$~AU).  The inner disk (i.e. radii up to 10~AU) has a puffed-up inner rim with an increased scale height between $\sim0.1$ to 0.2~AU ($H_0=30.5$~AU at a reference radius 100~AU) and decreased cavity mass density by a factor of $10^{-6}$.  Additionally, settling parameters for the inner disk limited the maximum grain size to 0.25~$\micron$.  The outer disk had an inner edge at $R_{in} \approx 50$\,AU and a characteristic radius $R_c = 85$\,AU.  The optically thin sub-mm/mm-wavelength dust emission from the outer disk is insensitive to the precise disk outer radius $R_{out}$ ($> R_{c}$), so A11 do not specify this parameter.  We, therefore, initially begin with a large outer radius so that $R_{out} >> R_{c}$, where $R_{out}=1000$~AU, for fitting the dust continuum spectral energy distribution (SED) for LkCa~15 as in A11.  We test varying outer radii below once we focus on fits to the HCO$^+$ emission.  The dust-to-gas ratio in the outer disk is held fixed at the ISM-value of $10^{-2}$ adopted by A11, and the settling parameters ($a_s = 0.1\,\mu$m and $\delta = 1.0$) are set as in \S\ref{disk_structure}.

\begin{figure}
\centering
\includegraphics[width=3in]{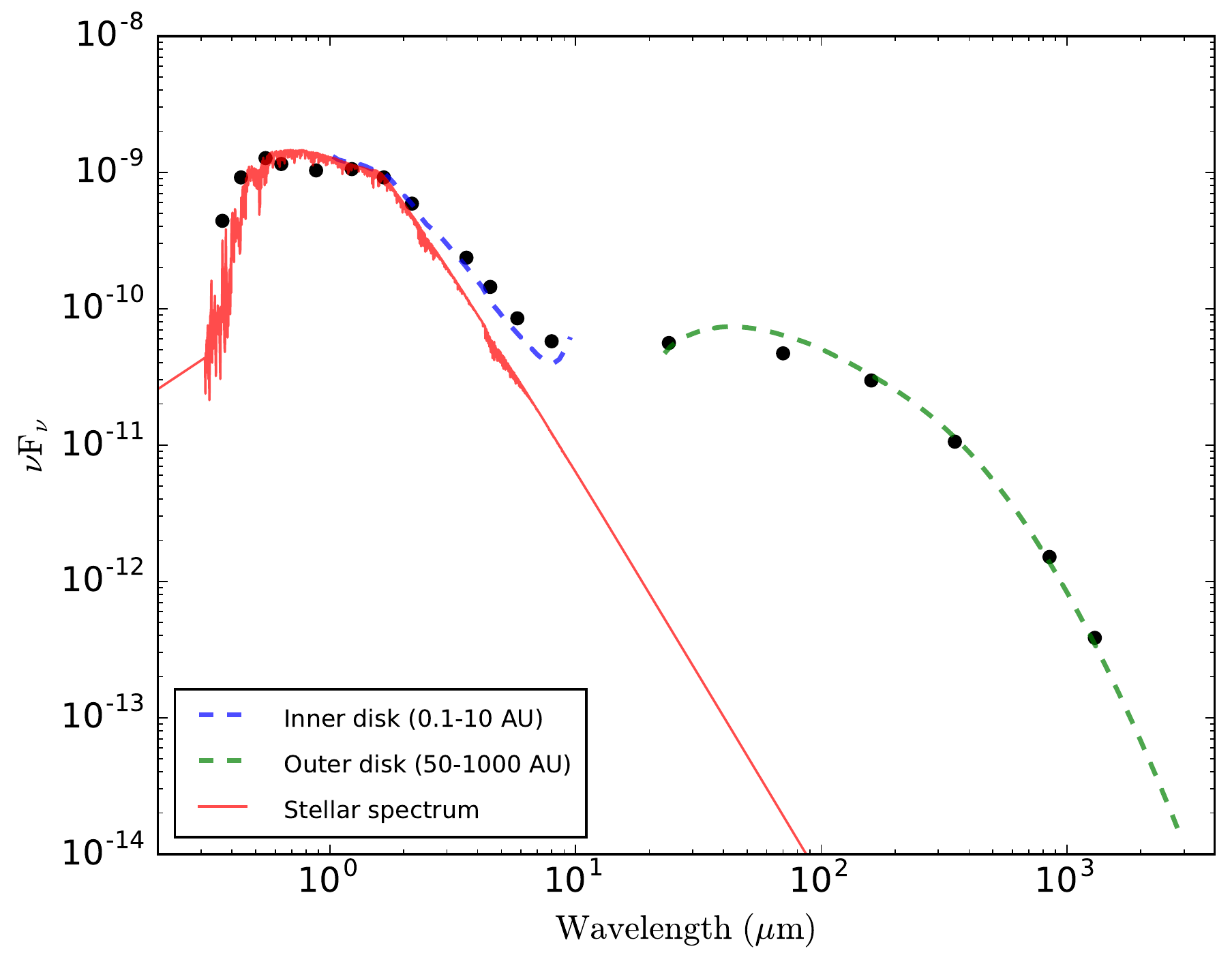}
\caption{The black points show the LkCa~15 SED taken from  \textsc{ProDiMo} benchmark with a two-component radial model, described in \S\ref{with_innercav} (from \citealt{2010ApJ...717..441E, 2007A&A...467..163P, 2009ApJ...703L.137I, 1995A&A...297..391N} in A11).  The innermost disk is set from 0.1 to 10~AU with $H_0=30.5$~AU at a reference radius $R_0=100$~AU and maximum grain size $a_{\mathrm{max}}=0.25\micron$.  The inner disk mass has been depleted by 10$^{-6}$.  The outermost disk extends from 50 to 1000~AU with $R_c=85$~AU.}
\label{fig:SED_andrews}
\end{figure}

Figure~\ref{fig:SED_andrews} shows the continuum SED produced by \textsc{ProDiMo}, where we are able to produce a good fit to observations of the LkCa~15 disk SED.  As discussed in \S\ref{disk_structure}, the innermost disk ($\sim0.1$ to 10~AU) does not contribute to the the HCO$^+$ observed from the LkCa~15 disk cavity.  In the following sections, we use a two-component disk model consisting of the cavity ($r<50$~AU) and the outermost disk ($r>50$~AU).  We reintroduce the innermost disk in \S\ref{final_model} for purposes of SED fitting in the final disk models.

\subsubsubsection{Outer Disk Only}
\label{outer_only_section}

Even though the dust continuum emission from sub-mm/mm-wavelengths is optically thin in the outer radii of the disk, the optically thick gas emission is sensitive to $R_{out}$; consequently, we now vary the outer radius to values greater than the characteristic radius $R_{out}>R_c$ to fit the HCO$^+$ emission.  We no longer set the innermost disk in the disk models since it will no contribute to the HCO$^+$ emission.  The dust-to-gas ratios are held fixed at the ISM-value (as above) and the settling parameters for the outer disk remain the same.

The results are shown in Figure~\ref{fig:tests_cavity} (top-left), with $R_{out}$ ranging from 250 to 1000\,AU (corresponding to a total disk mass of $M_d$ = 0.02--0.03\,M$_\odot$).  In all cases, the predicted HCO$^+$ line profile is much weaker than the observed emission in both the line peaks and line-wings: there is not enough emitting gas in the modelled disk to match the observed flux.  We note that increasing $R_{out}$ further does not affect this conclusion: $R_{out}\gtrsim 500$~AU has little impact on the HCO$^+$ line profile.  Thus, gas distributed in the same manner as the dust cannot explain the observed HCO$^+$ profile.

\subsubsection{Varying $R_c$}
\label{varying_rc}

Past studies \citep{2012ApJ...744..162A, 2013ApJ...775..136R, 2014ApJ...780..153B} suggest the dust and gas found in the disk are not co-located and the gas is likely more spatially extended than the dust, which is in agreement with our results from \S\ref{dust_gas_results_same}.  The characteristic radius $R_c$ determined from the dust continuum in A11 may not be accurate for the gas distribution in the disk.  Therefore, we examine whether gas with a characteristic radius $R_c>85$~AU can better fit the HCO$^+$ data.  The inner hole radius $R_{in}=50$~AU, disk surface density normalisation ($\Sigma_0$), d:g ($10^{-2}$) and grain settling parameters ($a_s=0.1\micron$ and $\delta_s=1.0$) are held fixed at the same values used above in \S\ref{dust_gas_results_same}. %From the dust continuum emission, we model a disk with an empty cavity with inner radius $R_{in}=50$~AU and $\Sigma_\circ=10.8$~cm$^{-2}$ so that our surface density normalisation is consistent with A11 from $50<r<85$~AU.  The outer radius $R_{out}$ can be varied to set the disk mass to match the HCO$^+$ data for a specific $R_c$.  

To determine the gas $R_c$ (and simultaneously $R_{out}$), we concentrate on modelling the low-velocity HCO$^+$ flux peak since the line peak is sensitive to the outer disk parameters.  Initially, we assume an arbitrarily large characteristic radius ($R_c=350$~AU) and set the outer radius to be equal to the characteristic radius ($R_{out}=R_c$) to produce a peak HCO$^+$ flux that exceeds the observed flux (where using $R_{out}>R_c$ would increase the modelled flux even further).  The characteristic radius is then reduced in increments of 50~AU continuing to assume $R_{out}=R_c$ at each step until the peak model flux falls below the observed value.  At this point, $R_{out}$ is increased in 50~AU increments until the predicted flux matches the data peak.  %The model with $[R_{c}, R_{out}]$ that best matches the observed flux in the region of the line peak (i.e. yields the smalls $\chi^2$ value) is then adopted as the initial best fit to the outer disk.

Our results are shown in Figure~\ref{fig:tests_cavity} (top-right) for the varying $R_c$ ranging from 250--350~AU and the best-fit models (bottom-left) $R_c=300$~AU and $R_{out}=300$ and 350~AU.  We find some degeneracy between $R_c$ and $R_{out}$, with estimated errors of $R_c\sim \pm50$~AU and $R_{out}\sim \pm 50$~AU.

%Varying $R_c$ and $R_{out}$ has no impact on the high-velocity HCO$^+$ line-wings, as shown by the relatively constant $\chi_{red}^2$ values in Table~\ref{table_param} for the corresponding models.  

These models do not have high enough flux in the line-wings to match the observed HCO$^+$ profile.  It appears the premise of a 50~AU cavity devoid of gas is incompatible with the high velocity line-wing data: there must be a non-negligible amount of gas within the dust cavity to explain the observed line-wings.

\subsection{Disk {\it {\bf With}} Gas in Inner Cavity}
\label{nohole}  

Since our analysis up to this point indicates the presence of gas in the cavity, we consider the amount of gas in the region and how much dust has been mixed into it.  The dust is an important factor in the gas heating/cooling processes.  Dust grains can shield the gas by absorbing UV photons, which cools the gas in the disk. This can generate higher levels of electrons (ejected from the grains) that can recombine with HCO$^+$ and decrease the HCO$^+$ density.  Conversely, dust can also contribute to gas heating in the disk due to the photoelectric effect.  Typical temperature ranges we model in the disk are $<10$ to $\sim1000$~K (dust) and $<10$ to $\sim5000$~K (gas), where the dust and gas tend to have the same temperatures in the midplane due to the energy exchange from inelastic collisions between the grains and gas particles (known as thermal accommodation).  The relation between HCO$^+$ and dust grains means that the line flux can be used as an independent probe of the dust properties in the inner cavity that have been modelled from dust continuum emission.  In this section, we demonstrate the HCO$^+$ data strongly implies the dust is severely depleted in the cavity.  We keep the outer disk parameters ($R_c$, $R_{out}$, $R_{in}$ $\Sigma_0$, $\delta$, $a_s$ and $\delta_s$) fixed at the best-fit values from Sections~\ref{disk_structure} and \ref{varying_rc} unless otherwise noted.

\subsubsection{Cavity Gas with ISM Dust-to-Gas Ratio (10$^{-2}$)}
\label{nohole_102}

From \citet{2010ApJ...717..441E} and A11, the dust continuum data is inconsistent with optically thick dust filling the cavity region ($r<50$~AU).  We now demonstrate using \textsc{ProDiMo} that the HCO$^+$ data is also inconsistent with such material within 50~AU, by extending our outer disk model -- which has a standard surface density profile and an ISM d:g ($10^{-2}$) -- inward to the sublimation radius $R_{in}=0.1$~AU.   We make minor adjustments to $R_{out}$ to maintain a good fit to the data line peak.  

\begin{figure*}
\centering
\includegraphics[width=2.7in]{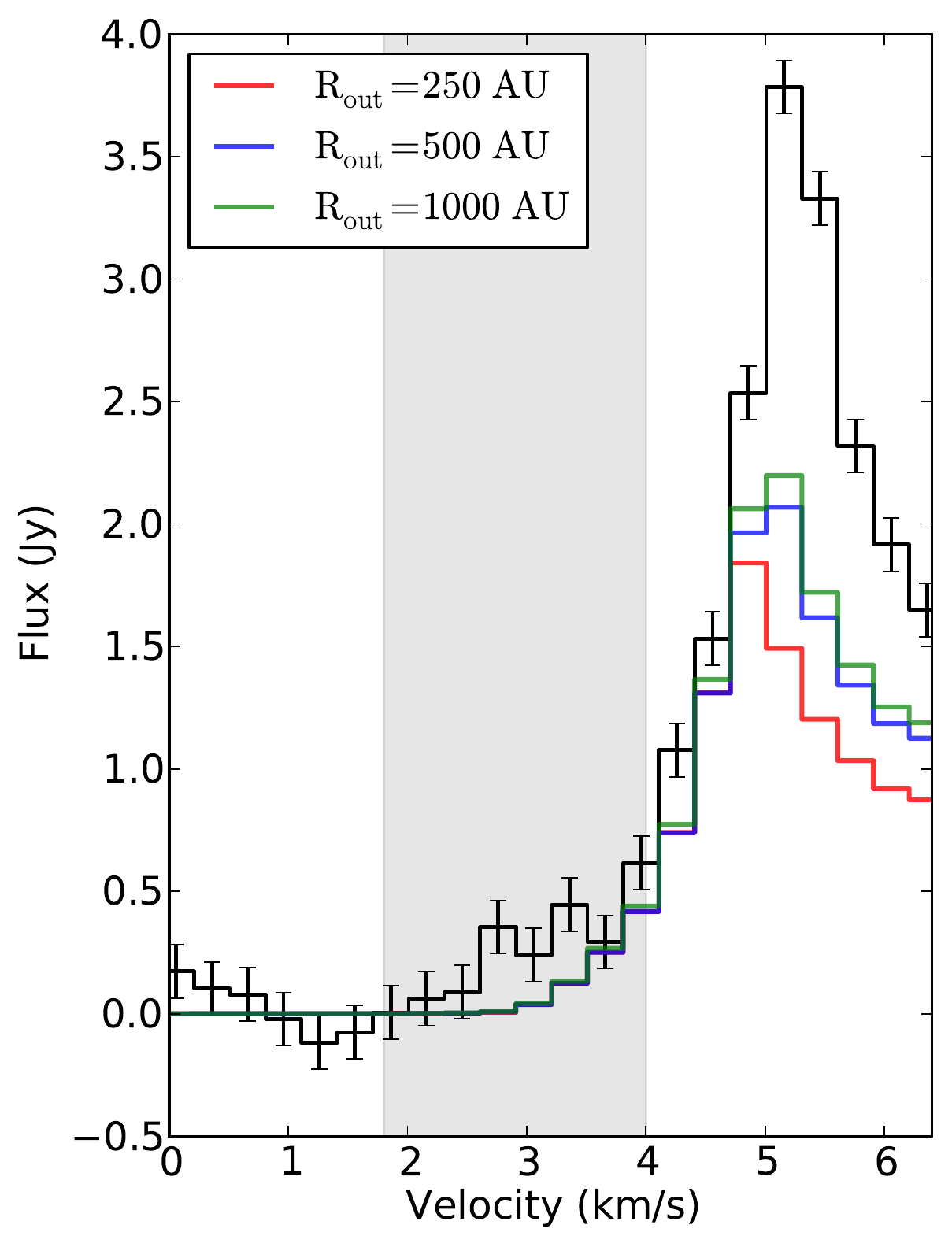}
\includegraphics[width=2.7in]{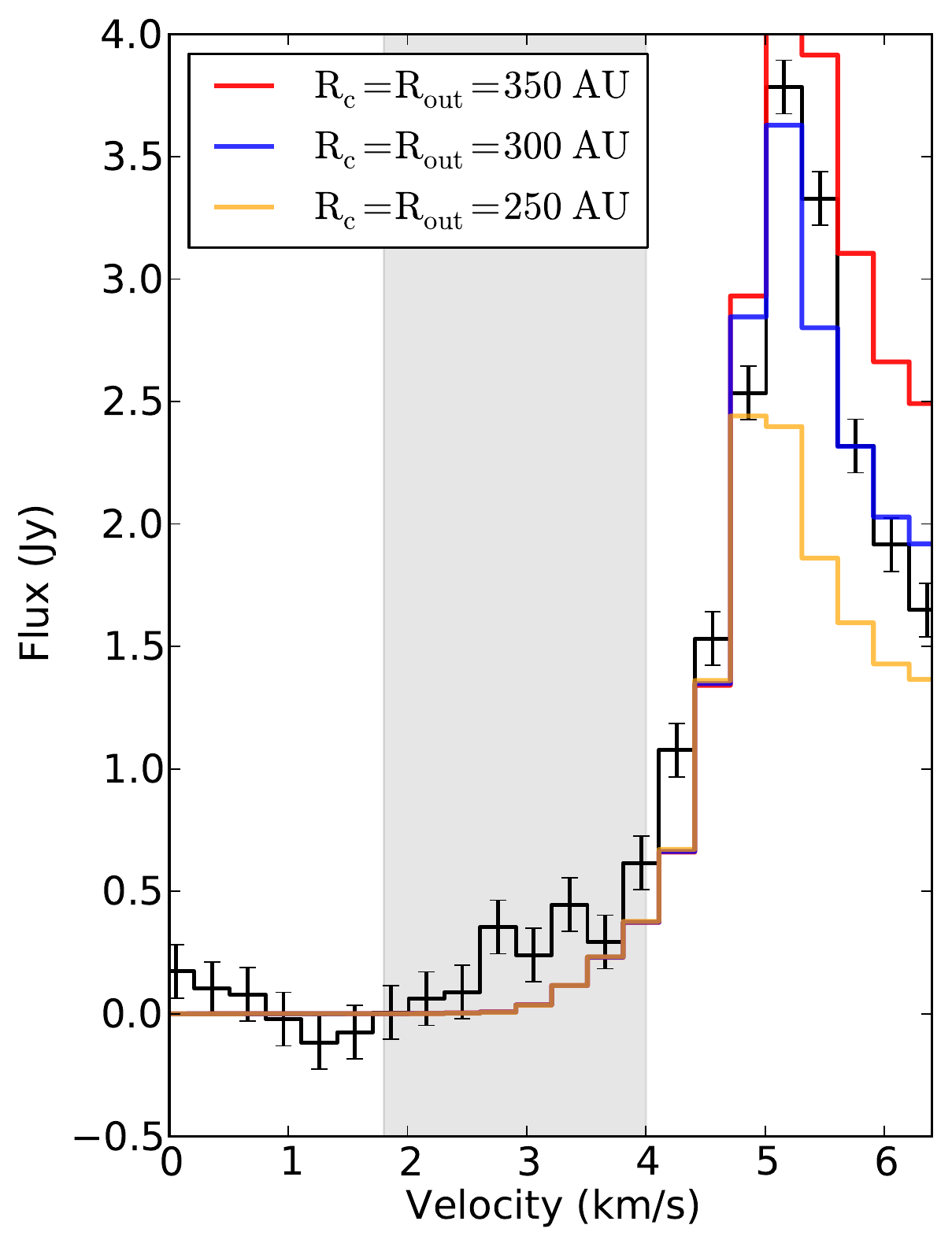}
\includegraphics[width=2.7in]{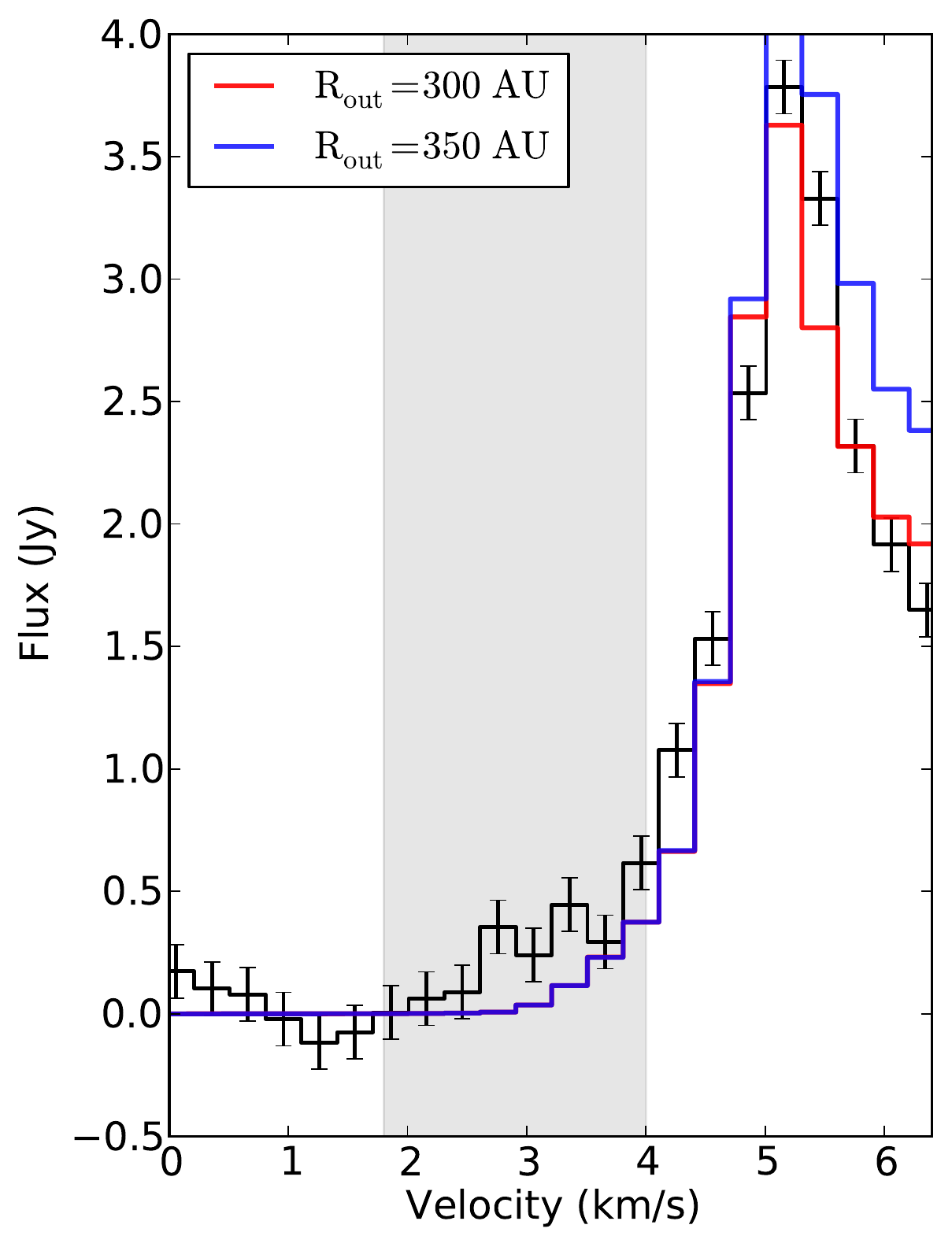}
\includegraphics[width=2.7in]{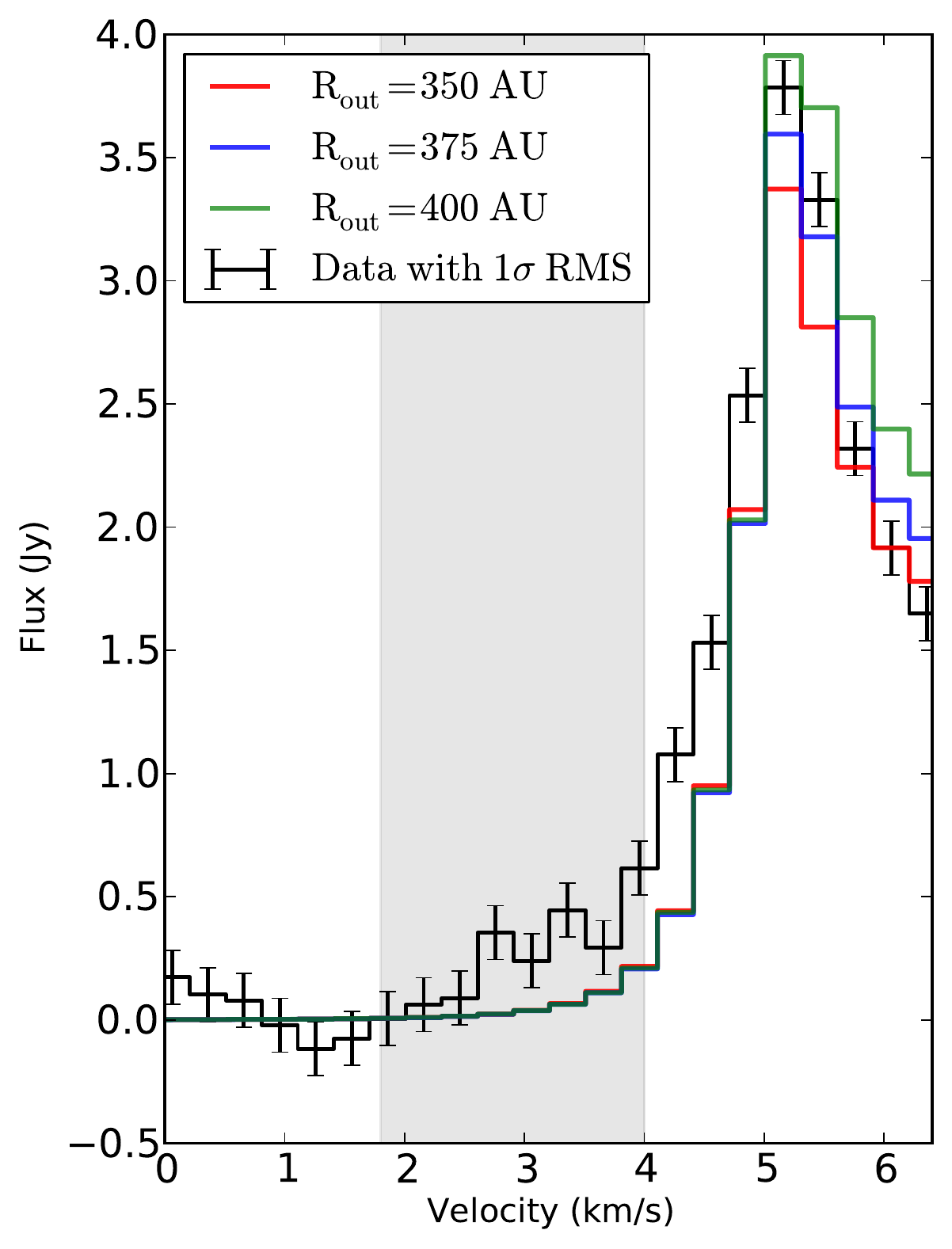}
\caption{Models are compared to the folded spectrum with $1\sigma$~RMS.  The grey boxes indicate velocity channels corresponding to the HCO$^+$ line-wings.  {\emph{Top left:}} Empty inner cavity model with the dust continuum parameters from A11 (see Section~\ref{dust_gas_results_same}).  {\emph{Top right:}}  Determining the characteristic radius $R_c$ with an empty inner cavity model (see Section~\ref{varying_rc}).  {\emph{Bottom left:}}  Empty inner cavity model with $R_c=250$~AU and standard ISM d:g$=10^{-2}$ (see Section~\ref{varying_rc}).  We note that models of the outer disk with an empty inner cavity and varying $R_c$ and $R_{out}$ have not been able to produce the observed HCO$^+$ high velocity line-wings.  {\emph{Bottom right:}}  Disk without an inner cavity and d:g$=10^{-2}$ (see Section~\ref{nohole}).  Even though there is some HCO$^+$ line-wing emission, the filled-cavity models do not fit the high velocity line-wing emission present in the HCO$^+$ spectrum either.}
\label{fig:tests_cavity}
\end{figure*}

Results are shown in Figure~\ref{fig:tests_cavity} (bottom-right) for $R_{in}=0.1$~AU and $R_{out}=$350--400~AU (corresponding to $M_d=$0.12 to 0.13~M$_\odot$).  We note that $R_{out}\leq350$~AU models now under-predict the line peak unlike the empty inner cavity models from \S\ref{with_innercav}.  The lower emission is caused by the lack of contribution from the cavity inner wall at $R_{in}=50$~AU due to direct stellar irradiation.  Increasing $R_{out}$ to 375--400~AU makes up for this deficit and increases the line peak.  

Even though we observe a small increase in the HCO$^+$ line-wing emission relative to the empty cavity models, none of the filled-cavity models show enough of an increase to fit the observed line-wing flux.  Since the gas within a radius of 50~AU is now optically thick (i.e. $\tau>1$ for radii $r>6$~AU), the only way to significantly enhance its flux is by altering the chemistry in the cavity (i.e. by constraining the dust settling/mixing parameters and dust-to-gas ratio) and increasing the gas scale height so that the molecular line emits over a larger area.  Depleting the dust within the cavity would not only suppress the dust continuum emission as previously observed, but would also remove the primary shielding mechanism for the gas, thereby changing the chemistry in the disk cavity (due to the increase in UV radiation).   

%increasing the gas temperature in this region.  In other words, the implication is that cavity gas with a standard surface density and ISM d:g is too cold to explain the HCO$^+$ emission: gas temperatures in the models fall below $40$~K at radii greater than 20~AU, compared to the 43~K upper state energy of HCO$^+$~$J=4\rightarrow3$ line.  Thus the HCO$^+$ data independently suggests a depletion of dust within the cavity: such depletion would not only suppress the dust continuum emission, as previously observed, but would also remove the primary shielding mechanism for the gas, thereby enhancing the model HCO$^+$ line-wing emission.

\subsubsection{Cavity Gas Model}
\label{cavity_fits}

In the fitting procedure for the LkCa~15 disk cavity, we constrain the dust settling and mixing ($a_s$ and $\delta_s$; discussed in Section~\ref{disk_structure}), gas (reference) scale height ($H_0$; for details see Section~\ref{varying_scale_height}) and dust-to-gas ratio (for details see Section~\ref{dust_to_gas}) parameters within the cavity (radii from 0.1 to 50~AU).  We first run models for a range of dust settling and mixing values and fix these parameters to values with the lowest $\chi^2 _{red}$ compared to the HCO$^+$ line-wings.  With the dust settling and mixing parameters fixed, we then repeat this procedure for a range of gas scale height values.  The final step is to additionally fix the gas scale height to the value with the lowest $\chi^2 _{red}$ and repeat fitting the HCO$^+$ line-wings for a range of dust-to-gas ratios.  

\begin{figure}
\centering
\includegraphics[width=4in]{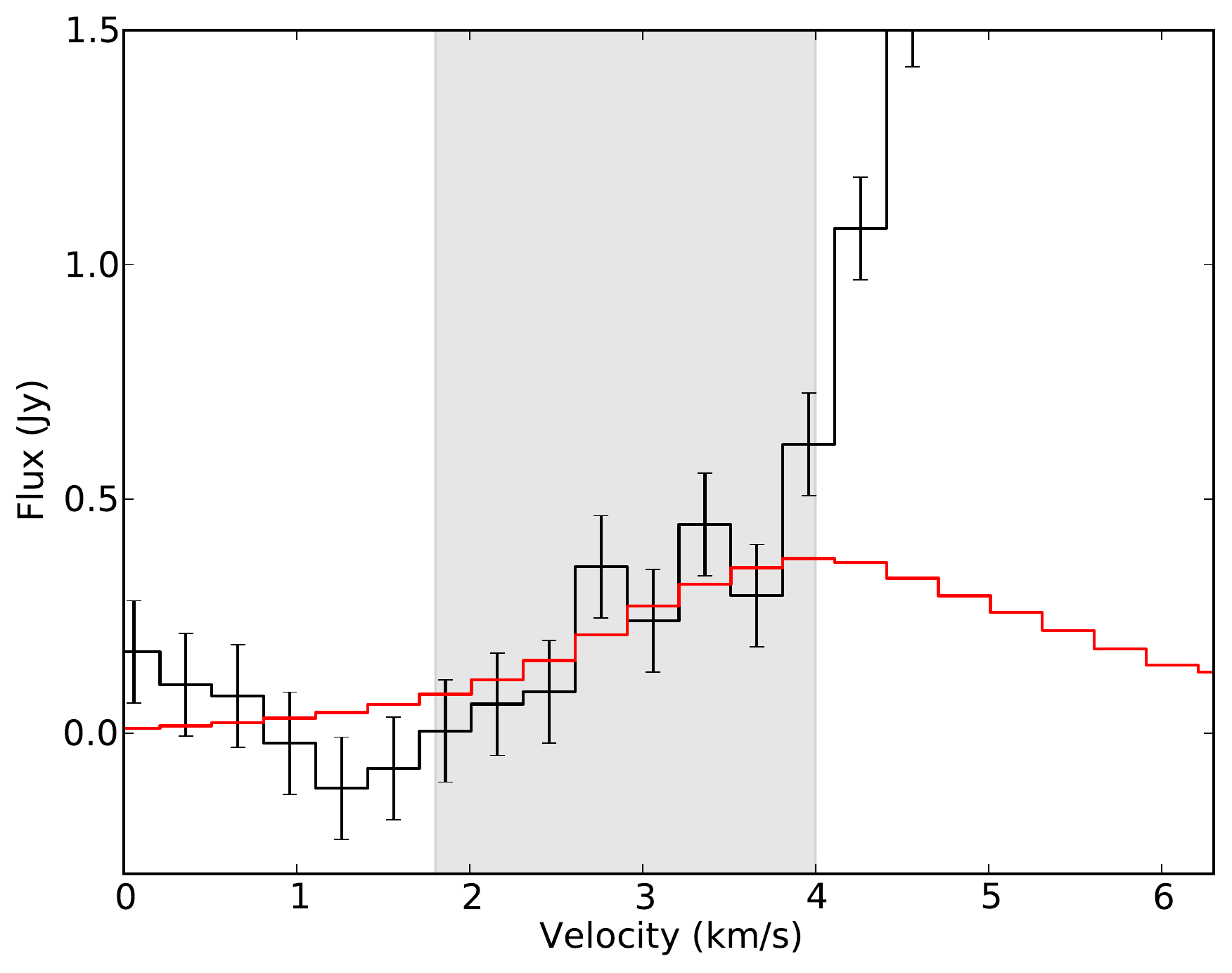}
\caption{Best-fit model to the emission from the disk cavity at radii from 0.1 to 50~AU, see Section~\ref{cavity_fits}.  Settling parameters are $a_s=0.01\micron$ and $\delta_s=0.01$, the gas reference scale height is $H_0=60$~AU and d:g$=10^{-6}$. }
\label{fig:cavity_fit}
\end{figure}

Figure~\ref{fig:cavity_fit} shows the best-fit model to the disk cavity with dust settling and mixing parameters  [$a_s$, $\delta_s$] = [0.01$\micron$, 0.01].  The dust settling and mixing parameters do not substantially affect the production of HCO$^+$ within the disk cavity, as discussed in Appendix~\ref{settling_effects}.  To fit the observed HCO$^+$ line-wing flux, we must increase the gas scale height within the cavity (from $H_0=10$~AU to 60~AU) and suppress the cavity dust (d:g$=10^{-6}$).  Increasing the cavity scale height also increases the molecular line emitting area of the disk, leading to more HCO$^+$ emission.  Furthermore, HCO$^+$ emission is at its maximum when there is a balance between high HCO$^+$ density and warm gas to produce line emission, which occurs at a dust-to-gas ratio of $\sim$10$^{-6}$.  Both a smaller scale height with a reduced dust-to-gas ratio and an ISM dust-to-gas ratio with a large cavity scale height under predict the HCO$^+$ line-wing flux.  Further details of the individual models are given in the appendix (\ref{settling_effects}--\ref{dust_to_gas}).

\subsubsection{Cavity Mass Constraints}
\label{lower_ten}

In all our fits so far, we have kept the mass inside the disk cavity fixed at $M_{cav}\sim$0.03~M$_\odot$, where the cavity gas mass is calculated by assuming the outer disk surface density extends inwards to 0.1 AU.  In other words when we change the dust-to-gas ratio in Section~\ref{dust_to_gas}, we decrease the amount of dust in the cavity while increasing the gas even though the total cavity mass is fixed.  We can now use the mass to change the amount of gas in the cavity using the best-fit model from Section~\ref{dust_to_gas}, i.e. keep the dust-to-gas ratio scale height constant at d:g$=10^{-6}$ and $H_0=60$~AU but vary the total disk mass (which changes both the gas and dust in the disk).  We increase and decrease the cavity mass by a factor of 10 ($M_{cav}\approx$ 0.3~M$_\odot$ and 0.003~M$_\odot$) to constrain the cavity mass.

\begin{figure}
\centering
\includegraphics[width=4in]{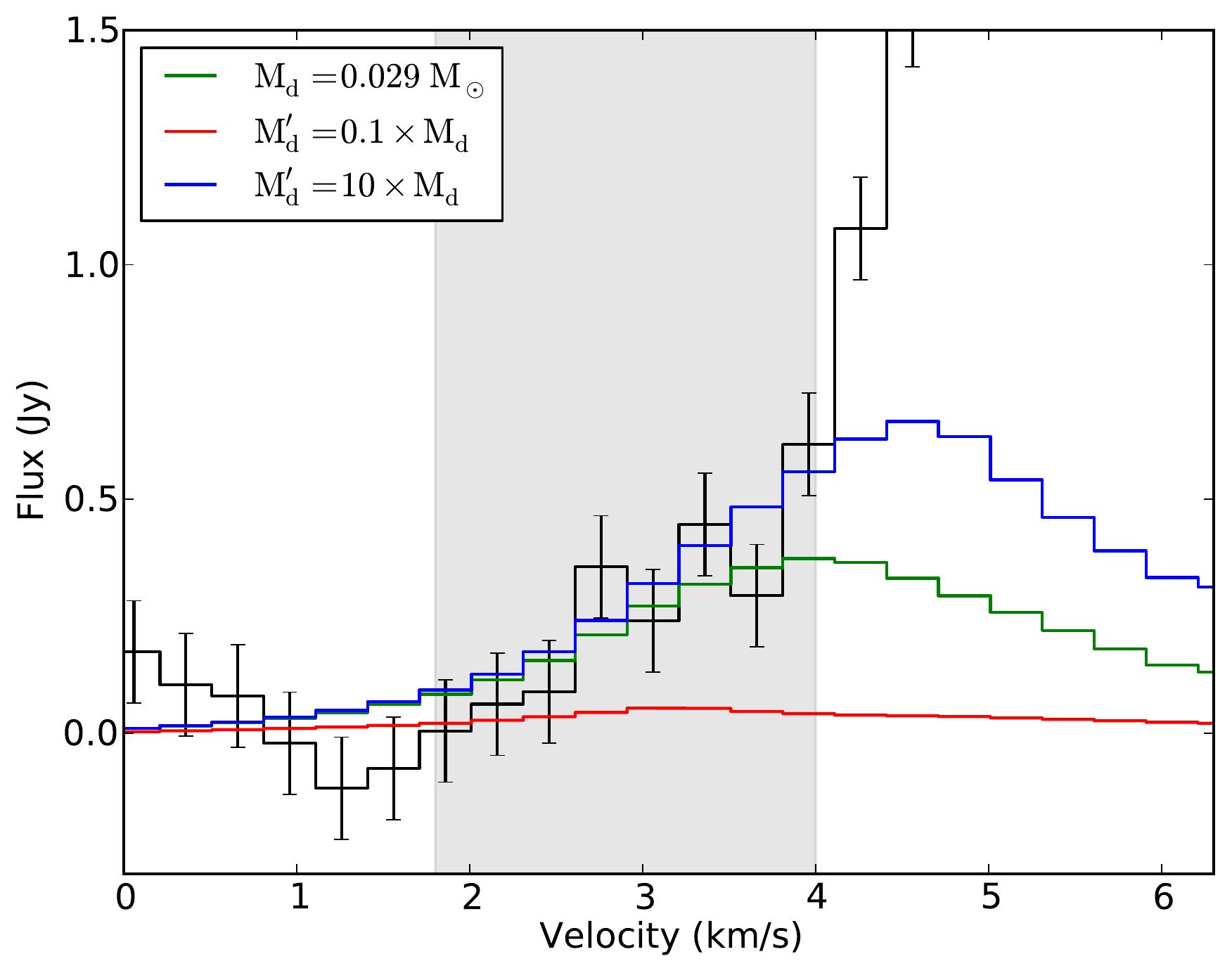}
\caption{Constraining the inner cavity mass by fitting the disk cavity emission, where the standard cavity mass $M_d$ (0.029~M$_\odot$) is in green, $0.1\times M_d$ (0.003~M$_\odot$) is in red and $10\times M_d$ (0.3~M$_\odot$) is in blue (see Section~\ref{lower_ten}).}
\label{fig:cavity_test}
\end{figure}

Figure~\ref{fig:cavity_test} shows the corresponding HCO$^+$ line profiles.  For the higher mass case, the HCO$^+$ line-wings have increased and still has a reasonable fit.  This scenario causes the cavity to be optically thick in dust continuum, eliminating the observed gap in the LkCa~15 disk.  However, the optically thick dust continuum emission can be fixed by simply lowering the dust-to-gas ratio in the disk cavity.  At cavity masses much larger than $\sim0.3$~M$_\odot$, the cavity soon becomes gravitationally unstable for a solar mass star like LkCa~15.  As a consequence, we conclude the upper cavity mass limit must be within an order of magnitude.

For the lower mass case, the HCO$^+$ line-wings disappear.  We emphasise this is not because the dust has gone down, which affects shielding and the corresponding chemistry of the disk.  Instead, there is not enough gas in the disk cavity.  For example, Figure~\ref{fig:dgratio_cavity_test} shows that there is HCO$^+$ line-wing emission still present when the disk has a low dust density (i.e. low dust-to-gas ratio of 10$^{-10}$).  We can, therefore, conclude that the LkCa~15 cavity lower mass limit is also accurate to within an order of magnitude. 

%to within an order of magnitude of the standard cavity mass $M_{cav}=$0.03~M$_\odot$ from the best-fit (d:g$=10^{-6}$ and $H_0=60$~AU) in \S\ref{dust_to_gas}.  Figure~\ref{fig:cavity_test} shows the comparison between the HCO$^+$ line profiles with disk masses a factor of 10 larger and smaller than the standard cavity mass (with settling parameters set as in \S\ref{varying_scale_height} and \ref{dust_to_gas}).  For the decreased disk mass, the HCO$^+$ cavity flux is greatly reduced due to the decrease in HCO$^+$ density and no longer reproduces the observed line-wings (further discussed in \S\ref{discussion}).  Conversely, increasing the cavity mass to 0.3~M$_\odot$ produces strong HCO$^+$ line-wings, similar to the line-wings modelled by the standard cavity mass.  The HCO$^+$ and dust continuum optical depths in the cavity, however, become optically thick at radii $r>10$~AU.  The increased optical depth for dust continuum emission is not only inconsistent with observations of an optically thin continuum cavity, but the increased cavity mass is unstable for a solar mass star like LkCa~15.  We can, therefore, conclude that the LkCa~15 cavity mass is accurate to within an order of magnitude.  

\subsection{Final Disk Model}
\label{final_model}

\begin{figure*}
\centering
\includegraphics[width=4in]{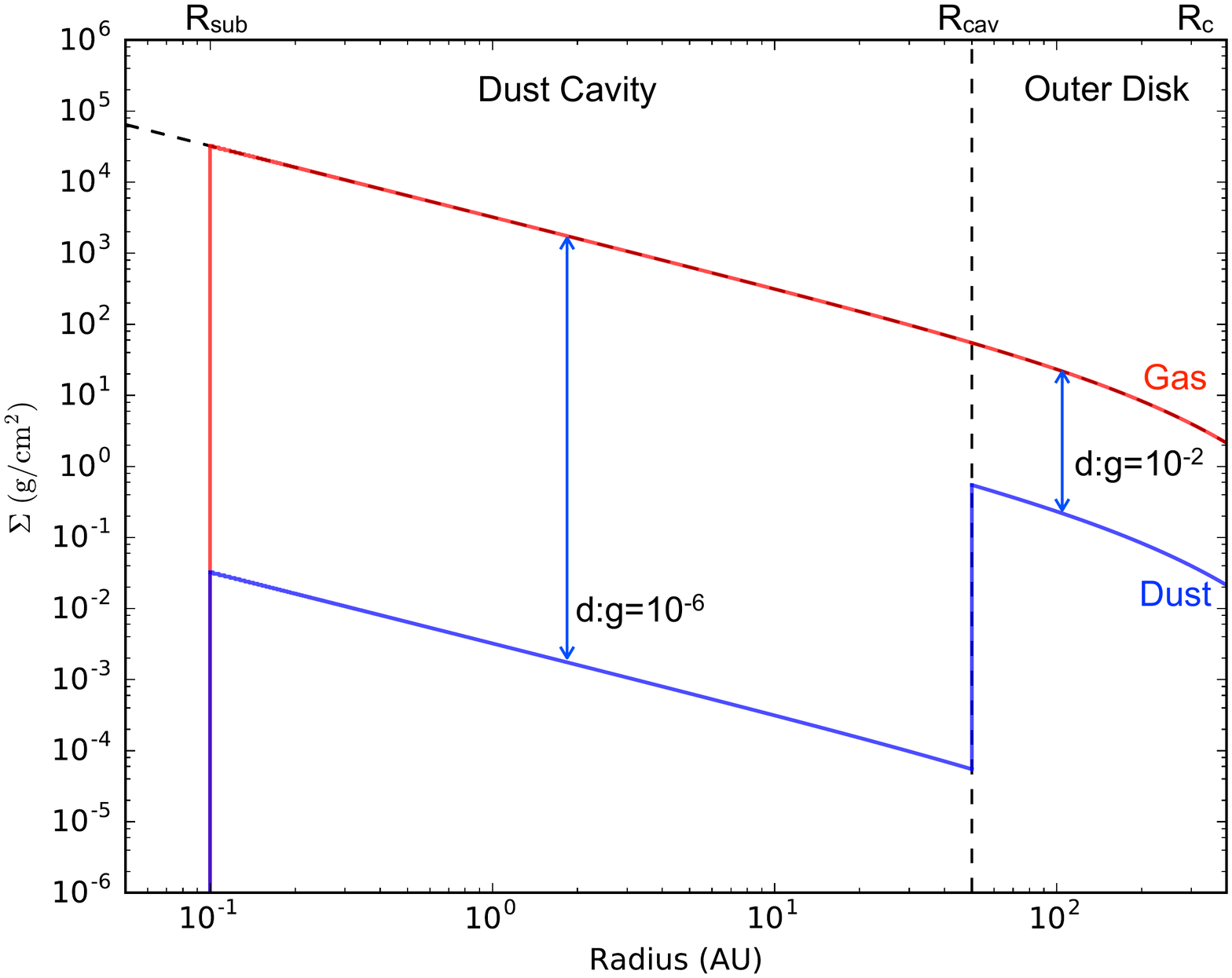}
\caption{Surface density profile of the gas (red) and dust (blue) of the final disk models.  $\mathrm{R_{sub}}$ denotes the sublimation radius (the inner radius for the dust cavity at 0.1~AU), $\mathrm{R_{cav}}$ denote the extent of the cavity (50~AU) and $\mathrm{R_{c}}$ denotes the characteristic tapering and scaling radius (300~AU). }
\label{fig:full_model_sigma}
\end{figure*}

\begin{figure*}
\centering
\includegraphics[width=2.5in]{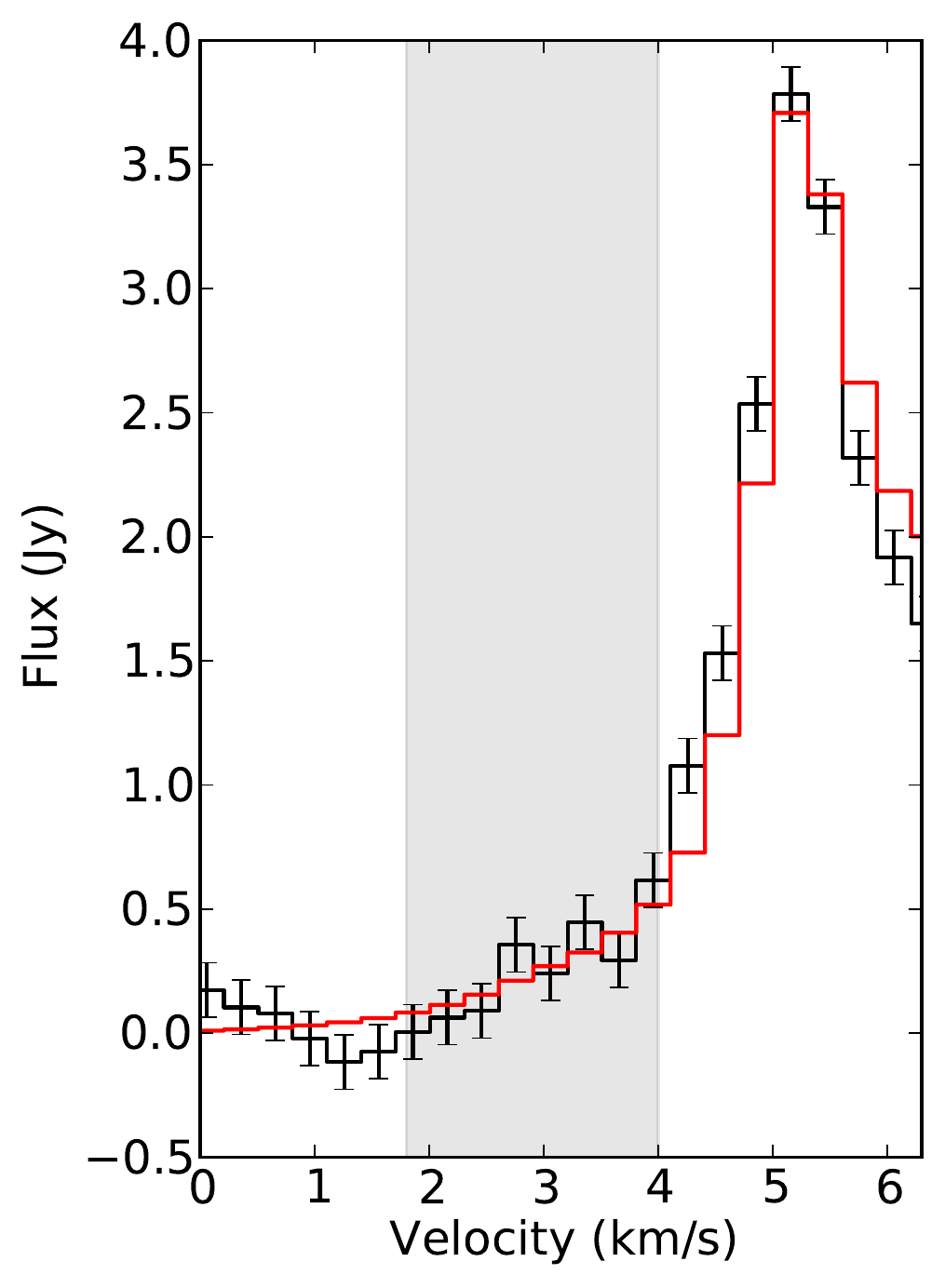}
\includegraphics[width=3.5in]{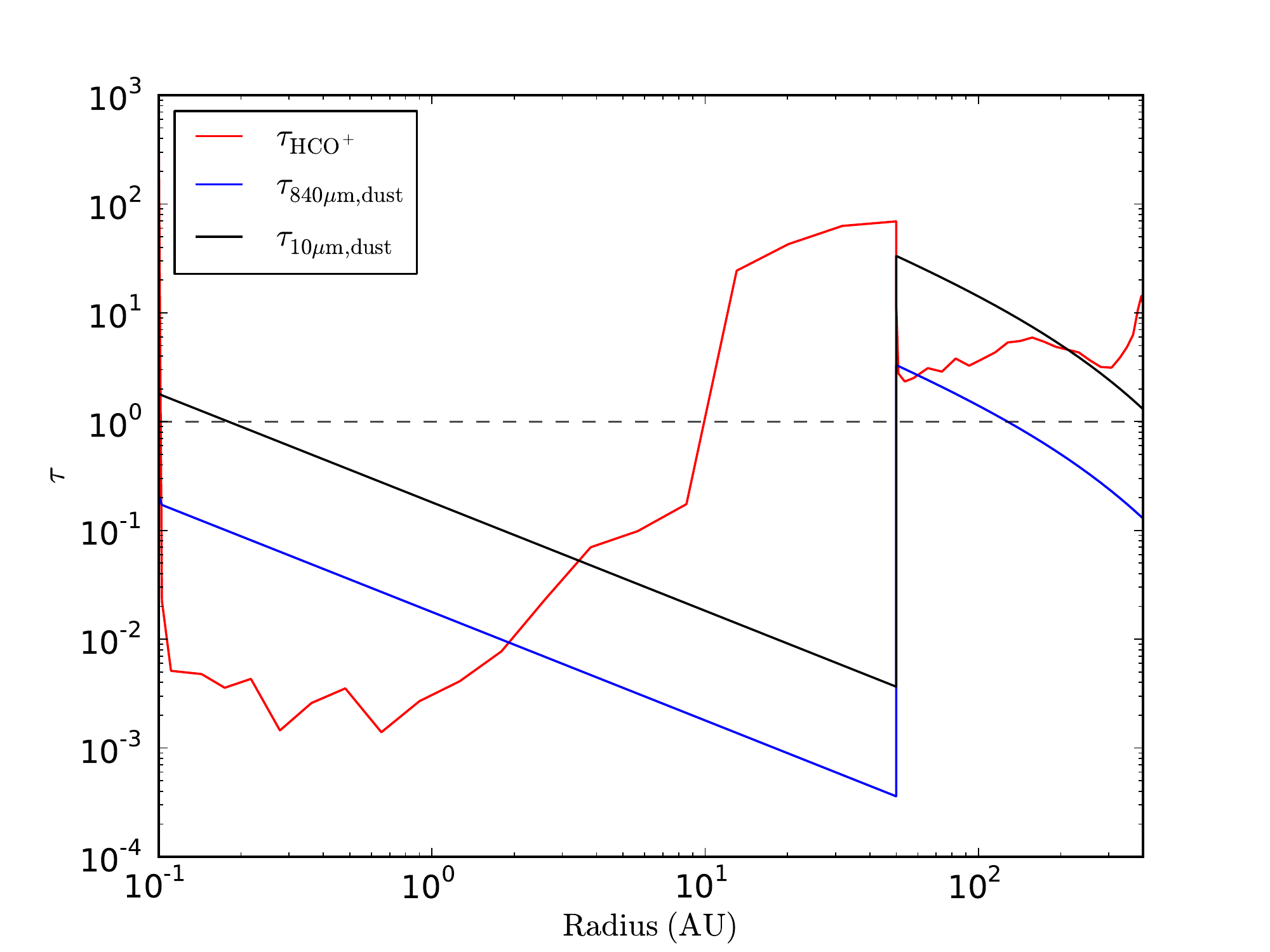}
\caption{{\emph{Left:}}  Best-fit model to both the inner cavity and the outer disk (see Section~\ref{dust_to_gas}) with total disk mass is 0.10~M$_\odot$.  {\emph{Right:}}  HCO$^+$~$J=4\rightarrow3$ and dust continuum (10 and 840~$\micron$) optical depths per radius for the best-fit models.  HCO$^+$ remains optically thin $\tau<1$ at radii $r\leq10$~AU.  The 840~$\micron$ dust continuum is optically thin at radii 50~AU$<r<100$~AU, consistent with the 880~$\micron$ dust continuum data from A11.  Similarly the $10\micron$ dust continuum is primarily optically thin at radii $r<50$~AU, consistent with SED fitting \citep{2008ApJ...682L.125E, 2010ApJ...717..441E}.}
\label{fig:full_model_end}
\end{figure*}

We model the dust and gas in the full disk using a model composed of two components:  (1) a radial model of the the dust-depleted cavity with large scale height between 0.1 to 50~AU (Section~\ref{dust_to_gas}) and (2) the outermost optically thick disk from 50 to $\sim$400~AU.  Figure~\ref{fig:full_model_sigma} shows the gas and dust surface density profiles of the full disk model and Figure~\ref{fig:full_model_end} shows the HCO$^+$ and dust continuum optical depths.  This best-fit cavity model leads to a line-wing fit ($a_s = 0.01\micron$ and $\delta_s= 0.01$) with $\chi^2 _{red}$ at 1.2 (see Table~\ref{table_param}).

Figure~\ref{fig:full_model_sed} shows the HCO$^+$ line-wings produced by the final disk model with a dust-to-gas ratio in the cavity of the disk at d:g$=10^{-5}$, 10$^{-6}$ and 10$^{-7}$.  We include a separate model of the innermost disk (i.e. as described in Section~\ref{andrews_sed}) from 0.1 to 10~AU in the Figure~\ref{fig:full_model_sed} SED since our the two-component cavity model alone is not designed to fit the shortest wavelength emission from the disk.  While this is not ideal, our current version of \textsc{ProDiMo} was not able to model a three-component disk with drastically varying gas scale heights.  Further discussion on how the innermost disk proposed by A11 will affect the HCO$^+$ found within the disk cavity can be found in Section~\ref{discussion}.  A cavity dust-to-gas ratio of $10^{-7}$ best fits SED wavelengths at $\sim2.2$ to 20~$\micron$.  However, HCO$^+$ line-wings are maximised with cavity d:g$=10^{-6}$.  We note our model is somewhat inconsistent with the SED at longer wavelengths, corresponding to the outermost disk.  This is due to the limitations in the characteristic scaling and tapering radius, discussed previously in Section~\ref{andrews_sed} and \ref{varying_rc}.

Lastly, Figure~\ref{fig:final_unfolded} shows the final disk model compared to the unfolded observed HCO$^+$ spectrum.  The final line-wing fit had $\chi^2 _{red}$ at 0.8 (see Table~\ref{table_final}).  The redshifted flanks and line-wings have a better overall fit compared to the blue side.  However, the entire unfolded spectrum is still consistent with our models.

%However, there is a noticeable difference in the shape of the blue- and red-shifted line velocities, where the blue-shifted portion of the line appears more asymmetric.   

\section{Discussion and Conclusions}
\label{discussion}

A number of models have been tested for fitting the HCO$^+$ line emission in the LkCa~15 disk, focussing on the disk cavity at a radius $<50$~AU.  We detect significant line-wing flux, indicating the presence of gas in the disk cavity up to 50~AU from the star.  We have been able to model the observed line-wing flux by suppressing the cavity dust (d:g$=10^{-6}$) and increasing the gas scale height substantially in this region ($H_0/R_0\sim0.6$ instead of the standard outer disk $H_0/R_0\sim0.1$).  Both an ISM-like d:g$=10^{-2}$ and/or a small scale height ($H_0=10$~AU) under-predict the HCO$^+$ line flux.  Lastly, the gas mass in the cavity is roughly what is expected in the absence of a cavity (0.03~M$_\odot$), where masses lower by a factor $\sim10$ under-predict line-wing flux.  Our study suggests that possible planets sculpting the LkCa~15 dust cavity appear to do so without greatly diminishing the amount of gas within it.  However, spatially resolved observations are needed to test this result.

The detected HCO$^+$~$J=4\rightarrow3$ line-wings in LkCa~15 are consistent with Greaves (2004) HCO$^+$ line-wings detected in GG~Tau, GM~Aur and DM~Tau, which are all known to have cavities in the dust continuum emission.  Indeed, at velocities $\mathrm{v_{lc}} \pm \gtrsim 3.0$~km~s$^{-1}$, each of these sources have similar high velocity line-wing flux, implying the cavities of all transitional disks potentially have similarly low dust-to-gas ratios and puffed-up inner rims in gas, as in LkCa~15.  

As discussed in \S\ref{disk_structure}, our models have focussed on fitting the disk cavity from 0.1--50~AU using a simple one-component model.  However, this model can be improved upon by using two-components, where an inner dusty disk is set from radii $\sim0.1$--10~AU (e.g. \citealt{2008ApJ...682L.125E, 2010ApJ...717..441E, 2011ApJ...732...42A, 2015A&A...579A.106V}) and dust gap between $\sim 10$--50~AU.  Past work from \citet{2013A&A...559A..46B} has suggested the inner dust disk (assumed to have a dust depletion factor $\delta_\mathrm{dust}=10^{-5}$ w.r.t. the dust density of the outer disk) can significantly influence chemistry in the gap, particularly for CO emission.  The inner disk can shield the cavity from direct stellar irradiation, decreasing gas temperatures in the gap and allowing CO and H$_2$ to survive at lower gas masses.  This could also potentially allow HCO$^+$ to survive at lower gas masses (where H$_2$ and CO are necessary for HCO$^+$ formation).  

A11 estimated the inner dusty disk to have a surface density (and total gas$+$dust mass) depleted by $\delta_\mathrm{inner}=10^{-6}$ w.r.t. the outer disk.  Similarly, \citet{2013A&A...559A..46B} used a dust depletion $\delta_\mathrm{dust}=10^{-5}$ to test the effects of an inner disk scenario (with and without gas depletion).  Our study has also tested dust and gas depletion within the complete cavity region at radii 0.1--50~AU by varying dust-to-gas ratios and the cavity mass.  We vary dust-to-gas ratios in \S\ref{dust_to_gas} from $10^{-2}$ down to $10^{-10}$ (without gas depletion), which corresponds to dust depletion factors $\delta_\mathrm{dust, cav} = 1$ down to $10^{-8}$ w.r.t. the dust surface density in the outer disk.  This range of dust-to-gas ratios (and thus dust depletion factors) investigates the dust content within the inner disk (and gap) region, where this range includes the inner disk depletion factors also used in A11 and \citet{2013A&A...559A..46B}.  

\begin{figure}
\centering
\includegraphics[width=3in]{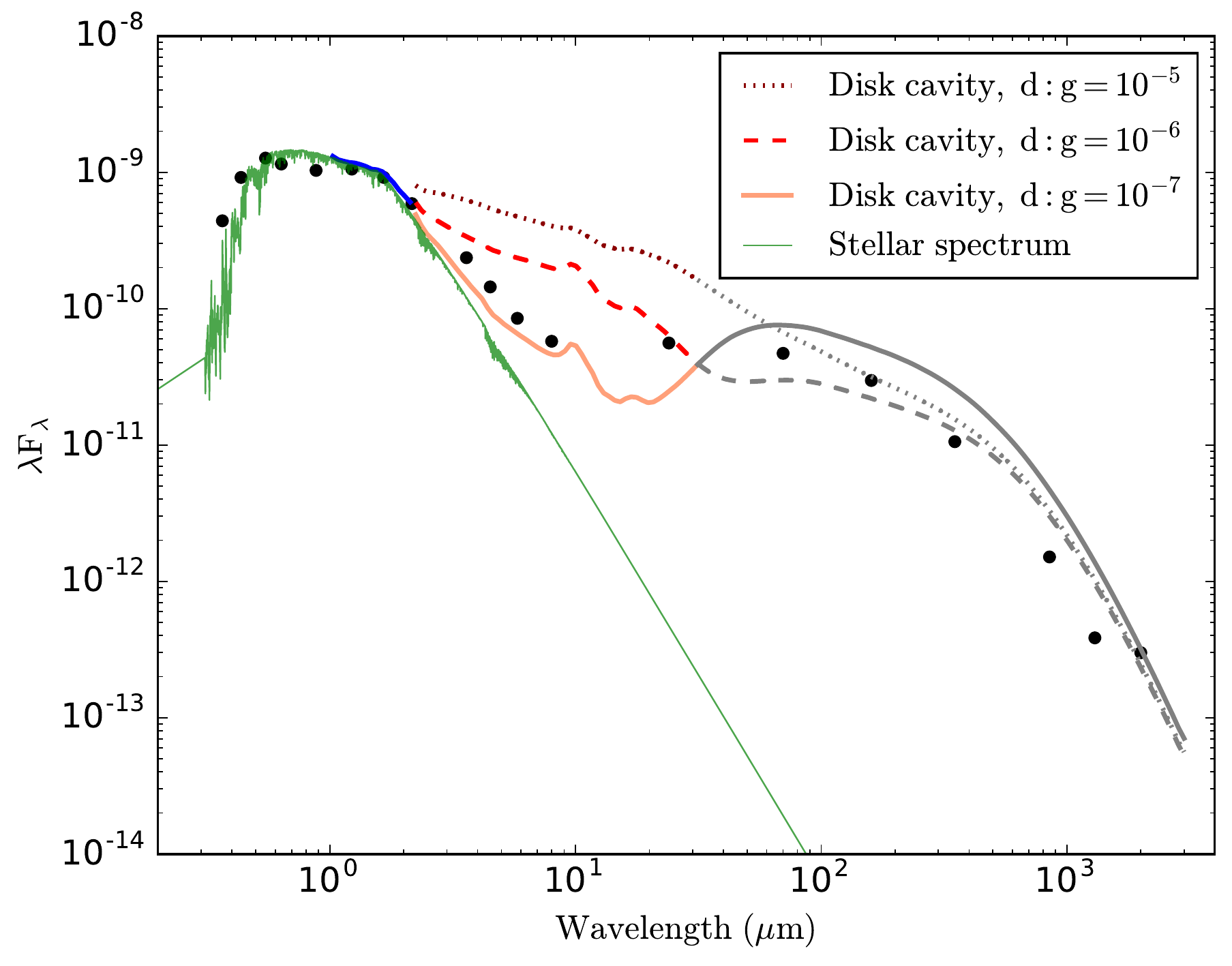}
\includegraphics[width=3in]{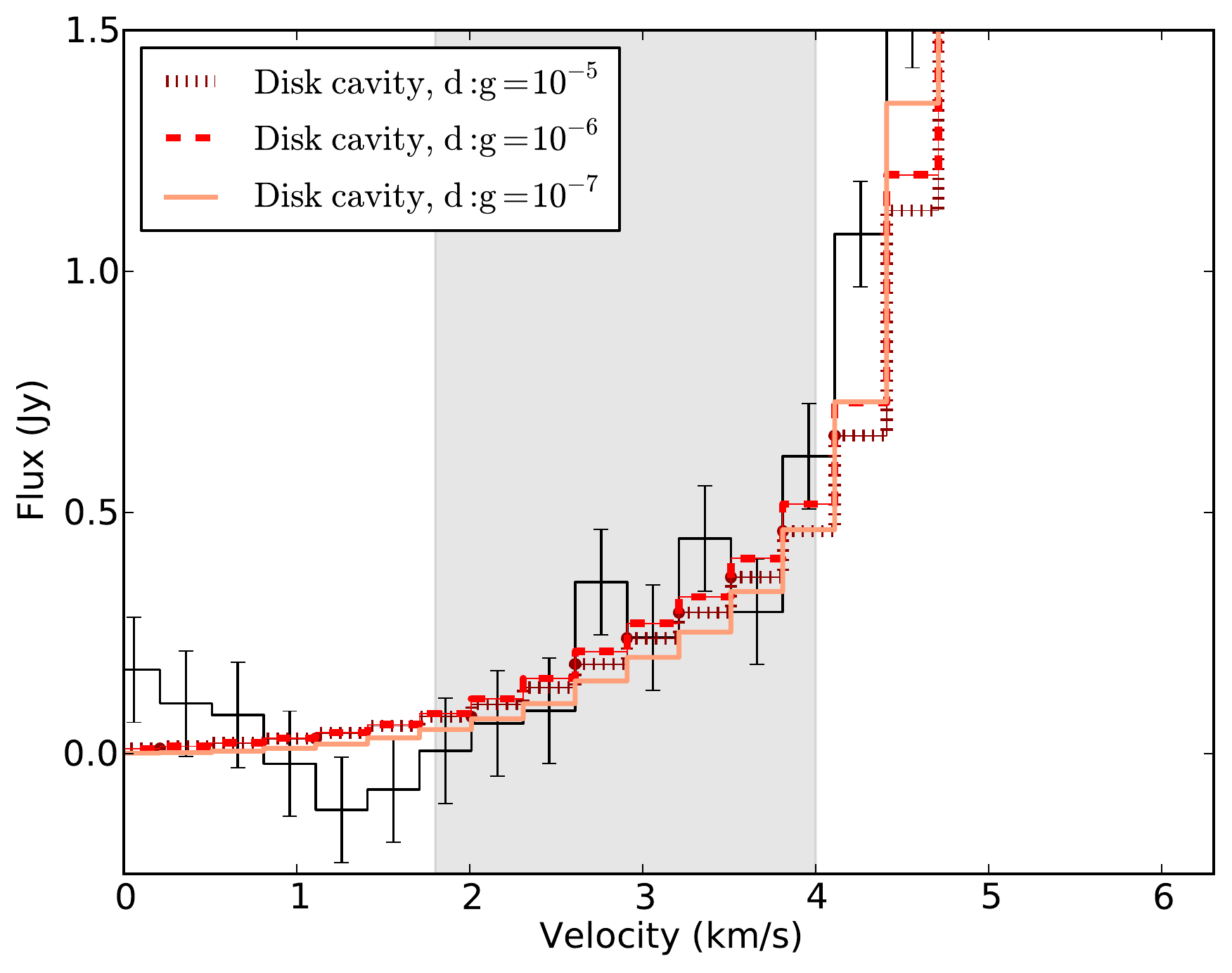}
\caption{Full-disk models generated from an innermost disk model to fit the inner disk described in A11 (radii 0.1 to 10~AU; Section~\ref{andrews_sed}) and a two-component disk model to fit the cavity (10 to 50~AU) and the outermost disk (50 to 400~AU).  We show the best-fit cavity scale height (60~AU at a reference radius of 100~AU) and d:g$=10^{-5}$, $10^{-6}$ and $10^{-7}$.  {\emph{Left:}}  Full-disk SED models with the stellar spectrum (green), innermost disk (blue), disk cavity (red) and outer disk (grey).  The dashed line depicts the two-disk model with a dust-to-gas ratio in the disk cavity at d:g$=10^{-5}$ (dark red), the solid line is at d:g$=10^{-6}$ (red) and the dotted line is at d:g$=10^{-7}$ (light red).   {\emph{Right:}}  HCO$^+$ spectra produced from the two-disk models using d:g$=10^{-5}$ (dark red, dotted line), $10^{-6}$ (red, dashed line) and $10^{-7}$ (light red, solid line).
  }
\label{fig:full_model_sed}
\end{figure}

\begin{figure}
\centering
\includegraphics[width=3in]{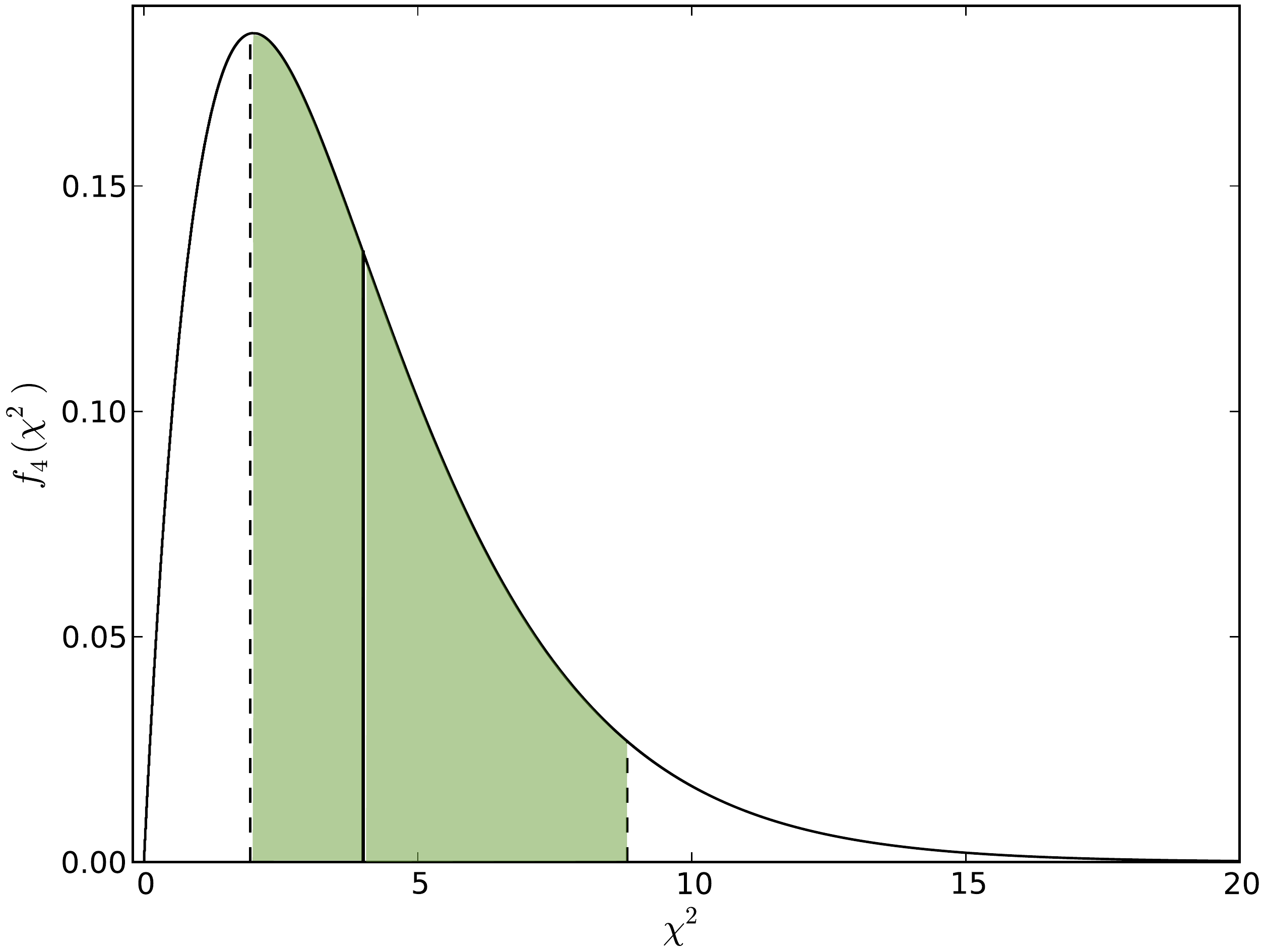}
\includegraphics[width=3in]{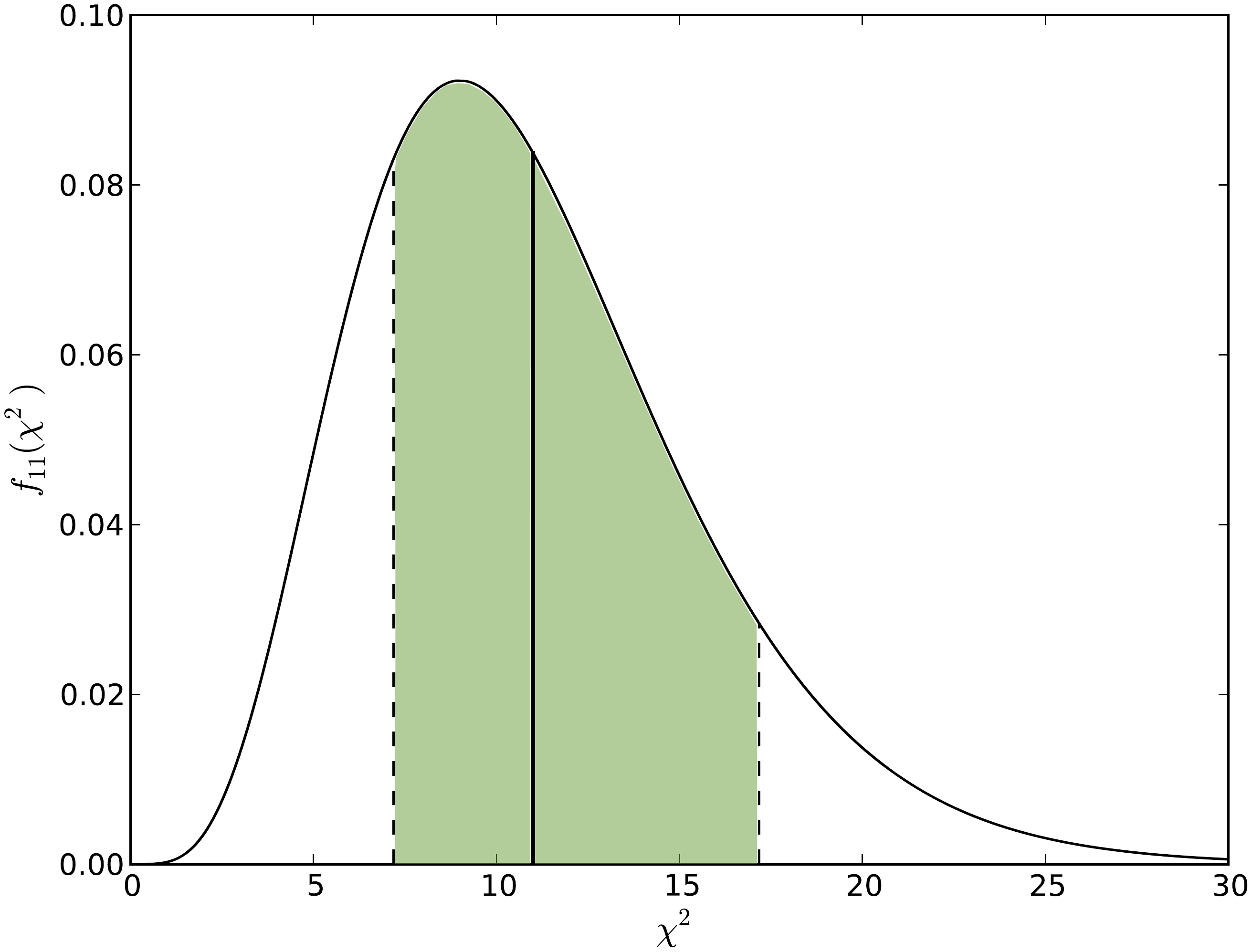}
\caption{{\emph Left:}  Probability density function (PDF) of a $\chi^2$ distribution with 4 degrees of freedom used to compare models to the folded, half spectrum line-wings of HCO$^+$.  The mean of the PDF ($x=4$) is shown as a solid line and 1$\sigma$ values are regions shaded in green within the dashed lines (where 1$\sigma \sim34 \%$).  {\emph Right:}  PDF of $\chi^2$ distribution with 11 degrees of freedom used to compare models to the unfolded, full-spectrum line-wings of HCO$^+$}.  
\label{fig:dof}
\end{figure}

\begin{figure}
\centering
\includegraphics[width=4in]{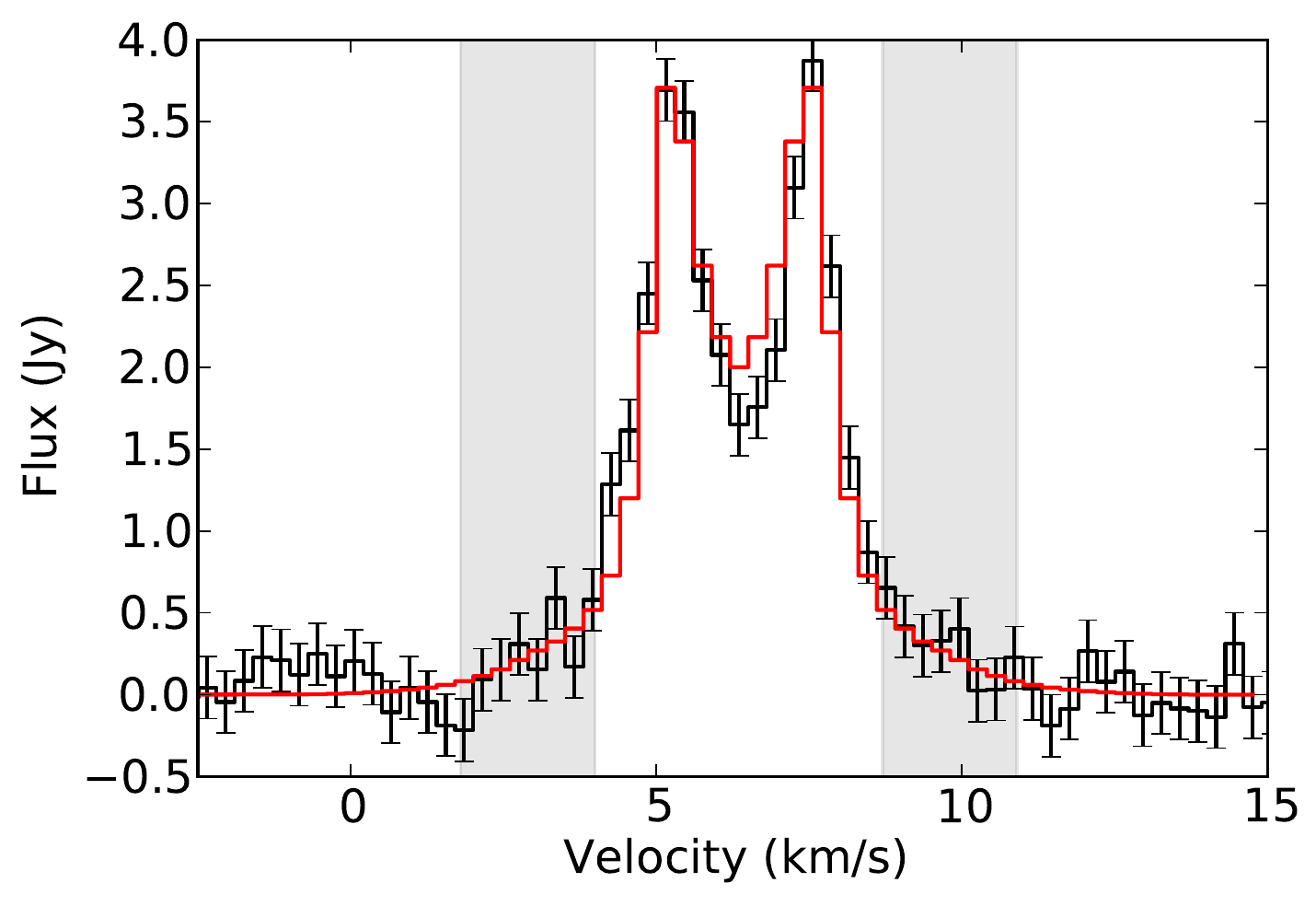}
\caption{Best-fit two-disk model compared to the unfolded LkCa~15 HCO$^+$ spectrum as in Table~\ref{table_final} and Section~\ref{final_model}.}
\label{fig:final_unfolded}
\end{figure}

Our best-fit cavity model (d:g$=10^{-6}$ or $\delta_\mathrm{dust, cav}=10^{-4}$) has a relative dust content that is a factor 100$\times$ higher than expected from the inner dusty disk in A11.  However, as shown in Figure~\ref{fig:full_model_sed} and discussed in \S\ref{final_model}, the SED for the best-fit model is inflated at near- to mid-IR wavelengths.  For the cavity model with a dust-to-gas ratio matching the inner disk dust depletion from A11  (i.e. $\delta_\mathrm{dust, cav}=10^{-6}$ or d:g$=10^{-8}$), the corresponding SED better matches the data but the HCO$^+$ line-wing emission is lower than the observations (likely because more UV emission is able to penetrate the disk with the lower dust content and dissociate CO and H$_2$ needed to form HCO$^+$).  These findings indicate that shielding from dust within the cavity region (either from an inner disk or a small reservoir of dust within the full cavity) is important for modelling HCO$^+$ within the disk.

A more surprising result is the lack of HCO$^+$ emission from the disk cavity when both the dust and gas are depleted in this region (see \S\ref{lower_ten}).  In \S\ref{lower_ten}, the best-fit cavity models are depleted by a factor $10$ in both dust and gas so that the gas depletion is $\delta_\mathrm{gas, cav} = 10^{-1}$ and dust depletion is $\delta_\mathrm{dust, cav} = 10^{-5}$ w.r.t. the gas and dust in the outer disk.  The relative dust content is still a factor 10$\times$ higher than what is expected from dust depletion in the inner disk in A11 and is equivalent to the dust depletion tested for an inner disk in \citet{2013A&A...559A..46B}.  Even though there is still a sizeable reservoir of dust available within the cavity to shield the remaining gas from UV emission, the factor 10 in gas depletion has caused the HCO$^+$ line-wing emission to drop significantly lower than the observed HCO$^+$ line.  Therefore, any depletion in gas density will not be widespread across the observed disk gap.

%We vary dust-to-gas ratios $10^{-2}$, $10^{-4}$, $10^{-6}$, $10^{-8}$ and $10^{-10}$, which corresponds to cavity dust depletion factors $\delta_\mathrm{dust, cav} = 1$, $10^{-2}$, $10^{-4}$, $10^{-6}$ and $10^{-8}$ w.r.t. the dust surface density in the outer disk.  Our range of dust-to-gas ratios (and dust depletion factors) tests dust content within the inner disk (and gap) region is comparable to A11 and \citet{2013A&A...559A..46B}.  Our best-fit cavity model (d:g$=10^-6$ or $\delta_\mathrm{dust, cav}=10^{-4}$) has a dust content that is higher than expected than in A11 and the higher dust content likely inflates the SED at near- to mid-IR wavelengths (see Figure ***).   For a dust-to-gas ratio that matches the expected dust depletion in the inner disk from A11 ($\delta_\mathrm{dust}=10^{-6}$, d:g$=10^{-8}$), we find the HCO$^+$ line wing emission to be lower than the observations (as in Figure *** and Table *** for $\chi^2$ values) and better fit the SED at these wavelength ranges.  

The modelled dust depletion is consistent with past work, including \citet{2015A&A...579A.106V} which suggests dust depletion in the LkCa~15 cavity is on scales $\sim10^{-4}$ w.r.t. the ISM dust-to-gas ratio and \citet{2011ApJ...729...47Z} which suggests similar dust-to-gas mass ratios in the inner portions of the GM~Aur disk (ranging from $10^{-2}$ to $10^{-5}$ with respect to the ISM dust-to-gas ratio).  Our dust-depleted fits to the LkCa~15 disk cavity support both observational and theoretical work that forming planets sculpt the cavity and affect dust grain evolution in the disk \citep{2012A&A...545A..81P, 2012A&A...538A.114P, 2013A&A...560A.105G, 2016A&A...585A..58V}.  From these past studies, there is evidence that the planet carves out a smaller cavity in small dust grains ($\leq10$~$\mu m$) and gas.  The pressure bump generated from the planet can filter larger grains at larger radii, creating the observed dust cavities or gaps in transitional and pre-transitional disks.  This gap in gas does not appear to be steep, gradually decreasing over several AU, allowing accretion to continue onto the star.  Recent results (e.g. \citealt{2015Natur.527..342S, 2012ApJ...745....5K}) suggest there are 2-3 accreting protoplanets at radii $\sim$15-19~AU in the disk cavity, though these planets are not necessarily sufficient to open the full 50~AU dust continuum hole.  However, \citet{2012A&A...538A.114P} suggest a single $\sim15$~M$_J$ planet at a radius of 20~AU can generate a pressure gradient at 54~AU which is in better agreement with current observations of the LkCa~15 disk.  As suggested above, spatially resolved observations, particularly from molecules like CO isotopologues (e.g.  $^{13}$CO and C$^{18}$O) are necessary to further study the structure of the LkCa~15 gap, particularly the size of the gas cavity in the disk.

% though these planets are not necessarily sufficient to open the full 50~AU dust continuum hole.

%Since our models indicate the HCO$^+$ emission is primarily produced at radii $>$****~AU, this study is not sensitive to the potential gas cavity within the LkCa~15 disk.  

%As suggested above, spatially resolved observations from other molecules (e.g. CO isotoplogues $^{13}$CO and C$^{18}$O) are necessary to further study the structure of the LkCa~15 gap, particularly the size of the gas cavity.

%The dust-depleted fits to the disk cavity in our work support theories of dust filtration caused by planets (e.g. \citealt{2011ApJ...729...47Z, 2014ApJ...788..129I, 2008A&A...480..859B}).  In this scenario, planetesimals and giant planets form from the coagulation and growth of dust grains in the inner portions of the disk \citep{2008A&A...480..859B}.  These forming planets will produce gaps in dust continuum emission that range from $\sim1$ to 70~AU in radius, where gas can continue to flow across the gap and accrete on the star.  Recent results (e.g. \citealt{2015Natur.527..342S, 2012ApJ...745....5K}) suggest there are 2-3 accreting protoplanets at radii $\sim$15-19~AU in the disk cavity, though these planets are not necessarily sufficient to open the full 50~AU dust continuum hole.%Our fits to the dust-depleted inner cavity suggest %and coagulation of smaller grains that can be found in the cavity 

Using the standard gas surface density derived in Section~\ref{modelling} based on A11, we calculate the LkCa~15 inner hole mass to be $\sim$0.03~M$_\odot$ or $\sim$30~M$_J$.  In Section~\ref{lower_ten}, we determine that depleting the hole of gas by an order of magnitude ($\sim$3~M$_J$) results in substantially lower HCO$^+$ line-wing flux, indicating the gas mass is too low to account for the high-velocity HCO$^+$ emission.  This result differs from \citet{2015A&A...579A.106V} which found a drop in the cavity gas surface density by a factor of 10 (in addition to the larger drop in dust density).  However, there are differences between our method and the method implemented by \citet{2015A&A...579A.106V} to fit the disk cavity mass that make it difficult for a direct comparison.  As described in Section~\ref{modelling}, our model relies on a surface density normalisation derived in A11, where fits were made to the SED and an 880~$\micron$ image.  To fit the HCO$^+$ profile, we had to not only vary the characteristic scaling and tapering radius $R_c$ to fit the line peak, but we also had to alter the cavity scale height and dust-to-gas ratio to fit the line-wings.  In contrast, \citet{2015A&A...579A.106V} used the SED and a 440~$\micron$ continuum image to fit the surface density normalisation and dust properties and then used $^{12}$CO~$6\rightarrow5$ to fit gas properties within the disk cavity.  In addition to the differences in fitting the disk,  \citet{2015A&A...579A.106V} uses optically thick $^{12}$CO~$6\rightarrow5$ emission from LkCa15, which makes the absolute gas density and mass uncertain.  Furthermore, the dusty inner disk is poorly constrained in LkCa~15, which can shield the cavity.  This can lower the gas temperature, allowing CO to survive down to lower gas masses \citep{2013A&A...559A..46B}.

%As a consequence, our models of the LkCa~15 disk differ.  

As explained above, our models show the HCO$^+$ line-wings can be fit using a standard gas surface density with increased scale height and decreased dust-to-gas ratio within the disk cavity.   The models in \citet{2015A&A...579A.106V} depicted a relatively large, flat disk, where the full disk size is consistent with our own radius $R_{out}=400$~AU, but the surface density normalisation is a factor $\sim3.4$ larger than A11 and the scale height and flaring angle are smaller than our best-fit models (particularly for the disk cavity at $H_0/R_0=$0.06 and $\psi=0.04$) in addition to the decreased cavity gas density.  Despite the structural differences in the disk models, our derived gas cavity mass ($\sim0.03$~M$_\odot$ constrained within an order of magnitude) is consistent with \citet{2015A&A...579A.106V} ($\sim0.007$~M$_\odot$) due to the discrepancies in the surface density normalisation.  

An important uncertainty in modelling disks is understanding the complex chemistry taking place, particularly with an ion like HCO$^+$.  Due to the differences between models of the LkCa~15 disk from past work (e.g. A11; \citealt{2015A&A...579A.106V}) in addition to our work, a more detailed analysis is required to test which models can fit the large number of molecular line observations of LkCa~15 (e.g. \citealt{2001A&A...377..566V, 2006A&A...460L..43P}).  This will not only better understand the detailed chemistry ongoing in the disk, but also place more rigid constraints on the disk structure and cavity mass.  Further studies can incorporate new methods in \textsc{ProDiMo} for better understanding the UV opacity and heating within the disk from PAH re-emission (see Appendix~\ref{disk_opacity} for full details).

%an inner hole mass an order of magnitude lower ($\sim$3~M$_J$) results in substantially decreased flux in the HCO$^+$ line-wings, indicating the gas mass is too low to account for the high-velocity HCO$^+$ emission.  This result disagrees with past work from van der Marel et al. (2015) using $^{12}$CO~$J=6\rightarrow5$, which suggests the gas surface density drops by a factor of 10 (along with the much larger drop in dust surface density).  Additionally, our models over predict the $^{12}$CO~$J=6\rightarrow5$ emission by a factor of a few according to the flux presented in van der Marel et al. (2015).  The observed difference between the gas surface density could be the result of the high optical depth of $^{12}$CO in the disk (making the molecule less sensitive to disk mass) and the potential for foreground absorption to affect the $^{12}$CO spectrum (leading to lower observed $^{12}$CO emission).  Past observations of $^{12}$CO~$J=3\rightarrow2$ from the HARP instrument on the JCMT indicate there is widespread CO emission from the ambient cloud, reaching 10-100 times the modelled CO~$J=3\rightarrow2$ emission from the LkCa~15 disk {\color{red}{(Greaves, private communication)}}, which could support the premise that foreground absorption from ambient material is affecting the $^{12}$CO~$6\rightarrow5$ emission.  

Past work has suggested accretion flows or gas streamers could contain the standard ISM dust-to-gas ratio while keeping the dust emission optically thin in the disc hole \citep{2011ApJ...738..131D}.  Even though our study models the disk with a typical morphology and standard gas surface density, the fits to the unresolved observations (which integrate over the entire disk) are unaffected by geometry.  Our analysis strongly indicates the gas must be hotter and at high velocities corresponding to the smaller radii of the disk cavity.  This can only be achieved if the dust is depleted, with a large gas scale height and a sufficient amount of gas present in the inner hole.  This dense gas in the disk cavity can then maintain the observed accretion rate onto LkCa~15.  Past work found accreting protoplanets LkCa~15b and c \citep{2012ApJ...745....5K, 2015Natur.527..342S} to have masses $ < 5$-10~M$_J$ between radii of $\sim15$--19~AU and accretion rates comparable to the star.  Our calculated disk cavity mass would allow a $\sim1$~M$_J$ protoplanet at a radius of 20~AU to accrete at least $\sim0.5$~M$_J$ (assuming a $\sim1$~AU Hill radius).  At a similar orbital radius to Uranus, the final protoplanet would be $>32$ times the mass of Uranus.  Uncovering the morphology and chemistry of the cavity, including forming planets, will only be accessible in future ALMA observations (reaching $5\times10^{-4}$~M$_J$ with high spatial resolution; \citealt{2014ApJ...788..129I}).

%Our calculated disk cavity mass suggests that accreting protoplanets like LkCa~15b and c \citep{2012ApJ...745....5K, 2015Natur.527..342S}, estimated to have a masses  $ < 5$-10~M$_J$ and accretion rates comparable to its star and other smaller planets.  Uncovering the morphology and chemistry of the cavity, including forming planets, will only be accessible in future ALMA observations (reaching a resolution of $5\times10^{-4}$~M$_J$; \citealt{2014ApJ...788..129I}). 

\acknowledgments{ED acknowledges support by the Science and Technology Facilities Council (STFC) of the United Kingdom.  The JCMT is operated by the Joint Astronomy Centre (JAC) on behalf of the STFC, the National Research Council of Canada and the Netherlands Organisation for Scientific Research.  We acknowledge the data analysis facilities provided by the Starlink Project maintained by the Joint Astronomy Centre.  The research leading to these results has received funding from the European Union Seventh Framework Programme FP7-2011 under grant agreement no 284405.  Lastly, we thank the anonymous referee for very useful comments and suggestions. }

\bibliography{references}

%\appendix
%\section{Model parameters}

%Need to insert table of parameters
\begin{table}
\centering
\begin{tabular}{l l c}
\hline
\hline
\multicolumn{3}{c}{Fixed Parameters} \\
\hline
\multicolumn{3}{c}{Stellar properties}\\
Stellar mass & $M_\ast$ (M$_\odot$) & 1.01\\
Stellar temperature & $T_\ast$ (K) & 4730 \\
Stellar luminosity & $L_\ast$ ($L_\odot$) & 1.2 \\
X-ray luminosity & $L_X$ (erg s$^{-1}$) & $3\times10^{30}$ \\
Solid material mass density & $\rho_\mathrm{dust}$ (g~cm$^{-3}$) & 2.3 \\
Turbulent velocity & $v_\mathrm{turb}$ (km s$^{-1}$) & 0.1 \\
Disk inclination & $i$ (degrees) & 49 \\
Flaring index & $\beta$ &1.2 \\
Fraction of PAHs w.r.t ISM & $f_{\mathrm{PAH}}$ & $10^{-2}$ \\
Cosmic ray flux & $\zeta$(s$^{-1}$) & $10^{-17}$ \\
\hline
\multicolumn{3}{c}{Outer disk parameters}\\
Surface density normalisation & $\Sigma_{0,85\mathrm{AU}}$ (g cm$^{-2}$) & 10.8 \\
Reference scale height & $H_0$ (AU) & 10 \\
Reference radius & $R_0$ (AU) & 100 \\
Minimum grain size & $a_{\mathrm{min}}\ (\micron)$ & 0.005 \\
Maximum grain size & $a_{\mathrm{max}} \ (\micron)$ & 1000 \\
Dust size distribution index & $p$ & 3.5 \\
Dust-to-gas ratio & d:g & $10^{-2}$ \\

\hline
\end{tabular}
\caption{Stellar and accretion properties for LkCa~15.  LkCa~15 stellar mass, stellar temperature, flaring index and outer disk parameters taken from \citet{2011ApJ...732...42A}.  X-ray luminosity from \citet{2013ApJ...765....3S}.}
\label{table1}
\end{table}

\label{appendix_param}
\begin{table*}
\centering
\caption{Model parameters.  Columns 1--3 are the radii parameters.  The asterisk ($^\ast$) in $R_{out}$ indicates that the model represents the inner disk only.  Columns 4 \& 5 are the mass of the outer disc ($>50$~AU) and the inner disc (0.1--50~AU).  Column 6 is the dust-to-gas ratio of the inner cavity (0.1--50~AU).  Columns 7 \& 8 are the settling parameters defined in \S\ref{modelling}.  Column 9 is the gas scale height $H_0$ at reference radius $R_0=100$~AU.  Column 10 is the $\chi^2$ values for the line-wing ($\sim$1.8 to 4.0~km~s$^{-1}$ or 8.8 to 11.0~km~s$^{-1}$) and column 11 is the reduced chi-square $\chi^2 _{red}$ assuming 4 degrees of freedom.  Lastly, column 12 gives the corresponding figure. }

\begin{tabular}{c c c c c c c c c c c c}
\hline

$R_{in}$ & $R_c$ & $R_{out}$ & $M_{d,out}$ & $M_{d,cav}$ & {\emph{d:g}}$_{cav}$ & $a_{s,cav}$ & $\delta_{s,cav}$  & $H_{0,cav}$  & $\chi^2$ & $\chi^2_{red}$ & Fig. \\
(AU) & (AU) & (AU) & (M$_\odot$) & (M$_\odot$) & & ($\micron$) & & (AU) & & & \\
(1) & (2) & (3) & (4) & (5) & (6) & (7) & (8) & (9) & (10) & (11) & (12) \\
\hline
\hline
\multicolumn{11}{l}{Empty inner cavity: Comparison to dust results of A11 (\S\ref{dust_gas_results_same})}\\
\hline
50 & 85 & 250 & 0.021 & -- & -- & -- & -- & -- & -- & -- & \ref{fig:tests_cavity} \\
50 & 85 & 500 & 0.028 & -- & -- & -- & -- & -- & -- & -- &\ref{fig:tests_cavity} \\
50 & 85 & 1000  & 0.030 & -- & -- & -- & -- & -- & -- & -- &\ref{fig:tests_cavity} \\
%50 & 85 & 1000 & 0.031 & -- & -- & -- & -- & -- & 24.5 & \ref{fig:tests_cavity} \\
\hline
\hline
\multicolumn{11}{l}{Empty inner cavity: Determining $R_c$ (\S\ref{varying_rc})}\\
\hline
50 & 350 & 350 & 0.113 & -- & -- & -- & -- & -- & -- & -- &\ref{fig:tests_cavity} \\
50 & 300 & 300 & 0.093 & -- & -- & -- & -- & -- & -- & -- &\ref{fig:tests_cavity} \\
50 & 250 & 250 & 0.073 & -- & -- & -- & -- & -- & -- & -- &\ref{fig:tests_cavity} \\
50 & 300 & 350 & 0.104 & -- & -- & -- & -- & -- & -- & -- &\ref{fig:tests_cavity} \\
%50 & 300 & 300 & 0.084 & -- & -- & -- & -- & -- & 24.6 & \ref{fig:tests_cavity} \\
\hline
\hline
\multicolumn{11}{l}{Disk without an inner cavity: Cavity gas with ISM d:g (Sections~\ref{nohole_102})}\\
\hline
0.1 & 300 & 350 & 0.134 & 0.030 & $10^{-2}$ & 0.1 & 1.0 & 10 & -- & -- &\ref{fig:tests_cavity} \\
0.1 & 300 & 375 & 0.138 & 0.030 & $10^{-2}$ & 0.1 & 1.0 & 10 & -- & -- &\ref{fig:tests_cavity} \\
0.1 & 300 & 400 & 0.143 & 0.030 & $10^{-2}$ & 0.1 & 1.0 & 10 & -- & -- &\ref{fig:tests_cavity} \\
%0.1 & 250 & 50$^\ast$ & -- & 0.029 & $10^{-2}$ & 0.1 & 1.0 & 10 & 24.6 & \ref{fig:tests_cavity}\\
\hline
\hline
\multicolumn{11}{l}{Effects of dust settling and mixing in the inner cavity (\S\ref{settling_effects})}\\
\hline
0.1 & 300 & 50$^\ast$ & -- & 0.030 & $10^{-2}$ & 0.1 & 1.0 & 10 & 31.9 & 8.0 & \ref{fig:asettle_dsettle_test} \\
0.1 & 300 & 50$^\ast$ & -- & 0.030 & $10^{-2}$ & 0.1 & 0.5 & 10 & 32.2 & 8.1 & \ref{fig:asettle_dsettle_test} \\
0.1 & 300 & 50$^\ast$ & -- & 0.030 & $10^{-2}$ & 0.1 & 0.1 & 10 & 34.9 & 8.7 & \ref{fig:asettle_dsettle_test} \\
0.1 & 300 & 50$^\ast$ & -- & 0.030 & $10^{-2}$ & 0.1 & 0.05 & 10 & 35.7 & 8.9 & \ref{fig:asettle_dsettle_test} \\
0.1 & 300 & 50$^\ast$ & -- & 0.030 & $10^{-2}$ & 0.1 & 0.01 & 10 & 36.6 & 9.1 & \ref{fig:asettle_dsettle_test} \\

0.1 & 300 & 50$^\ast$ & -- & 0.030 & $10^{-2}$ & 0.01 & 1.0 & 10 & 27.9  & 7.0 & \ref{fig:asettle_dsettle_test} \\
0.1 & 300 & 50$^\ast$ & -- & 0.030 & $10^{-2}$ & 0.01 & 0.5 & 10 & 29.2 & 7.3 & \ref{fig:asettle_dsettle_test} \\
0.1 & 300 & 50$^\ast$ & -- & 0.030 & $10^{-2}$ & 0.01 & 0.1 & 10 & 34.2 & 8.6 & \ref{fig:asettle_dsettle_test} \\
0.1 & 300 & 50$^\ast$ & -- & 0.030 & $10^{-2}$ & 0.01 & 0.05 & 10 & 35.4 & 8.8 & \ref{fig:asettle_dsettle_test} \\
0.1 & 300 & 50$^\ast$ & -- & 0.030 & $10^{-2}$ & 0.01 & 0.01 & 10 & 36.5 & 9.1 & \ref{fig:asettle_dsettle_test} \\

0.1 & 300 & 50$^\ast$ & -- & 0.030 & $10^{-10}$ & 0.1 & 1.0 & 10 & 22.8 & 5.7 & \ref{fig:asettle_dsettle_test} \\
0.1 & 300 & 50$^\ast$ & -- & 0.030 & $10^{-10}$ & 0.1 & 0.5 & 10 & 19.8 & 4.9 & \ref{fig:asettle_dsettle_test} \\
0.1 & 300 & 50$^\ast$ & -- & 0.030 & $10^{-10}$ & 0.1 & 0.1 & 10 & 16.5 & 4.1 & \ref{fig:asettle_dsettle_test} \\
0.1 & 300 & 50$^\ast$ & -- & 0.030 & $10^{-10}$ & 0.1 & 0.05 & 10 & 16.4 & 4.1 & \ref{fig:asettle_dsettle_test}, \ref{fig:dgratio_cavity_test} \\

\hline
\multicolumn{11}{l}{{\emph{cont.}}}\\
\end{tabular}
\label{table_param}
\end{table*}

\begin{table}
\caption*{}
\begin{tabular}{c c c c c c c c c c c c}
\hline
$R_{in}$ & $R_c$ & $R_{out}$ & $M_{d,out}$ & $M_{d,cav}$ & {\emph{d:g}}$_{cav}$ & $a_{s,cav}$ & $\delta_{s,cav}$  & $H_{0,cav}$  & $\chi^2$ & $\chi^2 _{red}$ & Fig. \\
(AU) & (AU) & (AU) & (M$_\odot$) & (M$_\odot$) & & ($\micron$) & & (AU) & & & \\
(1) & (2) & (3) & (4) & (5) & (6) & (7) & (8) & (9) & (10) & (11) & (12) \\
\hline
0.1 & 300 & 50$^\ast$ & -- & 0.030 & $10^{-10}$ & 0.1 & 0.01 & 10 & 16.5 & 4.1 & \ref{fig:asettle_dsettle_test} \\
0.1 & 300 & 50$^\ast$ & -- & 0.030 & $10^{-10}$ & 0.01 & 1.0 & 10 & 23.0 & 5.8 & \ref{fig:asettle_dsettle_test} \\
0.1 & 300 & 50$^\ast$ & -- & 0.030 & $10^{-10}$ & 0.01 & 0.5 & 10 & 21.6 & 5.4 & \ref{fig:asettle_dsettle_test} \\
0.1 & 300 & 50$^\ast$ & -- & 0.030 & $10^{-10}$ & 0.01 & 0.1 & 10 & 16.8 & 4.2 & \ref{fig:asettle_dsettle_test} \\
0.1 & 300 & 50$^\ast$ & -- & 0.030 & $10^{-10}$ & 0.01 & 0.05 & 10 & 16.4 & 4.1 & \ref{fig:asettle_dsettle_test} \\
0.1 & 300 & 50$^\ast$ & -- & 0.030 & $10^{-10}$ & 0.01 & 0.01 & 10 & 16.4 & 4.1 & \ref{fig:asettle_dsettle_test},\ref{fig:dgratio_cavity_test} \\
\hline
\hline
\multicolumn{11}{l}{Varying scale height in the inner cavity (\S\ref{varying_scale_height})}\\
\hline
0.1 & 300 & 50$^\ast$ & -- & 0.030 & $10^{-10}$ & 0.1 & 0.05 & 20 & 12.2 & 3.0 & \ref{fig:dgratio_cavity_test} \\
0.1 & 300 & 50$^\ast$ & -- & 0.030 & $10^{-10}$ & 0.1 & 0.05 & 30 & 9.7 & 2.4 & \ref{fig:dgratio_cavity_test} \\
0.1 & 300 & 50$^\ast$ & -- & 0.030 & $10^{-10}$ & 0.1 & 0.05 & 40 & 8.8 & 2.2 &  \ref{fig:dgratio_cavity_test} \\
0.1 & 300 & 50$^\ast$ & -- & 0.030 & $10^{-10}$ & 0.1 & 0.05 & 50 & 8.2 & 2.1 & \ref{fig:dgratio_cavity_test} \\
0.1 & 300 & 50$^\ast$ & -- & 0.030 & $10^{-10}$ & 0.1 & 0.05 & 60 & 8.2 & 2.1 & \ref{fig:dgratio_cavity_test} \\

0.1 & 300 & 50$^\ast$ & -- & 0.030 & $10^{-10}$ & 0.01 & 0.01 & 20 & 12.3 & 3.1 & \ref{fig:dgratio_cavity_test} \\
0.1 & 300 & 50$^\ast$ & -- & 0.030 & $10^{-10}$ & 0.01 & 0.01 & 30 & 10.0 & 2.5 & \ref{fig:dgratio_cavity_test} \\
0.1 & 300 & 50$^\ast$ & -- & 0.030 & $10^{-10}$ & 0.01 & 0.01 & 40 & 8.9 & 2.2 & \ref{fig:dgratio_cavity_test} \\
0.1 & 300 & 50$^\ast$ & -- & 0.030 & $10^{-10}$ & 0.01 & 0.01 & 50 & 8.7 & 2.2 & \ref{fig:dgratio_cavity_test} \\
0.1 & 300 & 50$^\ast$ & -- & 0.030 & $10^{-10}$ & 0.01 & 0.01 & 60 & 8.3 & 2.1 & \ref{fig:dgratio_cavity_test} \\
\hline
\hline
\multicolumn{11}{l}{Constraining the inner cavity d:g ratio (\S\ref{dust_to_gas})}\\
\hline
0.1 & 300 & 50$^\ast$ & -- & 0.030 & $10^{-8}$ & 0.1 & 0.05 & 60 & 7.7 & 1.9 & \ref{fig:dgratio_cavity_test} \\
0.1 & 300 & 50$^\ast$ & -- & 0.030 & $10^{-6}$ & 0.1 & 0.05 & 60 & 4.7 & 1.2 & \ref{fig:dgratio_cavity_test}\\
0.1 & 300 & 50$^\ast$ & -- & 0.030 & $10^{-4}$ & 0.1 & 0.05 & 60 & 6.3 & 1.6 & \ref{fig:dgratio_cavity_test} \\
0.1 & 300 & 50$^\ast$ & -- & 0.030 & $10^{-2}$ & 0.1 & 0.05 & 60 & 38.5 & 9.6 & \ref{fig:dgratio_cavity_test} \\

0.1 & 300 & 50$^\ast$ & -- & 0.030 & $10^{-8}$ & 0.01 & 0.01 & 60 & 7.8 & 1.9 & \ref{fig:dgratio_cavity_test} \\
0.1 & 300 & 50$^\ast$ & -- & 0.030 & $10^{-6}$ & 0.01 & 0.01 & 60 & 4.6 & 1.2 & \ref{fig:cavity_fit}, \ref{fig:cavity_test}, \ref{fig:full_model_end}, \ref{fig:dgratio_cavity_test} \\
0.1 & 300 & 50$^\ast$ & -- & 0.030 & $10^{-4}$ & 0.01 & 0.01 & 60 & 6.7 & 1.7 & \ref{fig:dgratio_cavity_test} \\
0.1 & 300 & 50$^\ast$ & -- & 0.030 & $10^{-2}$ & 0.01 & 0.01 & 60 & 39.0 & 9.7 & \ref{fig:dgratio_cavity_test} \\
\hline
\hline
\multicolumn{11}{l}{Constraining the gas mass in the inner cavity (\S\ref{lower_ten})}\\
\hline

0.1 & 300 & 50$^\ast$ & -- & 0.003 & $10^{-6}$ & 0.01 & 0.05 & 60 & -- & -- & \ref{fig:cavity_test} \\
0.1 & 300 & 50$^\ast$ & -- & 0.290 & $10^{-6}$ & 0.01 & 0.05 & 60 &  -- & -- & \ref{fig:cavity_test} \\

\hline
\hline
\end{tabular}
\end{table}

\begin{table}
\centering
\caption{Comparison between the best-fit two-disk model and the unfolded LkCa~15 HCO$^+$ spectrum.  The dust settling parameters used in the fit to the cavity are $a_{s}=0.01\micron$ and $\delta_{s}=0.01$.  Columns 1 and 2 are the scale height and dust-to-gas ratio for the inner cavity region.  Columns 3 and 4 are the $\chi^2$ and $\chi^2 _{red}$ values for full line-wings at velocities $\sim$1.8 to 4.0~km~s$^{-1}$ and 8.8 to 11.0~km~s$^{-1}$.  We assume 11 degrees of freedom for $\chi^2_{red}$ (where there are 14 data points and 3 fitted parameters).  Lastly, column 5 is the corresponding figure in the text.}
\begin{tabular}{c c c c c}
\hline
$H_{0,cav}$ & $ d:g_{cav}$  & $\chi^2$ & $\chi^2 _{red}$ & Fig. \\
(AU) & & & & \\
(1) & (2) & (3) & (4) & (5)  \\
\hline
%0.1 & 0.05 & 9.1 & 0.8 & \ref{fig:final_unfolded} \\
60 & 10$^{-6}$ & 9.0 & 0.8 & \ref{fig:final_unfolded} \\
\hline
\hline
\end{tabular}
\label{table_final}
\end{table}

\clearpage

\appendix

\section{\textsc{ProDiMo} Disk Opacity}
\label{disk_opacity}

The method \textsc{ProDiMo} uses to calculate opacities in the disk has been altered from that of previous publications.  Typically, the code is designed to assume the dust is more opaque in the UV than the gas in the disk.  This assumption is relatively valid for interstellar environments, where dust grains are small and dust opacities in the UV are large.  For protoplanetary disks, this isn't necessarily the case.  Grain growth can cause a substantial fraction of dust grains to grow to a few millimetres in size, which leads to lower dust opacities.  In evolved disks, dust has a tendency to settle in the midplane, which can lead to lower dust opacities in the upper layers of the disk.  Furthermore, the gas contained in the dust gap of a pre-transitional disk like LkCa~15 will have low UV dust opacities due to the low dust content in this region.  In these low dust opacity scenarios, the gas can become more opaque than the dust and act as the dominant source of UV opacity.  We, therefore, use UV opacities calculated from a combination of the gas (primarily), PAHs and dust in our models of LkCa~15.

The gas absorption coefficient is calculated as \begin{equation}  \kappa_\nu ^\mathrm{gas,abs} = \sum_i n_i \sigma_\mathrm{abs}(i,\nu), \end{equation} where the gas absorption cross-section $\sigma_\mathrm{abs}(i,\nu)$ is taken from the Leiden database \citep{2005A&A...432..369S} for only the continuous photodissociation and photoionisation of astrophysical relevant molecules using a range of photo-reactions (listed in Table~\ref{table_dust_opacity}).  A similar formula applies to the gas scattering coefficient $\kappa_\nu ^\mathrm{gas,sca}$ for Thomson scattering on free electrons and Rayleigh scattering on H, He and H$_2$ using cross-sections from Bues \& Wehrse (1976).  

\citet{2009A&A...501..383W} (Equation~13) shows how the radiative transfer equation is solved using UV dust opacities only.  Similarly, \citet{2016A&A...586A.103W} (Equation~7) shows how the radiative transfer equation is solved using both dust and PAH opacities.  For our method of incorporating UV gas opacities, the source function becomes \begin{equation} S_\nu = \frac{\kappa_\nu ^ \mathrm{dust,abs} B_\nu (T_\mathrm{dust}) + \kappa_\nu ^ \mathrm{PAH,abs} B_\nu (T_\mathrm{PAH}) + (\kappa_\nu^\mathrm{dust,sca} + \kappa_\nu ^\mathrm{gas,sca}) J_\nu}{\kappa_\nu ^ \mathrm{dust,ext} + \kappa_\nu ^\mathrm{gas,ext} + \kappa_\nu ^\mathrm{PAH,ext}},  \end{equation} where the dust and gas extinction coefficients (cm$^{-1}$) are $\kappa_\nu^\mathrm{dust,ext} = \kappa_\nu ^\mathrm{dust,abs} + \kappa_\nu ^\mathrm{dust,sca}$ and  $\kappa_\nu^\mathrm{gas,ext} = \kappa_\nu ^\mathrm{gas,abs} + \kappa_\nu ^\mathrm{gas,sca}$.  In this model, it is assumed that there is no re-emission from the gas that absorbs in the UV.  It is expected that the PAHs have insignificant scattering due to their small size.  Therefore, the PAH extinction coefficient can be estimated from absorption only, $\kappa_\nu^\mathrm{PAH,ext} = \kappa_\nu ^\mathrm{PAH,abs}$.

Even though the model uses a fixed density structure, we must iterate between the chemistry and radiative transfer.  \textsc{ProDiMo} uses the UV gas opacities in the 2D radiative transfer.  The radiative transfer depends on particle concentrations calculated by the chemistry and energy balance, which, in turn, depend on the mean intensities calculated by the radiative transfer.  Figure~1 of \citet{2009A&A...501..383W} describes the `global iterations' used in \textsc{ProDiMo} (though we do not adjust the density structure with each iteration).  

We note this method of calculating the opacities is a work in progress and further improvements are needed.  A caveat to implementing our current method is that the  gas opacities have been calculated from a chemical rate-network, i.e. only a small part of the opacities that cause chemical reactions.  It is likely the actual UV gas opacities are larger than what is calculated from \textsc{ProDiMo}.  With larger UV gas opacities, this could cause gas temperatures to decrease in the disk and potentially affect the  chemistry.  Furthermore in upcoming publications, we plan to test new improvements to the \textsc{ProDiMo} code (as outlined in \citealt{2016A&A...586A.103W}), where PAHs can now re-emit absorbed energy via strong PAH mid-IR resonances which heats the disk.  This process can affect the dust and gas temperature structure in the disk and in turn affect chemistry.

\section{Cavity Fits}
\label{appendix_cavity}

This section details the full set of models used to determine the fits to the HCO$^+$ line-wings (Section~\ref{cavity_fits}), including constraints on the dust settling and mixing (Section~\ref{settling_effects}), gas scale-height (Section~\ref{varying_scale_height}) and dust-to-gas ratio (Section~\ref{dust_to_gas}) in the disk cavity.

\subsection{Effects of Dust Settling and Mixing}
\label{settling_effects}

Until this point the model settling parameters have remained constant, where the minimum grain size affected by settling is $a_s = 0.1$~$\micron$ and the settling exponent is $\delta_s = 1.0$ (see \S\ref{disk_structure}).  The HCO$^+$ line emission is significantly affected by dust, i.e. cooling and heating, and possible recombination when electrons are released from grains in high UV environments.  Dust grains in the disk are controlled both by the dust-to-gas ratio and the dust settling/mixing.  To test the effects of dust on the HCO$^+$ emission, we first vary the minimum grain size affected by settling ($a_s=0.1$ and 0.01~$\micron$).  Decreasing the settling grain size increases settling for the dust grains $a> a_s$ present in the disk (see \S\ref{disk_structure}).  Additionally, we vary the settling exponent $\delta_s$ to values less than the original exponent from \S\ref{disk_structure} ($\delta_s=$1.0, 0.5, 0.1, 0.05, and 0.01).   Decreasing $\delta_s$ increases the scale heights of larger grains ($a>a_s$), mixing the dust with the gas.  Since the dust settling in the disk is likely linked to the modelled d:g, we compare models with the ISM dust-to-gas ratio (d:g$=10^{-2}$) to models with an arbitrarily small dust-to-gas ratio (d:g$=10^{-10}$).  We note that we only examine the inner cavity of the LkCa~15 disk and exclude the outermost region ($r>50$~AU) in the following sections (i.e. \S\ref{settling_effects}--\ref{dust_to_gas} and \S\ref{cavity_fits}--\ref{lower_ten}).  

\begin{figure*}
\centering
\includegraphics[width=3in]{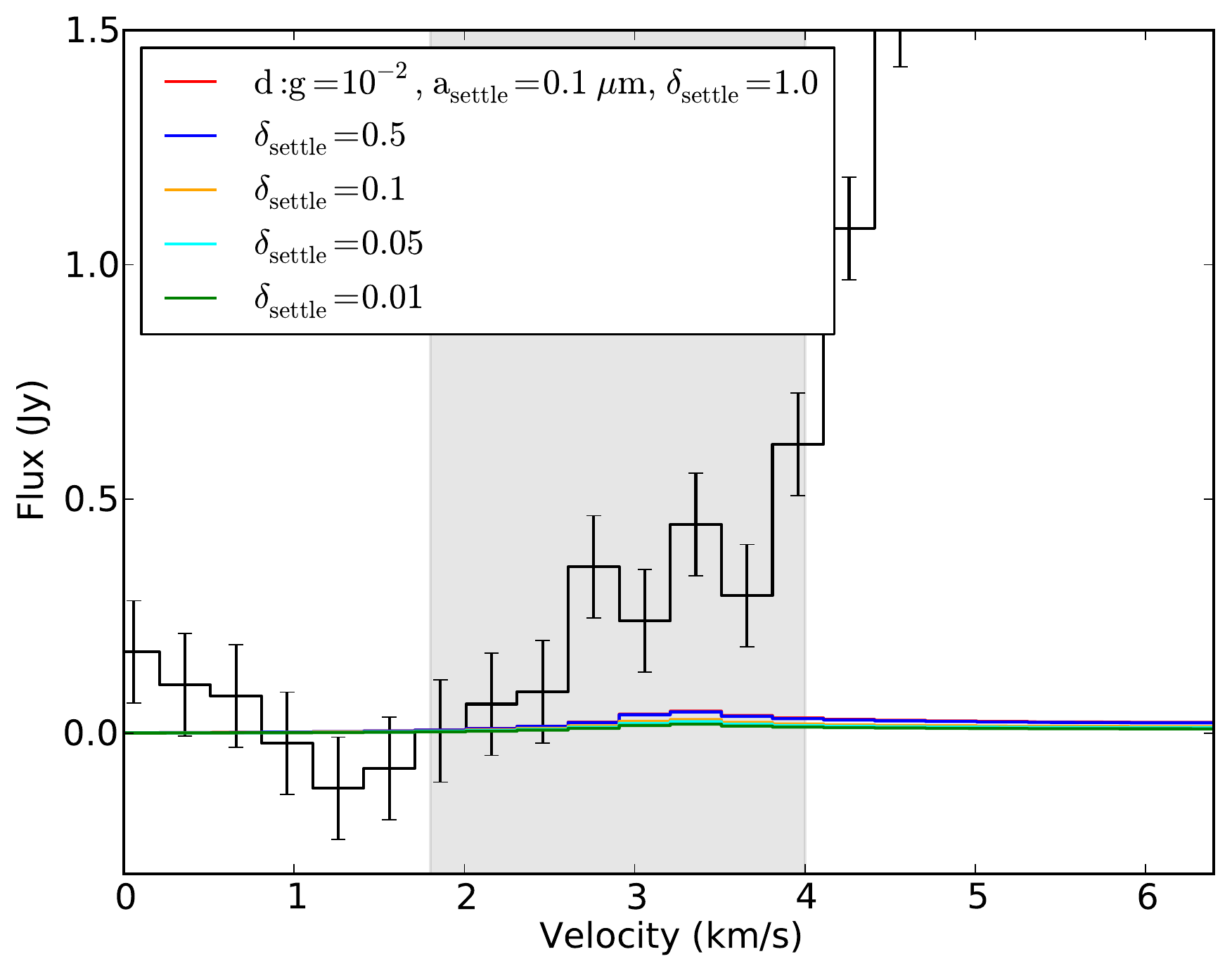}
\includegraphics[width=3in]{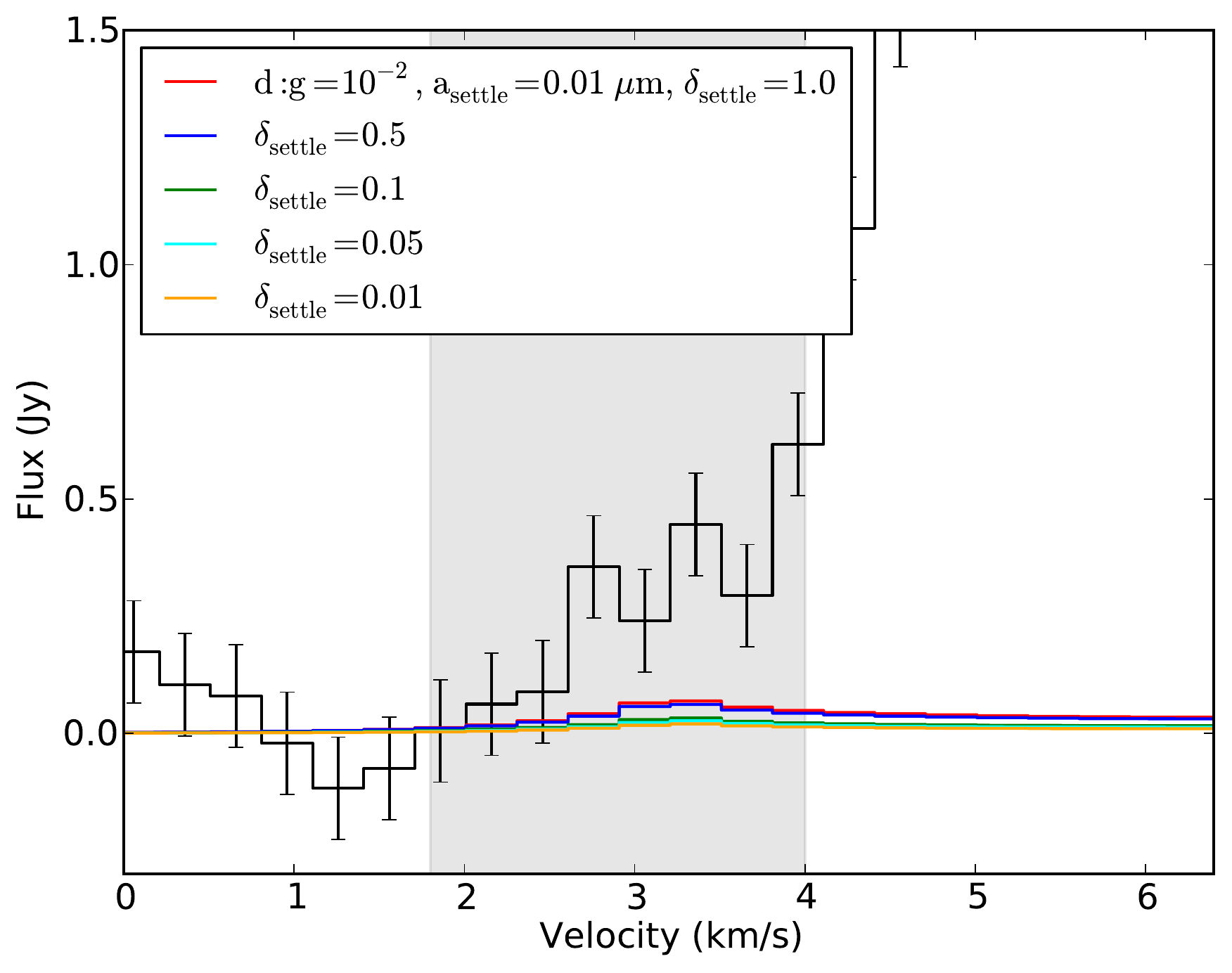}
\includegraphics[width=3in]{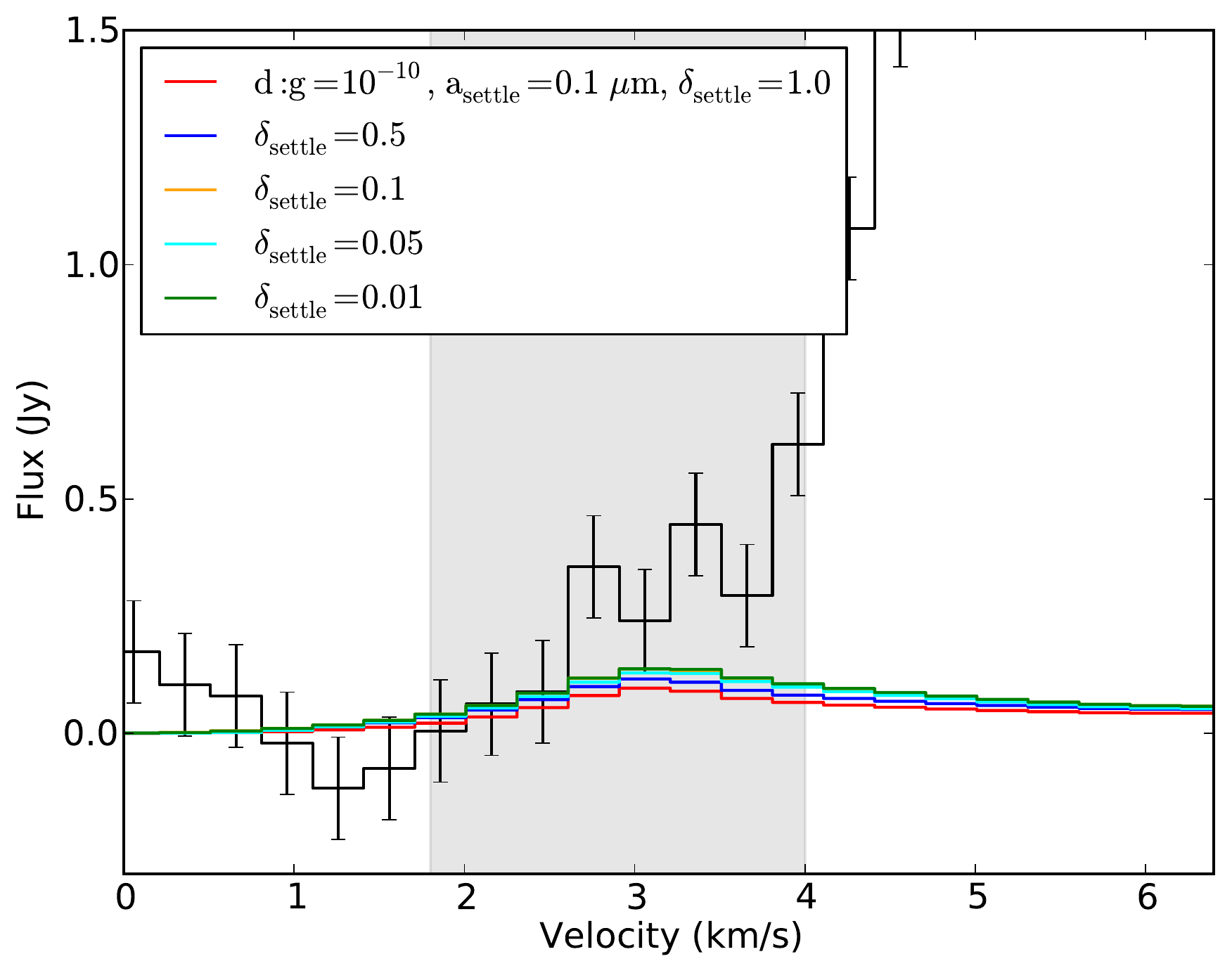}
\includegraphics[width=3in]{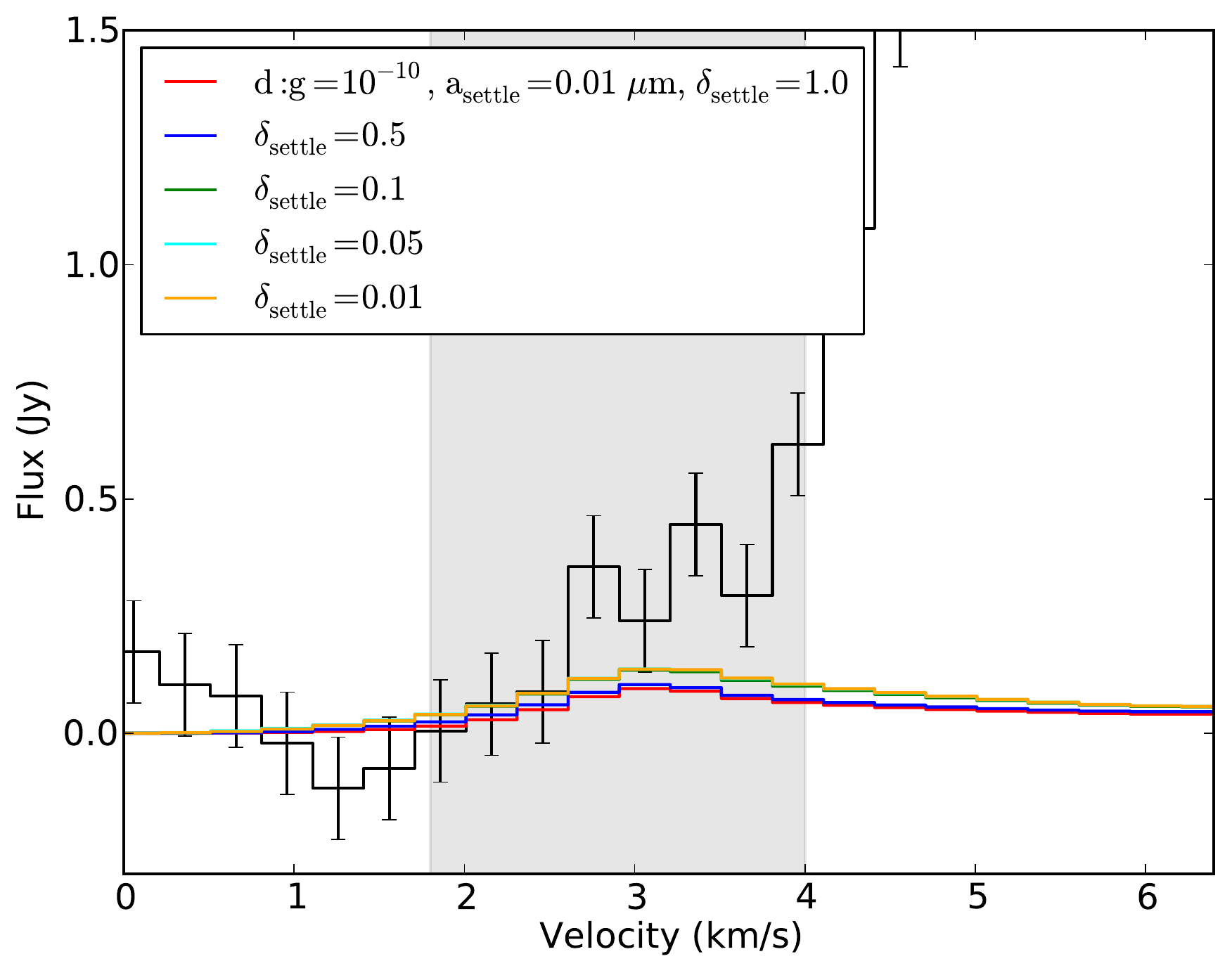}
\caption{Effects of dust settling and mixing in the inner disk cavity (see Section~\ref{settling_effects}).  Dust-to-gas ratios are set to $10^{-2}$ (top) and $10^{-10}$ (bottom).  Settling parameters are $a_s=0.1\micron$ (left) and 0.01$\micron$ (right) with varying settling exponent values ranging ($\delta_s= 1.0, 0.5, 0.1, 0.05$ and 0.01).  Models with lower dust-to-gas ratios (d:g$=10^{-10}$) have increased HCO$^+$ line flux compared to d:g$=10^{-2}$.  The highest flux is found when $a_s=0.1\micron$ and $\delta_s=0.05$ or $a_s=0.01\micron$ and $\delta_s=0.01$ for d:g$=10^{-10}$.}
\label{fig:asettle_dsettle_test}
\end{figure*}

Figure~\ref{fig:asettle_dsettle_test} is a comparison of the results for varying $a_s$ and $\delta_s$ in the disk cavity, assuming d:g$=10^{-2}$ and $10^{-10}$.  In general, dust settling and mixing parameters have little effect on the HCO$^+$ line-wing emission.  However, we do find minor correlations between the HCO$^+$ flux and $\delta_s$ for the different dust-to-gas ratios.  With an ISM d:g, the gas kinetic temperature and HCO$^+$ flux increase with increasing $\delta_s$ (i.e., with more settled dust), while with a lower d:g$=10^{-10}$, HCO$^+$ flux increases with decreasing $\delta_s$.  Dust cooling is the predominant effect for an ISM d:g, where increased settling (i.e., grains less well-mixed with the gas) reduces the cooling, and HCO$^+$ flux increases with the gas temperature.  With a much lower d:g$=10^{-10}$, cooling by grains is insignificant; instead, chemistry becomes the driving force behind HCO$^+$ emission.  Specifically, a higher $\delta_s$ means that the small amount of remaining dust becomes better mixed with the gas, thereby shielding the gas more from the incident UV and enhancing H$_2$ formation, which leads to higher HCO$^+$ production.  We note that models with a lower dust-to-gas ratio ($10^{-10}$) yield lower $\chi_{red}^2$ values (i.e., better fits) than the standard ISM d:g, with the best-fits for d:g$=10^{-10}$ obtained with [$a_s$, $\delta_s$] = [0.1$\micron$, 0.05] or [0.01$\micron$, 0.01].  The best-fits have similar settling parameters, where low $\delta_s$ indicates the grains are relatively well-mixed with the gas in the disk cavity.  For [$a_s$, $\delta_s$] = [0.1$\micron$, 0.05], $H_{d,0.1\micron}=10$~AU, $H_{d,1\micron}=9.9$~AU, $H_{d,10\micron}=9.9$~AU, $H_{d,100\micron}=9.8$~AU and $H_{d,1\mathrm{mm}}=9.8$~AU.  For [$a_s$, $\delta_s$] = [0.01$\micron$, 0.01], $H_{d,0.1\micron}=9.9$~AU, $H_{d,1\micron}=9.8$~AU, $H_{d,10\micron}=$9.7~AU, $H_{d,100\micron}=9.5$~AU and $H_{d,1\mathrm{mm}}=9.4$~AU.

%{\color{red}{We note the gas scale height and dust-to-gas ratio parameters, constrained in \ref{varying_scale_height} and \ref{dust_to_gas}, have more significant effects on the HCO$^+$ line wing emission corresponding to the disk cavity.     }}

\subsection{Varying Gas-Disk Scale Height in the Cavity}
\label{varying_scale_height}
In \S\ref{settling_effects}, there are only minor effects from dust settling and mixing on the HCO$^+$ line-wing emission corresponding to the disk cavity.  In following sections (\ref{varying_scale_height} and \ref{dust_to_gas}), we find the gas scale height and dust-to-gas ratio parameters have more significant effects on the HCO$^+$ flux.

From the dust settling analysis in \S\ref{settling_effects}, we detect a small increase in HCO$^+$ line flux with the arbitrarily small dust-to-gas ratio ($10^{-10}$).  If we continue to assume the disk cavity has little to no dust in the cavity, then it is possible the scale height of the gas in this region is different from the outer portion of the disk.  For example, the lower dust-to-gas ratio causes higher gas temperatures towards the cavity midplane (since the UV can penetrate further into the disk and heat the gas).  This could theoretically increase the gas scale height in the cavity.  A higher gas scale height would also increase the molecular line emitting area, which would increase the optically thick HCO$^+$ flux from the cavity.  The dust-to-gas ratio is later constrained within the disk cavity in Section~\ref{dust_to_gas}.

The gas scale height $H_g$ at radius $r$ follows the relation $H_g = H_0 (r/R_0)^\beta,$ where $H_0$ is the reference scale height (set at 10~AU; \S\ref{disk_structure}) at radius $R_0$ ($R_0=100$~AU) and $\beta$ is the flaring index ($\beta=1.2$; A11).  To test the effects of varying gas scale height in the inner disk on the HCO$^+$ line-wings, we increase the reference scale height by increments of 10~AU from 10 to 60~AU.   At the outer cavity radius $r=50$~AU, this corresponds to gas height $H_g$ ranging from 4 to 26~AU.  We continue to assume the dust-to-gas ratio is arbitrarily small ($10^{-10}$) from the increase in HCO$^+$ line-wing flux with the best-fit values for $a_s$ (0.1 and 0.01~$\micron$) and $\delta_s$ (0.05 and 0.01 respectively).

%The lower dust-to-gas ratio causes the cavity to have an increased gas temperature towards the disk midplane (since the dust can no longer shield the gas) and decreased temperature towards the outer portions of the disk, which can alter the chemistry
%The gas scale height $H_g$ at radius $r$ follows the relation $H_g = H_0 (r/R_0)^\beta,$ where $H_0$ is the reference scale height (set at 10~AU; \S\ref{disk_structure}) at radius $R_0$ ($R_0=100$~AU) and $\beta$ is the flaring index ($\beta=1.2$; A11).  To test the effects of varying gas scale height in the inner disk on the HCO$^+$ line-wings, we increase the reference scale height by increments of 10~AU from 10 to 60~AU.   At the outer cavity radius $r=50$~AU, this corresponds to gas height $H_g$ ranging from 4 to 26~AU.  We continue to assume the dust-to-gas ratio is arbitrarily small ($10^{-10}$) from the increase in HCO$^+$ line-wing flux with the best-fit values for $a_s$ (0.1 and 0.01~$\micron$) and $\delta_s$ (0.05 and 0.01 respectively).   

Figure~\ref{fig:dgratio_cavity_test} shows varying scale height $H_0$.  In both cases, the HCO$^+$ line flux steadily increases with increasing scale height, where the best-fit corresponds to $H_0=60$~AU ($H_g=26$~AU at $r=50$~AU) and $\chi_{red}^2$ values have improved from the standard $H_0=10$~AU scale height.  The larger scale height increases the molecular line emitting area of the disk, which results in the increase of HCO$^+$ high-velocity line-wing emission.

\subsection{Constraining the Disk Dust-to-Gas Ratio}
\label{dust_to_gas}

Since we have been able to model significant HCO$^+$ line-wing flux by increasing the scale height of the gas in the disk cavity, we can now constrain the dust-to-gas ratio in the gap.  We vary the dust-to-gas ratio from the standard ISM value (10$^{-2}$) to the lowest value we used in previous models (10$^{-10}$).  This effectively changes the amount of dust we find in the disk, where the dust will decrease with decreasing d:g.

\begin{figure*}
\centering
\includegraphics[width=3in]{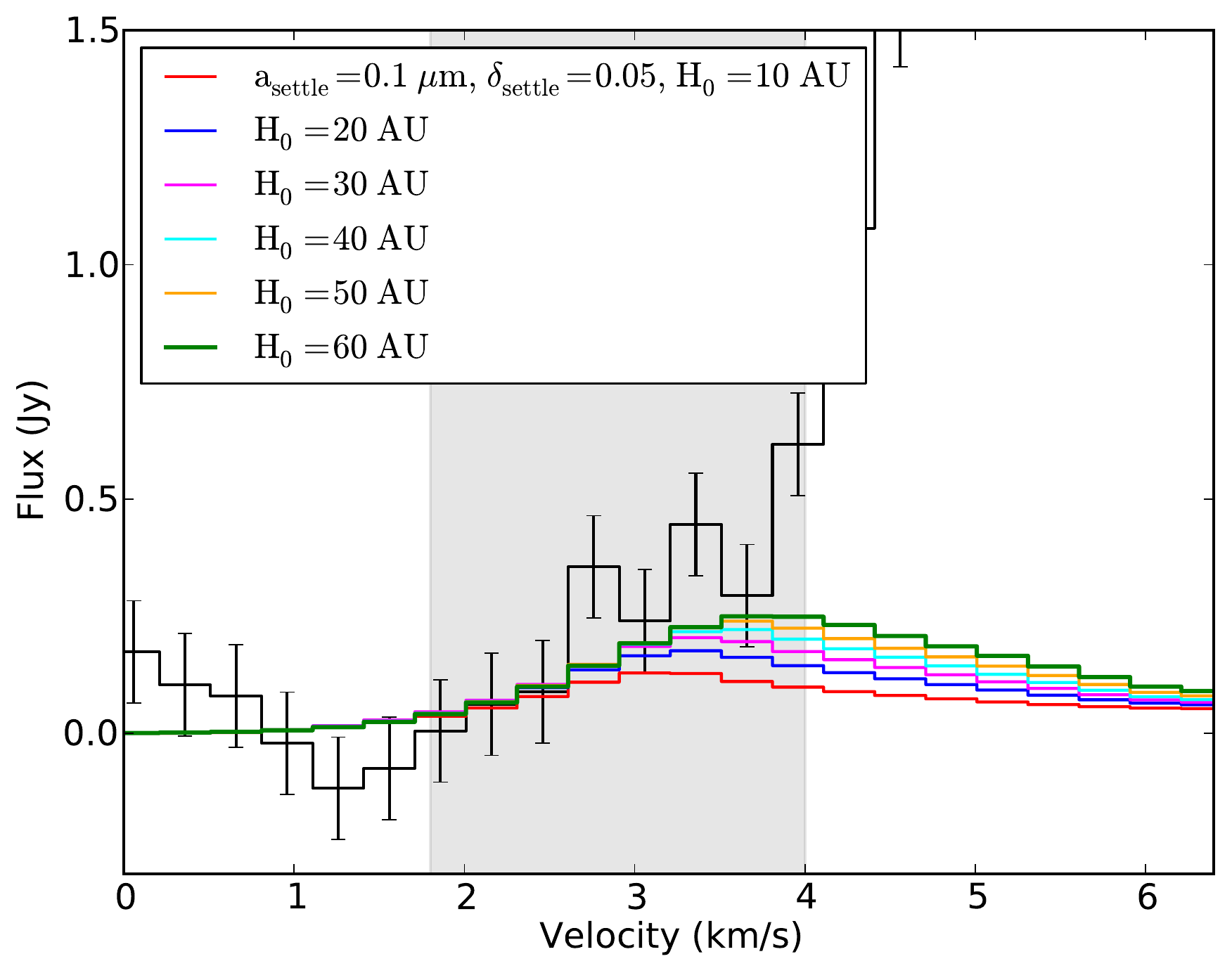}
\includegraphics[width=3in]{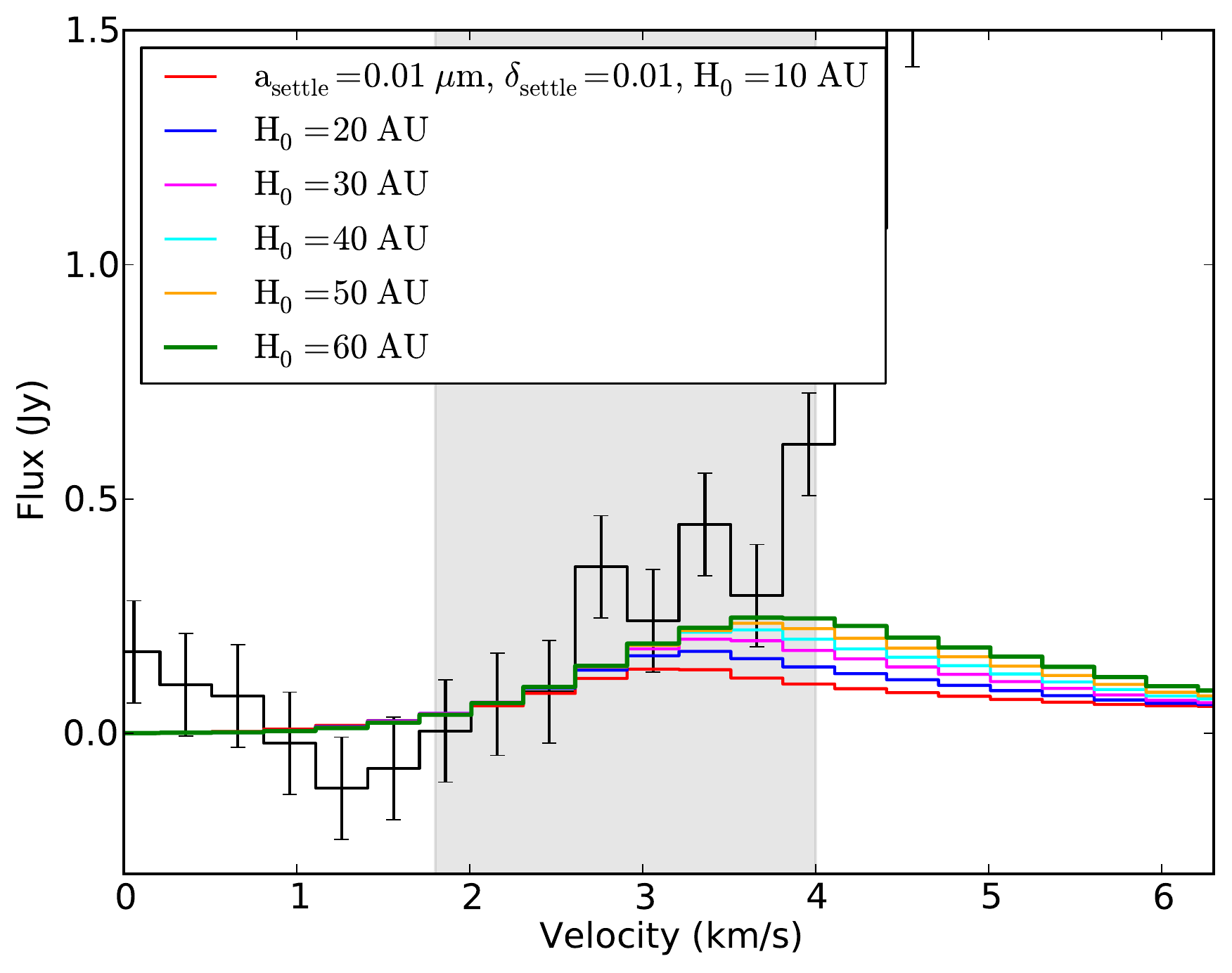}
\includegraphics[width=3in]{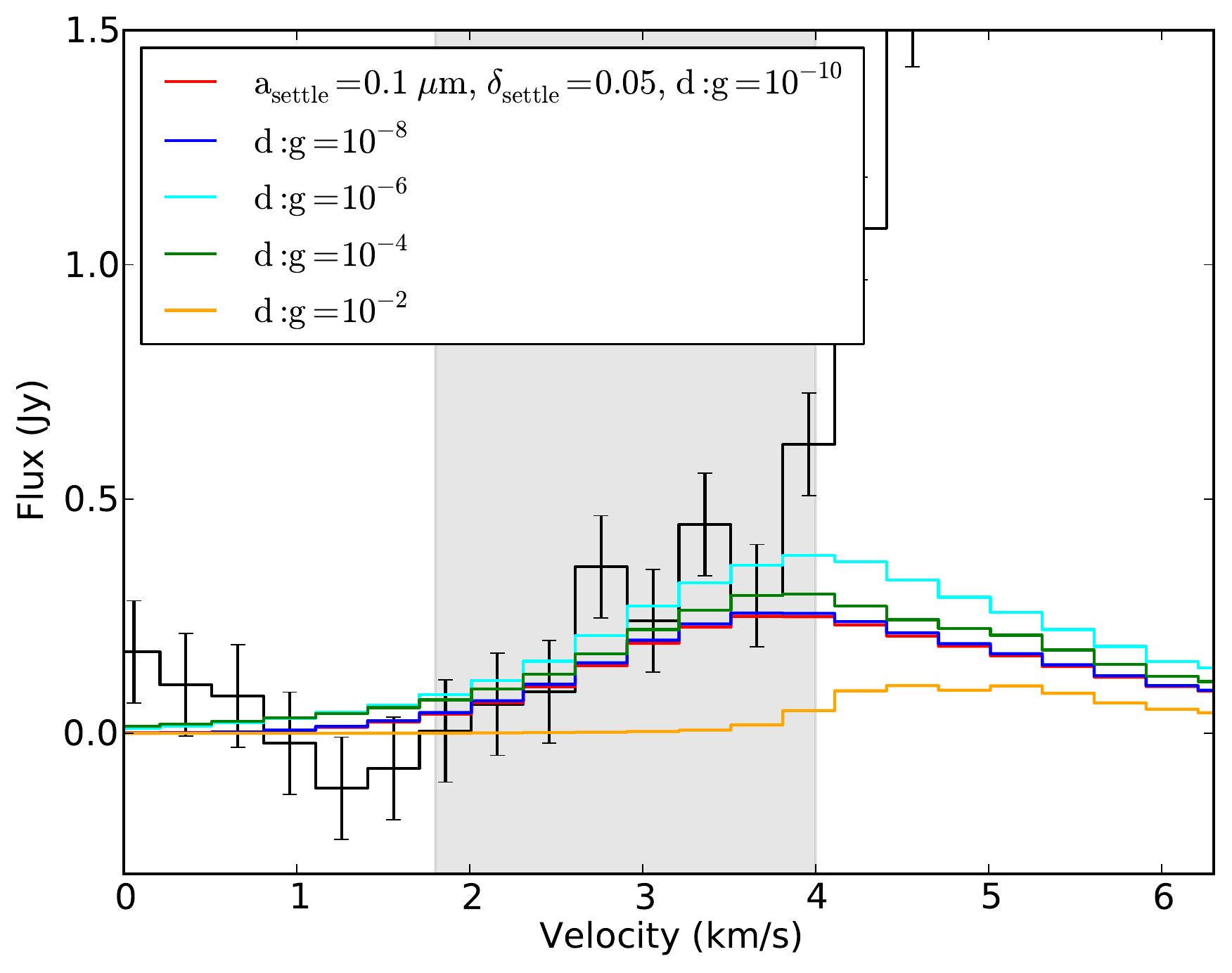}
\includegraphics[width=3in]{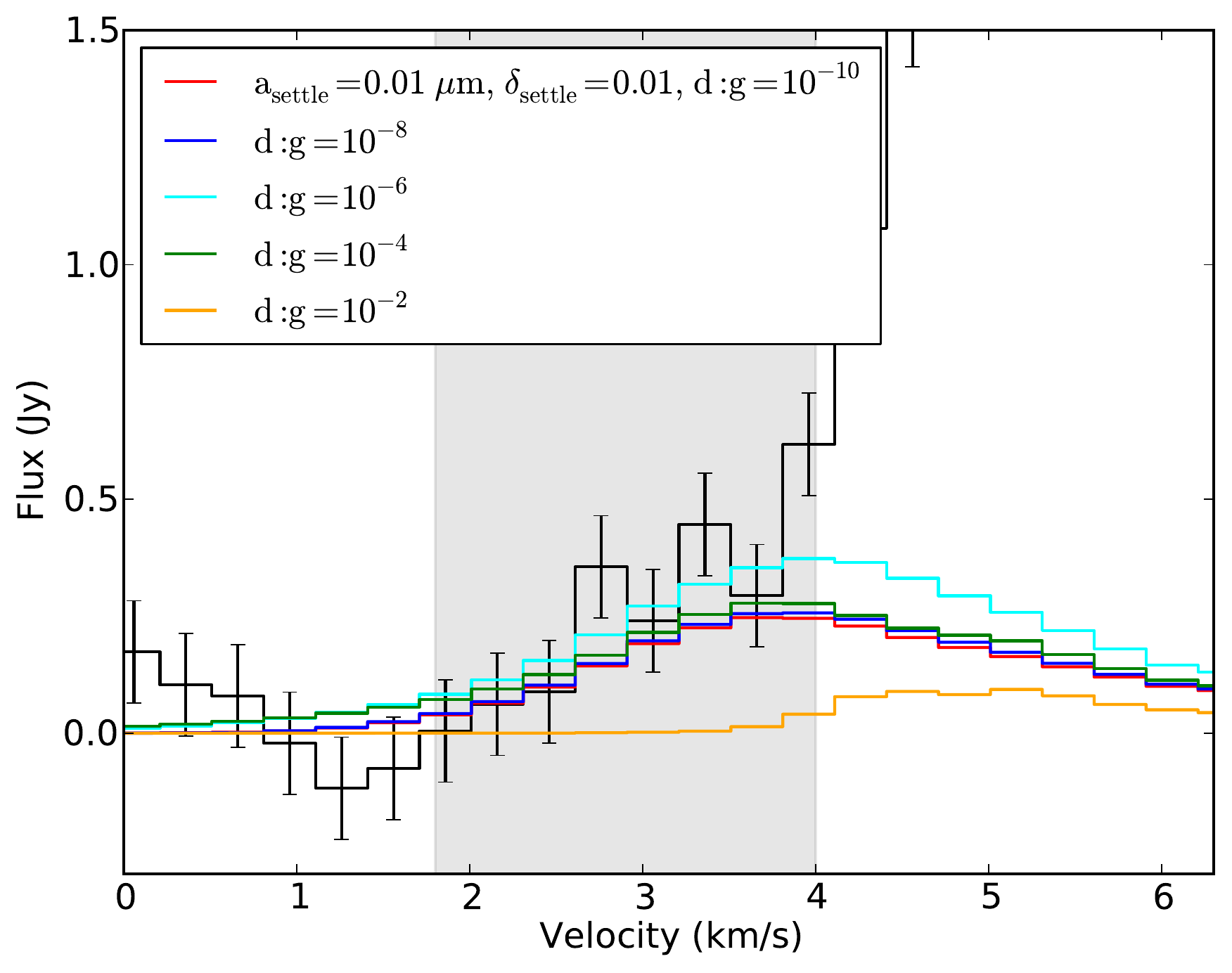}
\caption{{\emph{Top:}}  Inner cavity models with varying gas scale heights (see Section~\ref{varying_scale_height}) ranging from $H_0=10$ to 60~AU.  We find the HCO$^+$ line-wing flux increases with increase scale height.  {\emph{Bottom:}}  Varying d:g in the inner disk cavity from $10^{-10}$ to $10^{-2}$ with $H_0=60$~AU (Section~\ref{dust_to_gas}), where the best-fit inner cavity model has d:g$=10^{-6}$.  Settling parameters for the models are $a_s=0.1\micron$ and $\delta_s=0.05$ (left) and $a_s=0.01\micron$ and $\delta_s=0.01$ (right).}
\label{fig:dgratio_cavity_test}
\end{figure*}

\begin{sidewaysfigure}
\centering
\includegraphics[width=\textwidth]{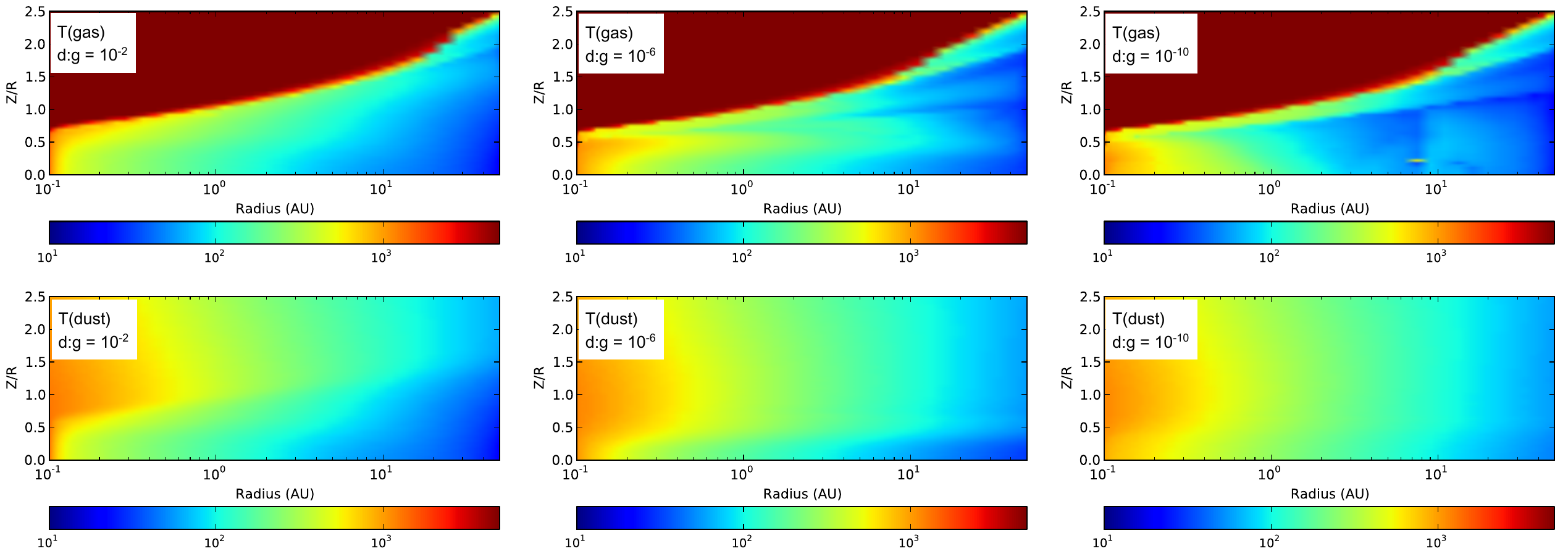}
\caption{Gas temperature (top) and dust temperature (bottom) in Kelvin for models in Section~\ref{dust_to_gas}.  Dust-to-gas ratios range from $10^{-2}$ (left), $10^{-6}$ (centre) and $10^{-10}$ (right) with gas scale height $H_0=60$~AU at reference radius $R_0=100$~AU. }
\label{fig:tempcompare}
\end{sidewaysfigure}

\begin{sidewaysfigure}
\centering
\includegraphics[width=8.5in]{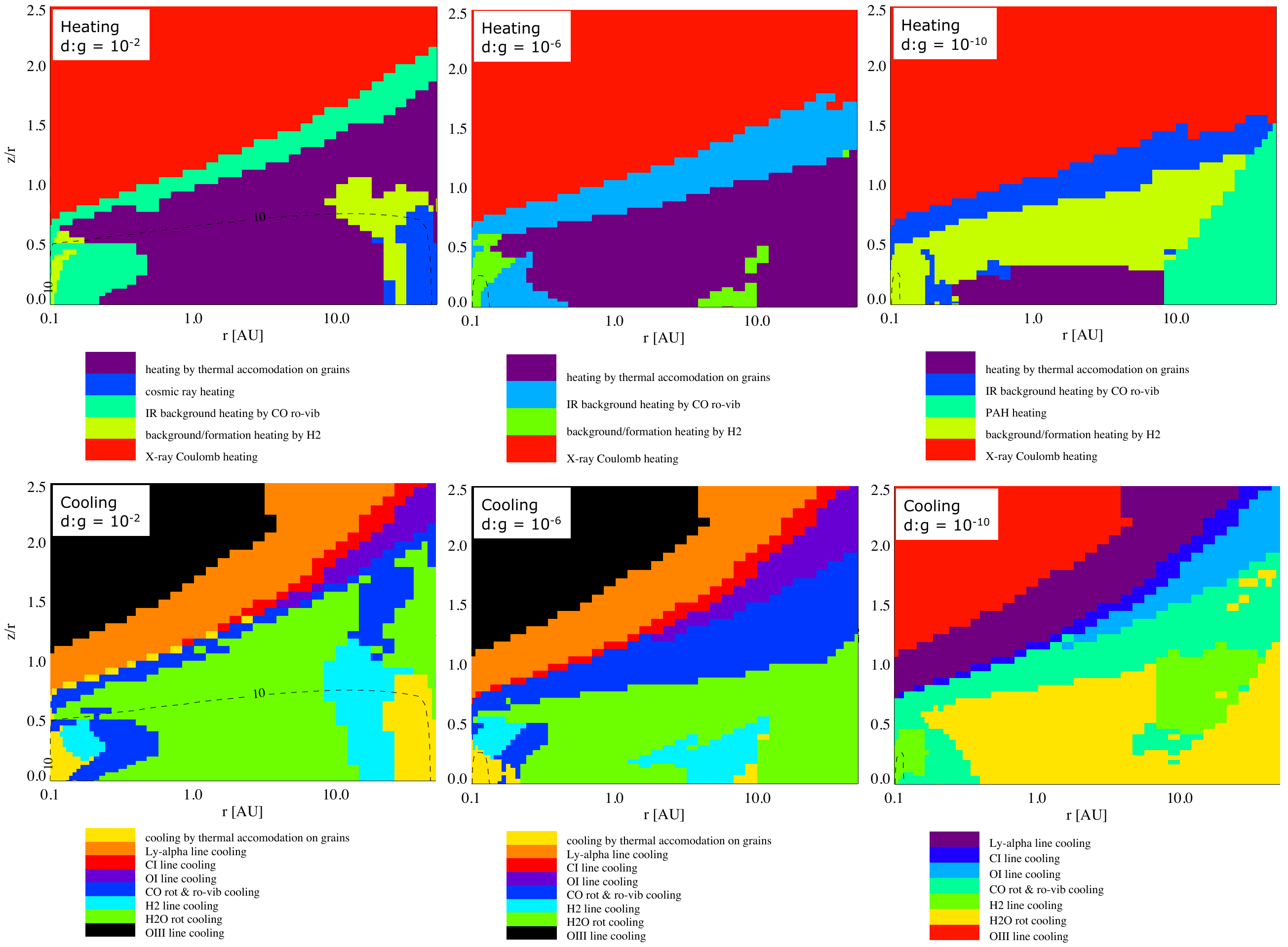}
\caption{Heating (top) and cooling (bottom) mechanisms for models in Section~\ref{dust_to_gas}.  Dust-to-gas ratios range from $10^{-2}$ (left), $10^{-6}$ (centre) and $10^{-10}$ (right) with gas scale height $H_0=60$~AU at reference radius $R_0=100$~AU. Note: Colour schemes vary for each individual plot.}
\label{fig:heating}
\end{sidewaysfigure}

From Figure~\ref{fig:dgratio_cavity_test}, we find models with dust-to-gas ratios $\leq10^{-4}$ show evidence of increased line-wings in the HCO$^+$ profile and have better $\chi_{red}^2$ values than d:g$=10^{-2}$.  Unlike the standard disk models from \S\ref{settling_effects} (with gas scale height $H_0=10$~AU),  the relationship between the dust and gas temperatures, molecular line densities and the dust-to-gas ratio is not straight-forward.  From Figure~\ref{fig:tempcompare}, we see the gas temperatures become colder at radii $\sim$10--50~AU with decreasing dust-to-gas ratio.  Figure~\ref{fig:heating} shows the modelled heating and cooling mechanisms in the disk.  For low dust-to-gas ratios, the dust is unable to absorb incoming stellar UV radiation and heat the gas, leading to lower gas temperatures towards the midplane of the disk.  The primary heating mechanisms with low d:g become background/formation by H$_2$ and PAH heating towards the midplane of the disk and X-ray Coulomb and IR background heating by CO r-vibrational lines towards the disk surface.

%{\bf UNCLEAR:} For low dust-to-gas ratios, the disk becomes colder primarily in regions typically heated by thermal accommodation on grains in the ISM d:g model.  Without the presence of grains, dust is unable to absorb the incoming stellar UV radiation and heat the gas.  The primary heating mechanisms with low d:g become background/formation by H$_2$ and PAH heating towards the midplane of the disk and X-ray Coulomb and IR background heating by CO r-vibrational lines towards the disk surface.

\begin{sidewaysfigure}
\centering
\includegraphics[width=8.5in]{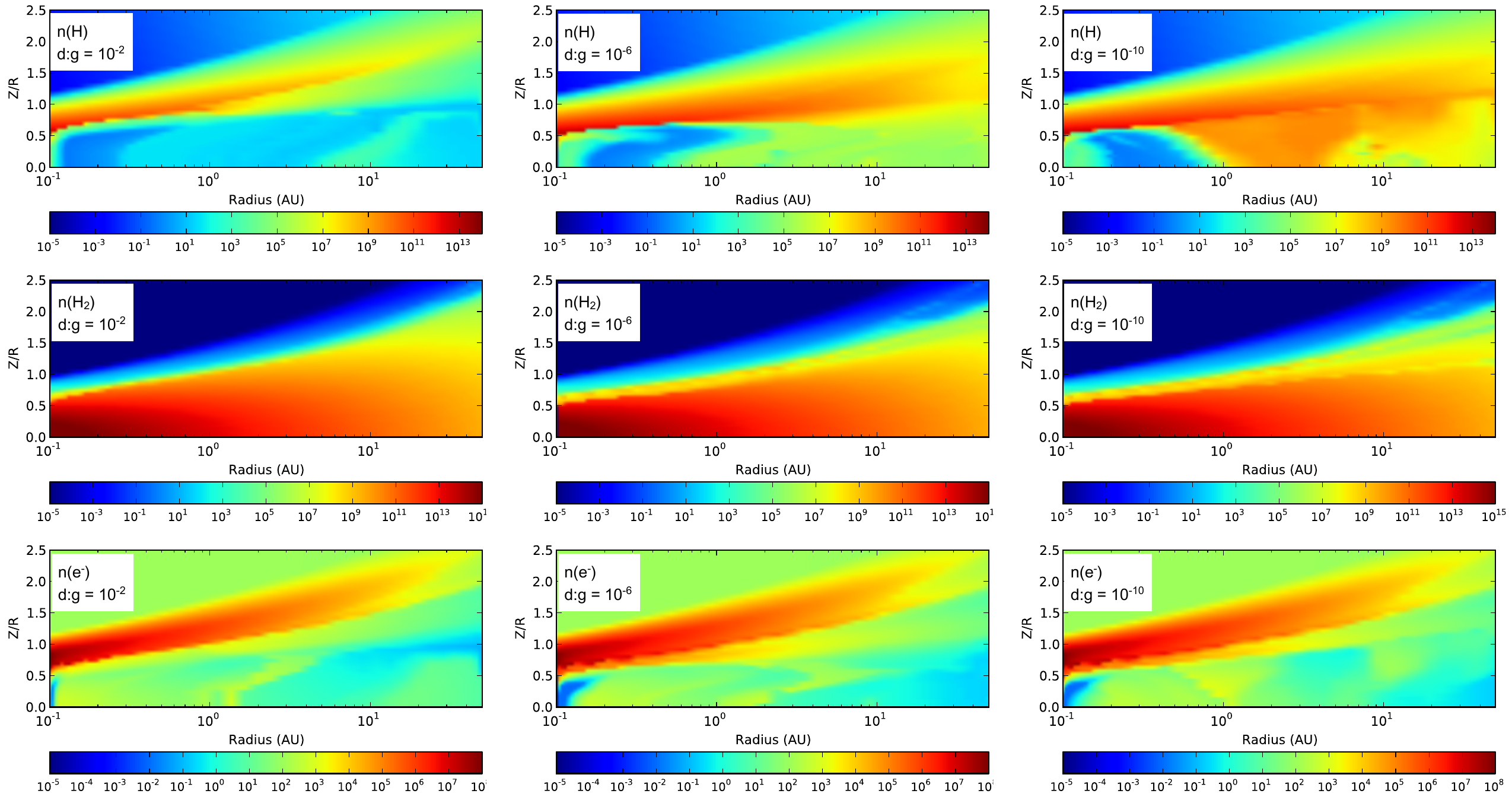}
\caption{Densities (cm$^{-3}$) for H (top), H$_2$ (centre) and e$^-$ (bottom) for models in Section~\ref{dust_to_gas}.  Dust-to-gas ratios range from $10^{-2}$ (left), $10^{-6}$ (centre) and $10^{-10}$ (right) with gas scale height $H_0=60$~AU at reference radius $R_0=100$~AU. }
\label{fig:h_h2_elec_compare}
\end{sidewaysfigure}

\begin{sidewaysfigure}
\centering
\includegraphics[width=8.5in]{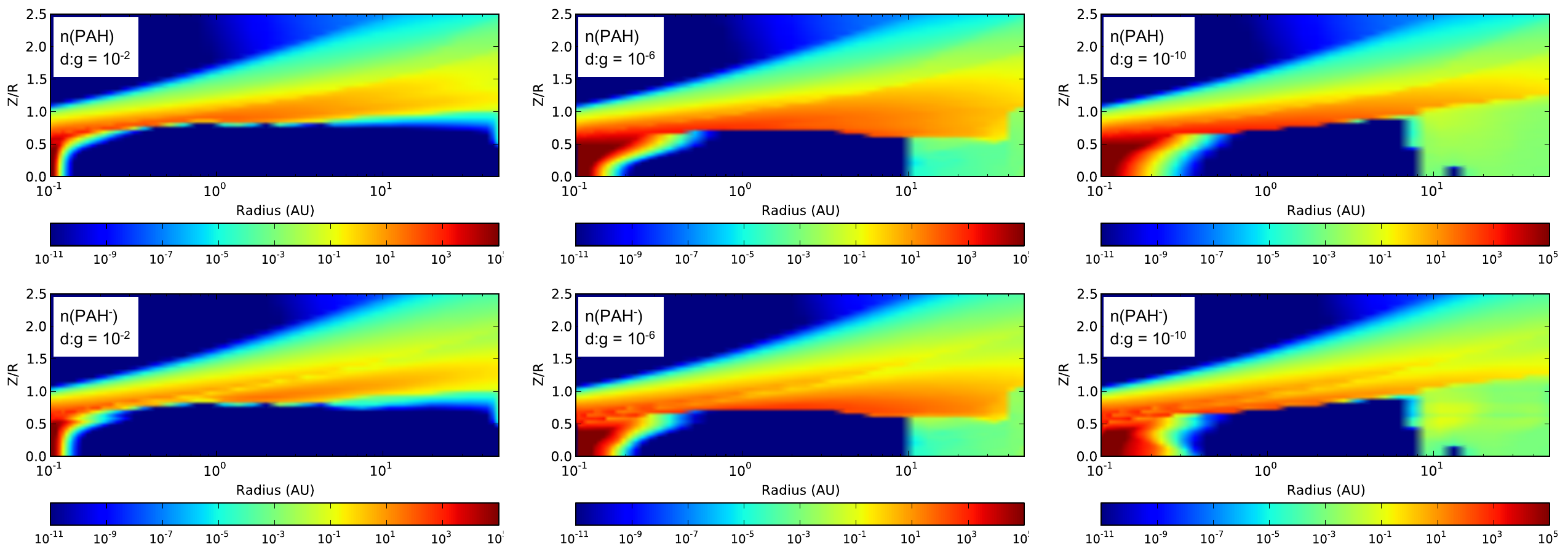}
\caption{Densities (cm$^{-3}$) for PAH (top) and PAH$^-$ (bottom) for models in Section~\ref{dust_to_gas}.  Dust-to-gas ratios range from $10^{-2}$ (left), $10^{-6}$ (centre) and $10^{-10}$ (right) with gas scale height $H_0=60$~AU at reference radius $R_0=100$~AU. }
\label{fig:pah}
\end{sidewaysfigure}

\begin{sidewaysfigure}
\centering
\includegraphics[width=8.5in]{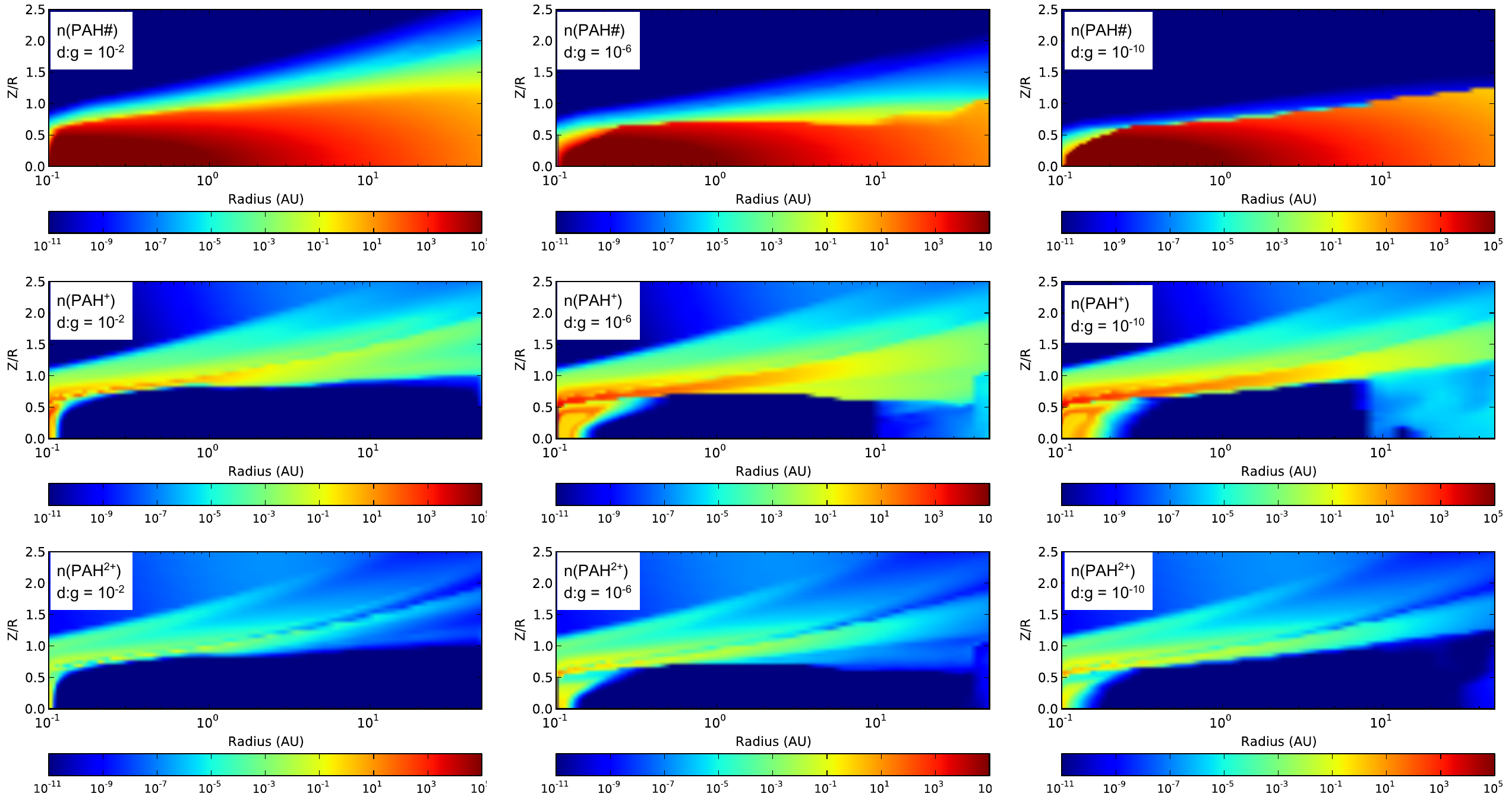}
\caption{Densities (cm$^{-3}$) for PAH\# (PAH ices; top), PAH$^+$ (centre) and PAH$^{2+}$ (bottom) for models in Section~\ref{dust_to_gas}.  Dust-to-gas ratios range from $10^{-2}$ (left), $10^{-6}$ (centre) and $10^{-10}$ (right) with gas scale height $H_0=60$~AU at reference radius $R_0=100$~AU. }
\label{fig:pah2}
\end{sidewaysfigure}

\begin{sidewaysfigure}
\centering
\includegraphics[width=8.5in]{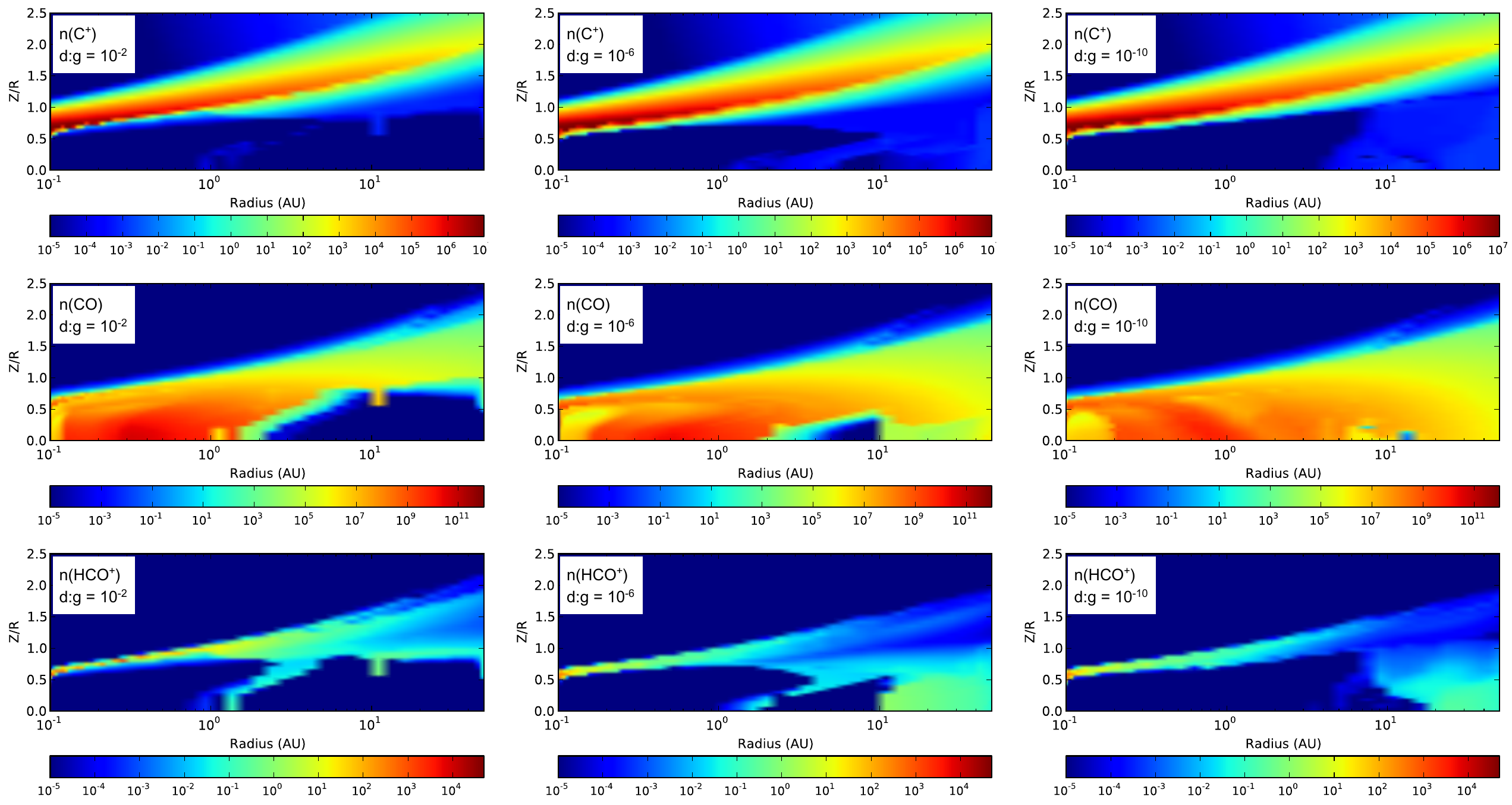}
\caption{Densities (cm$^{-3}$) for C$^+$ (top), CO (centre) and HCO$^+$ (bottom) for models in Section~\ref{dust_to_gas}.  Dust-to-gas ratios range from $10^{-2}$ (left), $10^{-6}$ (centre) and $10^{-10}$ (right) with gas scale height $H_0=60$~AU at reference radius $R_0=100$~AU. }
\label{fig:Cplus_CO_HCOplus_comparison}
\end{sidewaysfigure}

Figures~\ref{fig:h_h2_elec_compare}, \ref{fig:pah}, \ref{fig:pah2} and \ref{fig:Cplus_CO_HCOplus_comparison} show densities at d:g$=10^{-2}, \ 10^{-6}$ and $10^{-10}$ for H, H$_2$, electrons (e$^-$), C$^+$, CO, HCO$^+$, PAH, PAH$^-$, PAH ices (PAH\#), PAH$^+$ and PAH$^{2+}$.  With a smaller amount of dust, UV emission can penetrate further into the disk.  This process changes the disk chemistry by driving more photochemical reactions in the midplane and dissociating H$_2$ and CO in the upper portions of the disk.  This effect can be seen in our analysis, where models with lower dust-to-gas ratios have lower H$_2$, CO and PAH\# densities towards the disk surface and higher densities of H, C$^+$ and e$^-$.  In general, there are larger densities in the disk midplane at radii $\sim10$--50~AU as the d:g decreases, including densities for H, C$^+$, CO, PAH, PAH$^+$, PAH$^{2+}$ and PAH$^-$.  This supports the conclusion that the disk chemistry has been driven towards the midplane due to UV radiation.  HCO$^+$ density decreases at the surface of the disk, similar to CO and H$_2$, and tends to increase towards the midplane with decreasing dust-to-gas ratios.  However, HCO$^+$ has a slight decrease in density at the midplane for d:g$<10^{-6}$ at radii $\sim10$--50~AU.  Not only is the disk becoming colder as the d:g falls (i.e. $\leq40$~K), but the increased PAH$^-$ for the lower dust-to-gas ratio is likely destroying the HCO$^+$ molecule.  Taking these factors into account, we find the line-wing HCO$^+$ flux is maximised at a d:g$ = 10^{-6}$ where there is a balance between a high HCO$^+$ density and warm gas to produce line emission.

In \citet{2013A&A...559A..46B}, similar tests were done on the chemistry within the disk cavity, using a two-component cavity model with an inner disk and gap in dust.  The dust depletion factor was varied for the inner disk region for scenarios with an inner disk (i.e. a dust depletion $\delta_\mathrm{dust} = 10^{-5}$ w.r.t. the outer disk) and without an inner disk (i.e.  a dust depletion $\delta_\mathrm{dust} = 10^{-10}$).  Additionally, gas depletion factors were tested for the full cavity (both the inner disk and gap regions).  \citet{2013A&A...559A..46B} find that the dusty inner disk effectively shields the gas contained in the disk gap from UV emission, allowing for molecules like CO and H$_2$ to form at higher heights in this region.  When dust is depleted from the inner disk (i.e. without a dusty inner disk), both CO and H$_2$ are photodissociated at these heights and only exist in the midplane of the disk.  This is in agreement with the chemistry observed when we vary the dust-to-gas ratio in the full disk cavity, where dust acts as a shield for the gas contained in the disk.  

Since the dust settling and mixing parameters do not significantly affect the HCO$^+$ emission in the cavity as discussed in the above sections, cavity models in Sections~\ref{cavity_fits}--\ref{final_model} only have settling parameters $a_s=0.01~\micron$ and $\delta_s= 0.01$. 

\begin{table}
\centering
\caption{FUV photo processes with continuous cross sections.}
\begin{tabular}{c c c c c | c c}
\hline
\multicolumn{5}{c}{Reaction} & $\lambda_\mathrm{thr}$ [${\buildrel _{\circ} \over {\mathrm{A}}}$]  & $E_\mathrm{thr}$ [eV] \\
\hline

C & $\rightarrow$ & C$^+$ & $+$ & e$^-$ & 1102.0 & 11.25 \\
S & $\rightarrow$ & S$^+$ & $+$ & e$^-$ & 1197.0 & 10.35 \\
Si & $\rightarrow$ & Si$^+$ & $+$ & e$^-$ & 1522.0 & 8.14 \\
Fe & $\rightarrow$ & Fe$^+$ & $+$ & e$^-$ & 1576.0 & 7.86 \\
Mg & $\rightarrow$ & Mg$^+$ & $+$ & e$^-$ & 1630.0 & 7.60 \\
Na & $\rightarrow$ & Na$^+$ & $+$ & e$^-$ & 2413.0 & 5.13 \\
H$^-$ & $\rightarrow$ & H & $+$ & e$^-$ & 16420.0 & 0.75 \\
OH$^+$ & $\rightarrow$ & O$^+$ & $+$ & H & 2600.0 & 4.76 \\
CH$_4$ & $\rightarrow$ & CH$_3$ & $+$ & H & 1061.0 & 11.68 \\
CH$_4$ & $\rightarrow$ & CH$_2$ & $+$ & H$_2$ & 1061.0 & 11.68 \\
CH$_4$ & $\rightarrow$ & CH$_4 ^+$ & $+$ & e$^-$ & 980.0 & 12.65 \\
OH & $\rightarrow$ & O & $+$ & H & 1950.0 & 6.35 \\
CH & $\rightarrow$ & C & $+$ & H & 3650.0 & 3.39 \\
CH & $\rightarrow$ & CH$^+$ & $+$ & e$^-$ & 1200.0 & 10.33 \\
CH$^+$ & $\rightarrow$ & C & $+$ & H$^+$ & 3800.0 & 3.26 \\
O2 & $\rightarrow$ & O & $+$ & O & 1770.0 & 7.00 \\
O2 & $\rightarrow$ & O2$^+$ & $+$ & e$^-$ & 1008.0 & 12.30 \\
H$_2 ^+$ & $\rightarrow$ & H$^+$ & $+$ & H & 3000.1 & 4.13 \\
H$_3 ^+$ & $\rightarrow$ & H$_2 ^+$ & $+$ & H & 912.0 & 13.59 \\
H$_3 ^+$ & $\rightarrow$ & H$_2$ & $+$ & H$^+$ & 912.0 & 13.59 \\
SiH$^+$ & $\rightarrow$ & Si$^+$ & $+$ & H & 3816.0 & 3.24 \\
CH$_2$ & $\rightarrow$ & CH & $+$ & H & 2750.0 & 4.50 \\
H$_2$O & $\rightarrow$ & OH & $+$ & H & 2050.0 & 6.04 \\
H$_2$O & $\rightarrow$ & O & $+$ & H$_2$ & 1300.0 & 9.53 \\
H$_2$O & $\rightarrow$ & H$_2$O$^+$ & $+$ & e$^-$ & 985.0 & 12.50 \\
CO$_2$ & $\rightarrow$ & CO & $+$ & O & 1767.0 & 7.01 \\
NH & $\rightarrow$ & N & $+$ & H & 1700.0 & 7.29 \\
NH$_3$ & $\rightarrow$ & NH$_3 ^+$ & $+$ & e$^-$ & 1220.0 & 10.16 \\
CN & $\rightarrow$ & N & $+$ & C & 1150.0 & 10.78 \\
NO & $\rightarrow$ & O & $+$ & N & 1051.0 & 11.79 \\
NO & $\rightarrow$ & NO$^+$ & $+$ & e$^-$ & 1350.0 & 9.18 \\
H$_2$CO & $\rightarrow$ & CO & $+$ & H$+$H & 1505.0 & 8.23 \\
H$_2$CO & $\rightarrow$ & CO & $+$ & H$_2$ & 1505.0 & 8.23 \\
H$_2$CO & $\rightarrow$ & HCO$^+$ & $+$ & H & 1505.0 & 8.23 \\
PAH$^-$ & $\rightarrow$ & PAH & $+$ & e$^-$ & 3995.1 & 3.10 \\
PAH & $\rightarrow$ & PAH$^+$ & $+$ & e$^-$ & 1985.8 & 6.24 \\
PAH$^+$ & $\rightarrow$ & PAH$^{2+}$ & $+$ & e$^-$ & 1321.3 & 9.38 \\
PAH$^{2+}$ & $\rightarrow$ & PAH$^{3+}$ & $+$ & e$^-$ & 990.0 & 12.52 \\
\hline
\end{tabular}
\label{table_dust_opacity}
\end{table}

\end{document}